\documentclass[apj]{emulateapj}
\usepackage{lscape}
\usepackage{epsfig}

\def\arcsec{$^{\prime\prime}$}
\def\arcmin{$^{\prime}$}

\shorttitle{Spectroscopy of Globular Clusters in NGC 5128}
\shortauthors{Woodley et al.}
\begin{document}

\title{The Ages, Metallicities and Alpha Element Enhancements of Globular Clusters in the
  Elliptical NGC 5128:  A Homogeneous Spectroscopic Study with Gemini/GMOS}

\author{Kristin A.~Woodley\altaffilmark{1,2}, 
William E.~Harris\altaffilmark{1},
Thomas H. Puzia\altaffilmark{3,4},
Mat{\'i}as G{\'o}mez\altaffilmark{5}, 
Gretchen L. H. Harris\altaffilmark{6},
Doug Geisler\altaffilmark{7}
}
\altaffiltext{1}{Department of Physics \& Astronomy, McMaster
University,
  Hamilton ON  L8S 4M1, Canada; harris@physics.mcmaster.ca}
\altaffiltext{2}{Current Address: Department of Physics \& Astronomy,
  University of British Columbia, Vancouver, BC V6T 1Z1, Canada; kwoodley@phas.ubc.ca}
\altaffiltext{3}{Herzberg Institute of Astrophysics, 5071 West Saanich Road, Victoria, BC V9E 2E7, Canada; thomas.puzia@nrc.ca}
\altaffiltext{4}{Plaskett Fellow}
\altaffiltext{5}{Departamento de Ciencias Fisicas, Facultad de Ingenieria, Universidad Andres Bello, Chile; matiasgomez@unab.cl}
\altaffiltext{6}{Department of Physics and Astronomy, University of Waterloo, Waterloo, ON N2L 3G1, Canada; glharris@astro.uwaterloo.ca}
\altaffiltext{7}{Departamento de Astronom{\'i}a, Universidad de Concepci{\'o}n, Chile; dgeisler@astro-udec.cl}


\begin{abstract}
We present new integrated light  spectroscopy of globular clusters in NGC
5128, a nearby giant elliptical galaxy  less than 4 Mpc away, in order
to  measure  radial velocities  and  derive  ages, metallicities,  and
alpha-element  abundance  ratios.   Using  the  Gemini  South  8-meter
telescope with  the instrument GMOS,  we obtained spectroscopy  in the
range  of $\sim 3400-5700$  $\rm{\AA}$ for  72 globular  clusters with
signal-to-noise  greater  than 30  $\rm{\AA}^{-1}$  and  we have  also
discovered 35  new globular clusters  within NGC 5128 from  our radial
velocity measurements.  We measured and compared the Lick indices from
H$\delta_A$ through  Fe5406 with the single  stellar population models
of \cite{tmb03}  and \cite{tmk04} in order to  derive age, metallicity
and [$\alpha$/Fe] values.   We also measure Lick indices  for 41 Milky
Way globular  clusters from \cite{puzia02}  and \cite{schiavon05} with
the  same methodology for  direct comparison.   Our results  show that
68$\%$ of  the NGC  5128 GCs have  old ages  ($> 8$ Gyr),  14$\%$ have
intermediate ages ($5-8$ Gyr), and 18$\%$ have young ages ($< 5$ Gyr).
However, when we look at the metallicity of the globular clusters as a
function  of age,  we  find 92$\%$  of  metal-poor GCs  and 56$\%$  of
metal-rich GCs  in NGC 5128 have  ages $> 8$ Gyr,  indicating that the
majority  of  both  metallicity  subpopulations of  globular  clusters
formed early,  with a significant  population of young  and metal-rich
globular  clusters forming later.
Our   metallicity  distribution   function  generated   directly  from
spectroscopic  Lick  indices  is  clearly  bimodal, as  is  the  color
distribution of the same set  of globular clusters. Thus the  metallicity bimodality  is real and  not an artifact  of the
color to metallicity conversion. However, the metallicity distribution
function obtained  from comparison with the  single stellar population
models is  consistent with a unimodal, bimodal, or multimodal shape.  The
[$\alpha$/Fe]   values   are  supersolar   with   a   mean  value   of
$0.14\pm0.04$, indicating a fast  formation timescale.  However, the GCs
in NGC 5128 are not as [$\alpha$/Fe] enhanced  as the Milky Way GCs also examined in
this  study.  Our  measured indices  also indicate  that  the globular
clusters in NGC 5128 may  have a slight overabundance in nitrogen and
a wider range  of calcium strength compared to  the Milky Way globular
clusters.  Our results support a rapid, early formation of the
globular cluster system in NGC 5128, with subsequent major accretion
and/or GC and star forming events in more recent times.
\end{abstract}


\section{Introduction}
\label{sec:intro}

The  star formation history  of early-type  galaxies is  quite complex
with evidence supporting episodes of star formation at both high
and low redshifts.  At high redshifts (z $> 2$),  the formation of the
bulk of the  galactic stellar component is supported  by, for example,
the fundamental  plane relation \citep{djorgovski87,dressler87,treu99}
in          the          monolithic         collapse          scenario
\citep{eggen62,tinsley72,larson74,larson75,silk77,arimoto87}.       The
hierarchical                      merging                     scenario
\citep{toomre77,white78,peebles80,whitefrenk91,kauffmann93,baugh98,cole00,somerville01}
on the  other hand, predicts a  range of formation  timescales for the
stellar component down to much  lower redshift values.  In the extreme
case,  gas-rich  merging galaxies  are  seen  today  where young  star
clusters,  for  example in  NGC  4038/4039 \citep{whitmore95},  and/or
young  and  intermediate  aged   globular  clusters  (GCs)  have  been
detected, indicating  that GCs, as well  as a fraction  of the stellar
component of the galaxy can form at z $\leq 1$.

The  key to  understanding  the  formation epochs  of  these types  of
galaxies  is through  their star  formation history.   Most early-type
galaxies are  too distant to resolve  stars from which  we can measure
their ages  directly. Unresolved light studies are  difficult as well,
as they  are hindered  by the age-metallicity  degeneracy \citep[][and
references  therein]{mould80} as  well as  internal extinction  in the
galaxies.

GCs, on  the other hand,  are essentially coeval structures  that form
with nearly single  age and single metallicity.  They  have been used,
therefore,  as powerful  tools  to trace  the  complex star  formation
history of  early-type galaxies.   Their internal properties,  such as
ages  and metallicities  are used  as representations  of  the overall
stellar component of the galaxy.  We  also benefit from the use of GCs
as   they   do   not   suffer   from   problems   involving   internal
reddening.   GCs can be used as good tracers of the time period  of
early-type galaxy formation because we can determine their ages,
metallicities, and alpha-enhancements through their spectra.  By
examining the ratio of GCs that are old
to GCs with younger ages, as well as
comparing the age difference between metal-rich and
metal-poor GCs, we can determine which formation scenario is dominant.

Previous studies using photometry  to determine ages and metallicities
of  GCs are  plagued with  age-metallicity degeneracies,
especially in the optical regime \citep{wortheyg94}.  
One way  to avoid the  age-metallicity degeneracy using  photometry is
through the use of (near-)infrared  data, which is mainly sensitive to
the metallicity of the red  giant branch, combined with optical data, sensitive
to both age  and metallicity.  This technique
has been shown to be effective at  distinguishing between  GC populations
that   appear  to   be  single-aged   and  old  such as NGC  7192
\citep{hempel03},   NGC    1399   \citep{kundu05},   and    NGC   3115
\citep{puzia02}, versus those with a wide spread of ages  such as NGC    4365
\citep{puzia02,kundu05} and  NGC 5846 \citep{hempel03, hempel07}.   
However, a sample of GCs in NGC 4365 with
intermediate ages were found
to have old ages using spectroscopic techniques \citep{brodie05}.
These color indices are also not precise enough to be useful for individual GCs
or to derive genuine age distribution functions.

Spectroscopy allows us  to determine the internal properties
of  individual GCs  as  long  as we  have  high signal-to-noise  data.
The
measurement  of  the  strength  of  absorption line  features  in  the
spectra enables us to obtain ages,
metallicities, and [$\alpha$/Fe] through comparison with SSP  models.
This technique enables the  collection of a
significant fraction of both metal-rich  and metal-poor GC data in one
environment that can show the formation epoch(s) and
chemical composition at the time of formation of the GCs
within one early-type galaxy.

The proximity of NGC 5128 makes this early-type galaxy extremely
attractive to conduct a large, homogeneous age and  metallicity study
\citep[$3.8\pm0.1$  Mpc][]{harris09}.  
Features within this galaxy (the  dust lane and faint shells
\citep{malin78}, visible  star formation  from the interaction  of the
radio jet  with the shells of  HI \citep{graham98}, and  a young tidal
stream in the  halo \citep{peng02}) are recent phenomena which will
have little bearing on the much older GC population.

Previous studies have been able to obtain spectroscopy for GCs in many
different galaxies  to determine  global properties of  the metal-poor
versus             metal-rich clusters       \citep[][among
others]{puzia05,strader05,sharina06}.  Within  NGC 5128, we  should be
able to determine whether there  is a significant fraction of young or
intermediate-aged GCs,  or if the majority are  old, similar to
their Milky Way  GC counterparts.  Our primary  goal is to  use a large
and high signal-to-noise (S/N) sample of  GC spectra within NGC 5128  to obtain the  relative age difference
between  the metal-rich  and  the metal-poor  clusters.  Currently, we
have the advantage of knowing over 415 GCs in NGC 5128 which have been
confirmed with  either radial  velocity measurement and/or  HST images
cataloged in  \cite{woodley07}.  This  catalog has provided  the
database from which we selected our target objects.

We present our results in the sections below:  section \ref{sec:data} describes our data set and
its reduction; section \ref{sec:rv} describes  the radial velocity
measurements  obtained  and   the  confirmation  of  newly  discovered
GCs; section   \ref{sec:results}   describes   the
calibration of  our data and our resulting
ages, metallicities,  and [$\alpha$/Fe]; section \ref{sec:dis}
discusses our findings; and we conclude
in Section \ref{sec:conclusions}.

\section{Data Sample and Reduction Techniques}
\label{sec:data}

\subsection{Gemini/GMOS Dataset}
\label{sec:dataset}
We  obtained low-resolution  spectroscopy of  known and  candidate GCs
with  the Gemini  South 8-meter  telescope  using the Gemini Multi-Object Spectrograph (GMOS).   We placed  8
fields around the center of NGC 5128, covering the inner 15 kpc of the
halo over two  observing periods, program IDs GS-2005A-Q-28 (PI:
William E.   Harris) and GS-2007A-Q-55  (PI: Doug Geisler).   The GMOS
instrument  allows us to  obtain spectroscopy  of multiple  objects in
each 5.5\arcmin\ field of view.  The 8 masks each contained between 18
and  25  objects observed  through  0.5\arcsec\  slits, totalling  176
objects (see Table~\ref{tab:gmos}).  Typical exposures were taken
in multiples of 1800 second intervals,  totalling $3-3.75$  hours per  field.  The
central wavelengths  were split among  447 nm, 450  nm, and 453  nm in
order to interpolate across the  chip gaps between the CCDs.  The mean
wavelength coverage  of the data is $\sim  3400-5700 \rm{\AA}$, safely
including the  major absorption features, in the  blue region of
the spectra  from Ca H+K  to redward of  Mgb, for the majority  of our
target  objects.   All  exposures  used  no  filter  and  the  grating
B600+G5232 with a blaze wavelength of 461 nm.  
The 2005 data were binned 4 times in the spectral and 2 times
in the spatial  direction and a spectral resolution of 
$1.8\rm{\AA}$/px.  The 2007 data were binned 2 times in the spectral
and  unbinned  in  the  spatial  directions and a spectral resolution
of $0.9 \rm{\AA}$/px.  A gain of 2.07  e$^-$/ADU and a readout noise of 3.69
e$^-$ for all data were chosen.

Each science exposure was bracketed by flat field exposures and a CuAr
calibration arc  with the  same configuration and  central wavelength.
However, the  2005 data had arcs taken  only at 450 nm,  so we shifted
the  arcs  using  the  $5577  \rm{\AA}$  night-sky  emission  line  to
calibrate the  447 and 453  nm exposures.  The biases  were downloaded
from the Canadian Astronomy Data Centre in the Gemini archive and were
taken within the same month as each science exposure.

\subsection{Data Reduction}
\label{sec:datareduction}

The   data  reduction  was   completed  with the Gemini package in IRAF\footnote{IRAF  is
  distributed by  the National Optical  Astronomy Observatories, which
  is  operated by  the  Association of  Universities  for Research  in
  Astronomy,  Inc.,  under  cooperative  agreement with  the  National
  Science Foundation}.  Initially, the data were prepared with the task
  {\it  gprepare} followed  by the  bias and  trimming of  the science
  fields  and flat  fields as  well as  overscanning and  trimming the
  calibration  arcs ({\it gsreduce}).   The  3 CCD  chips of  the
  calibration arcs were mosaiced ({\it gmosaic}) and the slits were
  individually cut ({\it gscut}).  The calibration  arcs were then
  used  to calibrate  the  data from  pixel-to-wavelength using  known
  spectral  features ({\it gswavelength})  and  the flat
  fields  were  prepared by  fitting  the  response  functions of  the
  detectors in each CCD chip ({\it gsflat}).  A bad pixel mask was
  created for each CCD chip which replaced the dead pixels and the hot
  pixels found in the flat fields and bias frames, respectively 
({\it imreplace}).  The masks were then used to replace the bad pixels
  in  the science and  flat fields ({\it  fixpix}). The  task {\it
  gsreduce} was used  again to flat field, interpolate  over the gaps,
  and mosaic  the 3 CCD chips  in the science spectra.   A task called
  {\it mosproc}, written by Bryan  W.  Miller, and explained in detail
  in  \cite{trancho07}   was  then  used  to   perform  the  remaining
  tasks. The cosmic ray subtraction in {\it mosproc} was done with the
  \cite{vandokkum01}  Laplacian edge  detection  algorithm.  The  task
  {\it gstransform} was then used to rectify and apply the calibration
  to each spectra  which were then traced, background
  sky subtracted, and then extracted into one-dimensional spectra
  with {\it  gsextract}.  Lastly, a quantum  efficiency correction was
  applied to the data ({\it gecorr}).

Following  the  pre-reduction described  above,  on  every exposure  we
isolated   the  segments  containing   each  Lick-index   feature  and
surrounding continuum.  We  used  20  Lick  features  from
H$\delta_A$  to Fe5406.   Each index  was then  cross-correlated using
{\it fxcor} in the IRAF package {\it rv} with a template spectrum of a
globular cluster in M31, 158-213 \citep{puzia02}, which was shifted to
a velocity of 0 km  s$^{-1}$.  This yielded a measured radial velocity
for  each of  the 20  features in  each exposure  (to be  discussed in
greater detail in Section~\ref{sec:rv}).  We then flux calibrated each
individual  science  spectrum  using  standard  stars  taken  for  our
observations and obtained  from the CADC.   Our standard stars  were EG21
for the 2005 data and LTT4364 for the 2007 data.  The sensitivity file
as a function of wavelength was generated for the standard stars ({\it
gsstandard}) and  applied to the science spectra  with {\it calibrate}
which also corrected for  extinction.  Following the flux calibration,
each  individual exposure  was then  shifted  to 0  km s$^{-1}$  ({\it
dopcor}) and then sum combined appropriately ({\it scombine}) for each
GC.

\section{Radial Velocity Measurements and Newly Confirmed Globular Clusters}
\label{sec:rv}

We used any spare  space on our masks not occupied by a known  GC in NGC 5128 
to search  for previously unidentified clusters.   Our candidate GCs  were selected
from  color, magnitude,  and  point spread  function subtraction  from
Gemini pre-images for the 2005 dataset  ({\it i} filter) and from $1.2 \times
1.2$ deg$^2$  Magellan/Inamori Magellan Areal  Camera and Spectrograph
(IMACS)  images   ({\it  R}   filter)  taken  in   0.5\arcsec\  seeing
\cite[Harris  et  al.  2009,  in  preparation,  and described  briefly
in][]{gomez07} for  the 2007 dataset.  At  the low
galactic latitude of NGC 5128 ($b=19^o$), we expect contamination from
both background  galaxies and foreground  stars in the Milky  Way that
can have colors and magnitudes expected for normal GCs, but a powerful
way to  securely classify candidate  objects as GCs is  through radial
velocity measurements.   It has  been shown in  previous spectroscopic
studies  that  GCs in  NGC  5128, have typical  $v_r  = 200-1000$  km
s$^{-1}$ and are thus distinguishable  from foreground  stars,  $v_r \lesssim
200$ km  s$^{-1}$ and  background galaxies, $v_r  > 1000$  km s$^{-1}$
\cite[see][for examples]{peng04b,woodley05,beasley08} at a very high level of confidence.

We  obtained spectra  for 101  known GCs (13 of  which  were measured
twice)  and 60  new candidate  GCs.  Of  our candidates,  35
satisfied the velocity  criterion  of $v_r  = 200-1000$  km
s$^{-1}$ with two  of these clusters measured twice.
Of the remaining  candidates, we ended up with  18 stars, 5 background
galaxies,  and  2 objects  for  which we  were  unable  to obtain  any
correlation  with  our template  GC.   It  is  important to  note  the
difference in confirmation of candidates as genuine GCs from the 2005
and 2007 observations.  The
typical selection  of GCs from  GMOS pre-images with  1\arcsec\ seeing
yielded a $56\%$  hit rate of confirmed GCs.  In 2007, we used high
quality IMACS images for candidate selection.  With the higher spatial
resolution of these images, taken in excellent
seeing (0.5\arcsec),  we achieved an $85\%$ hit  rate.

The  S/N of  the  GC  data,  measured between  $4700
 \rm{\AA}    -  4830    \rm{\AA}$,  ranged   from  
$\sim   10   -  200 \rm{\AA}^{-1}$,  with a  mean  and median of  
$56 \rm{\AA}^{-1}$ and $44 \rm{\AA}^{-1}$  for the 2005 data and 34
$\rm{\AA}^{-1}$  and 32 $\rm{\AA}^{-1}$    for  the   2007 data,
respectively.     Tables~\ref{tab:rv_known}     \&
~\ref{tab:rv_new} list  the results for  the previously known  GCs and
the  35  newly  confirmed  GCs.  The  tables  list  the
cataloged name  of the  GC, the  old literature name  of the  GCs (for
previously known  GCs), the  {\it R.A.} and  {\it Decl.} in  the J2000
epoch, the mask/field  number and slit number (designated  by m and s,
respectively), the approximate S/N per $\rm{\AA}$ measured around 4765
$\rm{\AA}$,    and   the   radial    velocity   obtained    from   our
cross-correlation technique.   For Table~\ref{tab:rv_known}, the final
column  is  the  weighted  mean  of all  previously  published  radial
velocity measurements for that  GC {\it including} the radial velocity
determined    in    this   study.     The    remaining   columns    in
Table~\ref{tab:rv_new} are  the C, M, and  T$_1$ Washington photometry
with associated uncertainties from \cite{harris04}.

Figure~\ref{fig:vel}  shows the  radial velocity  distribution  of our
known and newly confirmed GCs. The  duplicate velocities obtained
for the GCs follow a 1:1 correlation quite closely with an rms error $\pm32$ km
s$^{-1}$  from the  least  squares fit.   This  tight correlation
indicates a strong reproducibility in our radial velocity
measurements.  The velocities that we  obtained for the previously measured GCs
matched well  with those  in the literature,  except for  three cases.
GC0049     \citep{peng04b},    GC0188,    and     GC0226    \cite[both
from][]{beasley08} (see the bottom panel of Fig.~\ref{fig:vel}) have
had only  one previous radial velocity measurement and 
the latter two cases  had high radial velocity uncertainties.
All the measurements are still consistent with being
genuine GCs.   There appears  to be a potential
scale difference between our measured velocities and the cataloged
velocities.  For the GCs in common, there is an average radial
velocity uncertainty of 34 km s$^{-1}$ in our study and 39 km
s$^{-1}$ in the catalog.  We have  not been  able to  find any
reason for this difference, however, the offset or scale
error between the two datasets is not significant within the these
measurement uncertainties.

Of our candidate objects, 4 had velocities  between $150-250$ km
s$^{-1}$ indicating these objects may either be GCs or Milky Way
foreground stars. 
With the quality of most spectra obtained for radial velocity studies,
it  is not a  simple task  to  distinguish between  these two
types of objects.  We decided  to measure the structural parameters of
these 4 objects using our IMACS images of NGC 5128 in combination with
our velocity and photometry  information to help distinguish between either
foreground stars  or GCs.  To do  this, we used  the code
ISHAPE  \citep{larsen99,larsen01}  which  convolves  the  stellar  point
spread  function with  an
analytical King  profile \citep{king62}  and compares the  result with
the  input candidate  image achieving  a best  match.   The structural
parameters  measured from  the  models are the  core
radius, $r_c$,  the tidal radius, $r_t$,  the concentration parameter,
$c=r_t/r_c$,  and  ellipticity.   The  half-light radii  can  also  be
obtained  from  the  transformation, $r_e/r_c  \simeq  0.547c^{0.486}$
which  is  good  to  $\pm2\%$  for $c>4$  \citep{larsen01},  which  is
satisfied for  GCs in NGC 5128 \cite[see][]{gomez07}.   Based on their
structural  parameter  values  compared  to  normal GCs  in  NGC  5128
\citep{gomez07} (as  well  as  their velocities  and  photometry)  we
classify 3  of the 4 candidates  as GCs (GC0463,  GC0467, and GC0471).

The  luminosity  function of  our  newly  confirmed  GCs is  shown  in
Figure~\ref{fig:lum_N}, along with  the entire GCS and the  GCs with ages,
metallicities, and [$\alpha$/Fe] obtained  in this study.  On average,
our newly confirmed GCs are fainter than the globular cluster system
(GCS) 
turnover magnitude.  This is
not surprising  as our  long GMOS exposures  permitted us to  get good
radial velocity  measurements on  fainter objects.  Our  targetted GCs
for measurement  of age, metallicity, and  [$\alpha$/Fe] were brighter
than  the average  cluster in  the  system.  This  is also  expected as  we
targetted the brightest GCs to  meet the high S/N requirement of this
study.

\section{Results}
\label{sec:results}

\subsection{The Lick Index System}
\label{sec:index}

The     standardized    Lick    index     system    \citep{burstein84,
worthey94,worthey97,  trager98}  provides a  framework  of recipes  to
measure  the strength  of absorption  line features  in low-resolution
spectra. These  calibrated Lick index measurements can  be compared to
SSP  models  in  order  to  estimate their  ages,  metallicities,  and
[$\alpha$/Fe].

The Lick indices were measured using the code called GONZO described in
full in \cite{puzia02}.  This code measures the line index of each
Lick feature according to the observer's definition of equivalent
width \citep{worthey94}.  This measures the difference between  the flux of the feature compared
to its  pseudo-continuum.  For  this study, we have used the  passband and
pseudo-continuum  definitions  of  \cite{worthey94} and \cite{worthey97}.   The
input spectra and  variance spectra are used within  GONZO to generate
the uncertainty of the Lick index measurements through the addition of
Poisson  noise  via  100  MonteCarlo  simulations.   The  indices  are
measured  on the  noise enhanced  spectra and  the  $1\sigma$ standard
deviation of the measured index is the total Lick index uncertainty.

The  calibrated indices  and their  bootstrapped uncertainties  are
listed in
Tables~\ref{tab:index}  \&  \ref{tab:index_uncert}, respectively.   For
GCs  that were  measured  more than  once,  only the  indices for  the
highest-S/N spectra and/or the  spectra which had the most measureable
indices  are  listed.  For  some  red  indices  there is  no  recorded
measurement due to  the placement of the GC on the  mask.  For most of
these  latter objects, it  is  was  therefore not  possible  to obtain  ages,
metallicities and [$\alpha$/Fe] from the SSP models.

\subsection{Data Calibration to the Lick System}
\label{sec:cal}

Our GMOS  data were calibrated onto  the Lick system  with a secondary
standard star method.   \cite{beasley08} have obtained spectroscopy of
GCs in NGC  5128 using the Anglo Australian Telescope  (AAT) + 2dF and
have  calibrated their  measured indices  to the  standard  Lick index
system with 12 Lick standard stars.  We have a number of GCs in common
with \cite{beasley08} from  both our 2005 and 2007  GMOS data sets.
We show  the comparison of  the indices in  Figure~\ref{fig:cal05} for 
the  2005 data  and  Figure~\ref{fig:cal07} for  the  2007 data.  
The number of calibrating GCs in common  (once deviants $> 2\sigma$ have been removed)
ranged from
$14-22$ with an average of $\sim 18$ for the 2005 data and $9-16$ with
an average of $\sim 14$  for the 2007 data. From Figs.~\ref{fig:cal05}
\& \ref{fig:cal07},  it is  clear that a  simple zeropoint shift  is a
valid correction of our system  to the standard Lick system.  The mean
shift between the  AAT and GMOS data sets was then  applied to the GMOS
data in order to calibrate it to the Lick system. Our typical shifts were $0.41 {\rm{\AA}}$ and 0.10 mag.

We  also took integrated  light spectroscopy  for 40
Milky Way GCs (excluding NGC  6254 from the 41 available spectra) from
\cite{schiavon05}  available  online.  We measured their absorption features and calibrated them
to the Lick system using 11 GCs in common with \cite{puzia02}.  We
added one  additional GC, NGC  6981, from \cite{puzia02} that  was not
included in  \cite{schiavon05}.  The  comparisons of the  Lick indices
and  the  adopted shifts  are  shown  in Figure~\ref{fig:calMW}.   The
indices Fe4383 and  Fe5015 were not used in  these results because the
spectroscopy from  \cite{schiavon05} had defects from the  CCD or poor
sky subtraction in the regions of these two features.  We obtained
ages,   metallicities,  and  [$\alpha$/Fe] of the Milky Way GCs with the same
methodology as we did for GCs in  NGC 5128.  In this way, we permit
the most direct possible comparison between the two galaxies.

\subsection{The Measured Lick Indices}
\label{sec:indices}

In Figure~\ref{fig:diagnostic_hydrogen} we  show the measured indices
of the  GCs in NGC 5128  and the Milky Way  for H$\beta$, H$\delta_A$,
H$\delta_F$,   H$\gamma_A$,  and   H$\gamma_F$   versus  [MgFe]$^\prime$.
[MgFe]$^\prime$  is  a  composite  index  that has  been  shown  to  be
insensitive  to varying  [$\alpha$/Fe].   It  is defined  as
[MgFe]$^\prime  = \sqrt{\rm{Mg}_b  \times\ (0.72  \times  \rm{Fe5270} +
0.28  \times \rm{Fe5335})}$  \citep{tmb03}.  Overplotted  are  the SSP
models   of   \cite{tmb03}  and   \cite{tmk04}   for   the  grids   of
[$\alpha$/Fe]$=0$.   

We see the majority of GCs for both systems lie off the model
grids with extremely old ages that would be inconsistent with the WMAP age of the
Universe \citep{spergel03}.  This is a general problem for all studies
of integrated light GC  spectroscopy applied to the presently
available model grids.  The internal uncertainties for indices 
from  high-S/N  spectroscopy  are   very  low,   as  seen  in
Fig.~\ref{fig:diagnostic_hydrogen}, and these measured uncertainties will not  get much
lower for  extragalactic GCs because S/N$  > 100$ is  hard to achieve.
However,  we still find  that they  generally do  not match  the model
grids of  SSPs.  The SSP  models are calibrated on the
integrated-light optical photometry of Milky Way GCs and
high-resolution spectroscopy of individual member stars.  There
must be further astrophysics,  therefore, that needs to be incorporated
into these models in order to directly match the indices of GCs.

Also
shown in Fig.~\ref{fig:diagnostic_hydrogen} are the plotted indices of
$<$Fe$>  =  $(Fe5270 +  Fe5335)/2  and  Mg$_2$,  which allow  a  clear
separation of  the 3  different model [$\alpha$/Fe]  grid lines  of 0,
0.3,  and  0.5.   The  majority  of  GCs  in  NGC  5128  fall  between
[$\alpha$/Fe]$=0-0.3$, as do the Milky Way GCs.

We  can also  look at  the diagnostic  plots of  indices  that provide
information  on  the  abundances  of  calcium,  carbon,  and  nitrogen
\citep{tripicco95} in Figure~\ref{fig:diagnostic_cn}.  The strength  of the calcium index is
shown in  the diagnostic plot of  Ca4227.  There is a  large amount of
scatter,  both overabundant and underabundant for the GCs in NGC 5128, unlike the  Milky Way GCs
which  follow  the  SSP  model  grids  quite  closely.   However,  the
calibrations  of Ca4227  to  the Lick  index  system,  shown  in
Figs.~\ref{fig:cal05}  \&  \ref{fig:cal07},  have  a  large  scatter  
which  could artificially produce the calcium  scatter seen in
Fig.~\ref{fig:diagnostic_cn}.  Looking at the plot of CN$_2$, we see 
clear indications for systematically stronger index strengths (i.e. a
carbon/nitrogen overabundance) for the GCs in NGC 5128 and some GCs in
the Milky Way with respect to the model grids.   This cannot be accounted  for solely by
variations in [$\alpha$/Fe] or by age-metallicity degeneracy of the grids.
Rather this indicates an overabundance in carbon and/or nitrogen.

The index  strengths of  G4300 and C$_2$4668  have been shown  to have
sensitivity to carbon abundance but only weak sensitivity to the
nitrogen abundance \citep{tripicco95}.
In Fig.~\ref{fig:diagnostic_cn}, we see a mild
overabundance in  G4300,  but not  in  C$_2$4668.  
The C$_2$4668 index appears to have  a small scatter centered on the model
grids.  This indicates that the GCs in NGC
5128 may have a slight overabundance  in nitrogen and a wider range of
calcium index strength compared to the Milky Way GCs.  The implication
of this  result, however,  is unclear.  \cite{puzia08}  found evidence
for a  varying nitrogen  abundance with little  evidence for  a carbon
enhancement in  their recent  study of GCs  in low  surface brightness
dwarf galaxies.  Our abundance  trends for these elements do, however,
agree with the  general results of GCs in  M31 \citep{puzia05b}, which
as a giant galaxy of comparable luminosity to NGC 5128 is likely to be
a closer analog than the dwarfs.  More
evidence is needed  to understand if the abundance trends found
in GCs have environmental dependences.

The  ages,   metallicities,  and  [$\alpha$/Fe]   were  determined  by
iterating  between  the  measured   indices  and  the  SSP  models  of
\cite{tmb03} and \cite{tmk04} \citep[described in][]{puzia03,puzia05}.  
The  indices used in the iteration process are
the Balmer lines, Mg$_2$,  Mg$_b$, Fe5270, and Fe5335.  For
the GCs  that fall outside of  the model grids, a value  of 15 Gyr  in age is
assumed.   Typically these  GCs  have a  higher  uncertainty in  their
measured index and thus a  lower weighting in the final generated age
value attributed to the  GC.  The measurement routine iterates between
the  diagnostic plots of indices  and  once  convergence is reached it  provides  an age,  a
metallicity  and an  [$\alpha$/Fe] estimate  and their uncertainties for each  input  GC 
(see Table~\ref{tab:agemetafe}).    With our  Lick index measurement routines,
we are able to obtain low internal statistical uncertainties on the ages ($\Delta$t/t $\sim
0.3$),  metallicities ($\Delta$Z  $\sim0.15$  dex), and  [$\alpha$/Fe]
($\Delta$[$\alpha$/Fe]  $\sim  0.1$  dex) crucial for 
constraining different formation scenarios from a differential analysis.  There were 4 GCs in NGC
5128 that had ages, metallicities, and [$\alpha$/Fe] measured 
from two different spectra obtained in this study.  The root mean
square of the 4 multiple measurements are $\pm2.4$ Gyr for age,
$\pm0.2$ dex for
metallicity, and $\pm0.5$ dex for [$\alpha$/Fe] (see Table~\ref{tab:agemetafe} for the multiple measurements).

The  ages for  the  Milky Way  GCs  determined in  this  study can  be
directly compared  to relative ages recently derived by \cite{marin-franch09},
obtained from main sequence  fitting on HST/ACS  color magnitude
diagrams of 64  Milky Way GCs.
This  comparison provides an  excellent external
validation on the Lick index ages that we derive for the NGC 5128 GCs.
The comparison of the 24 overlapping  Milky Way GCs with this study is
shown in Figure~\ref{fig:MW_age_comp}.  Clearly, the two techniques do
not produce  a precise  1:1 correlation in  their results,  however the
ages derived by both studies  are consistently old.  The scatter shows
the Lick index routine underestimates the ages of the Milky Way GCs by
about 1 Gyr for the younger GCs, while it
overestimates the ages for the oldest GCs.  This could lead to an
artificial spread in our determined ages for the GCs in NGC 5128 and
may overestimate the number of young GCs.
We note that the horizontal spread in data is about
twice as large as the vertical spread, indicating that the internal scatter of
the main  sequence fitting routine is  about $1/2$ of  the Lick system
method.

\subsection{Ages of the GCs in NGC 5128}
\label{sec:ages}

Figure~\ref{fig:agemetafe}    shows   the   age,    metallicity,   and
[$\alpha$/Fe]  distributions for 72  individual GCs  in NGC  5128 with
S/N$> 30$  $\rm{\AA}^{-1}$ and for 41 Milky  Way GCs.
The  histograms have been  fit with  Gaussian distributions  using the
statistical code  RMIX \footnote{ The  complete code, available  for a
variety of platforms, is publicly available from Peter MacDonald's Web
site at  http://www.math.mcmaster.ca/peter/mix/mix.html. The same site
gives links to an extensive  bibliography with further examples of its
use.}, a library written  with the statistics programming language, R.
RMIX allows  a range of  multimodal fitting, including  the functional
form of the curve, and  hetero- and homo-scedastic fitting cases. 
The histograms have been fit with unimodal, bimodal, and trimodal
Gaussian distributions, with the fitted values as well as reduced $\chi^2$ listed in
Table~\ref{tab:rmixfits}. We have also included in Table~\ref{tab:rmixfits} the probability, 
listed as the p-value, that the data are drawn from this distribution. 
When there is no result
listed  in the table, we were unable to obtain a fit using RMIX.  
The fits with the lowest reduced $\chi^2$ are shown  as solid
lines in Fig.~\ref{fig:agemetafe} along with its  reduced $\chi^2$
value. For the GCs in NGC 5128 we were able  to obtain additional bimodal and/or  trimodal fits with nearly
equal  reduced $\chi^2$  values for [Z/H]  and [$\alpha$/Fe]
indicating that they are also fit well by
these alternate modalities. 

The age estimates provided by the iterative code  enable us to
distinguish among old ($> 8$ Gyr), intermediate ($5-8$ Gyr), and young
($< 5$ Gyr,  forming at z $\leq 1.2$))  stellar populations within our
uncertainties.  The Gaussian fits to the NGC 5128 data suggest
two or three major bursts  of cluster formation within this galaxy and
a rejection of the single burst solution.  More
specifically,  68$\%$  of  our  sample  have  old  ages,  14$\%$  have
intermediate ages, and 18$\%$ have young ages.  For the Milky Way GCs,
interestingly,  the best  statistical fit  is a  bimodal distribution,
with   both   peaks   at   nearly   the  same   formation   time in
agreement with the results of \cite{marin-franch09}.    A
Kolmogorov-Smirnov   statistical   comparison   test   indicates   the
normalized  age  distributions of  NGC  5128  and  Milky Way  GCs  are
different at greater than a $99\%$ confidence level.

Also in Fig.~\ref{fig:agemetafe} is the histogram of all 64 Milky
Way  GCs  obtained  by  \cite{marin-franch09}.  Their results are $\sim  1$ Gyr  older on  average  than our
ages obtained using the Lick index method (see Fig.~\ref{fig:MW_age_comp}). Our spectroscopic Lick
index study also shows a small extension to younger ages for the Milky
Way clusters, not seen in \cite{marin-franch09}.  However, out of the
6 Milky Way GCs for which we have measured ages $< 10$ Gyr, only 3
of these were measured by \cite{marin-franch09}.  

In contrast  to NGC 5128, it appears  nearly all GCs in  the Milky Way
have formed primordially, with a small fraction  of younger GCs,
perhaps obtained
from the accretion of satellite galaxies. We do detect a
positive age-metallicity slope  relation in the Milky Way clusters, also
noted  by \cite{mendel07}.   The 41  Milky Way  GCs have a mean of $11.3\pm0.1$ Gyr and only one GC (NGC 6553) has
a determined  age less  than 8 Gyr.   This particular result  could be
illustrating  one of  the generic  problems with  using SSP  models to
derive internal properties of  GCs from integrated light spectroscopy.
NGC 6553 has been shown to  be an old GC \citep{ortolani95} and one of
the most metal-rich \citep{barbuy98} in the Milky Way.  Our results do
provide  a metal-rich  [Z/H] value  of  $0.14\pm0.01$, but  it is  the
youngest of  our sample ($6.7\pm0.2$ Gyr).  The  result   indicates  once  again  that  this
methodology could  be underestimating the  ages of the  metal-rich GCs.  
This could, in part, be attributed to horizontal
branch  stars and  blue stragglers  that have  been shown  to  mimic a
younger  stellar  population  \citep{burstein84,cenarro09}.  These  hot  stellar
populations are incorporated into  SSP models, however are still based
on individual star spectroscopy within Milky Way GCs, whereas here, we
are  using integrated  spectroscopy.  However,  NGC 6553  is projected
onto  the bulge  stellar population  of the  Milky Way  as it  sits in
Baade's  Window.  In  this region,  a good  background  subtraction is
crucial  as to  not  artifically enhance  the  measured Lick  indices,
particularly   the  Balmer  lines.    Alternatively,  or   perhaps  in
conjunction  with,  NGC  6553  has  a  high  blue  straggler  specific
frequency with evidence for  spatial variation \citep[see Figure 12
of][]{beaulieu01}. NGC  6553 also  has an anomalously  blue horizontal
branch  for its high  metallicity that could  influence the
Lick  measurements depending  on where  the  study focused  in the  GC
itself,  leading  to a  over  or undersampling  of  the  true GC  blue
straggler contribution.

\subsection{Metallicities of the GCs in NGC 5128}
\label{sec:metals}

The  GCs in  NGC 5128 used in this study were biased  towards  more enriched
metallicities  based   on  our   selected  field  location
concentrated towards the inner 10\arcmin. 
We find over 65$\%$ of GCs
have [Z/H]$ > -1$, while larger samples of GCs in NGC 5128 show a near
50-50$\%$ split  between the  metal-rich and metal-poor cluster population
\citep[see][for example]{woodley05}.  This  stems from the metallicity
gradient where the metal-rich GCs are more centrally concentrated than
the   metal-poor  \citep[see][]{woodley05,woodley07,beasley08}.   This
bias  leads  to  an  artificial  inflation of  young,  metal-rich  GCs
relative to the  entire halo.  In addition, we  would expect the inner
few  arcmins of  the galaxy  to  show the  largest range  of ages  and
formation  timescales because accreted  material will  sink to  the central
regions  of  the  galaxy.

We  compare  the  different  Gaussian distributions of  color  and  metallicity
obtained from  this study in Figure~\ref{fig:cmd} using  our sample of
68 GCs  in NGC 5128 that  have both metallicity obtained  from the SSP
models  as   well  as  available   color  information.  The color
distribution,  (C-T$_1$) \citep{harris04}, and the  metallicity,
[Fe/H]$_{C-T_1}$, obtained from a color transformation, are plotted.  This transformation of
color-to-metallicity has been calibrated through Milky Way GC data
\citep{harris02}.  
We use a  foreground reddening value of E(B - V) =
0.11  for NGC  5128, corresponding  to  E(C-T$_1$) =  0.22 for  the
transformation.   This  transformation is  slightly  nonlinear so  the
uncertainty in [Fe/H] is a function of metallicity.  The uncertainties
for  a typical  color  uncertainty of  0.1  are $\pm0.07$  dex in  the
metal-rich  and   $\pm0.2$  dex  in  the   metal-poor  regimes.   Both
distributions are clearly bimodal.  

We also plot  the synthetic [MgFe]$^\prime$ index for  the same set of
GCs,  which is  an  almost  pure metallicity  indicator,  shown to  be
weakly sensitive  to  age  and  insensitive to [$\alpha$/Fe]  \citep{tmb03}. 
The  [MgFe]$^\prime$ index is measured directly from our spectroscopy and thus, model independent.  The
[MgFe]$^\prime$  distribution  is adequately  fit  by a  bimodal
model, as  are the Milky Way  GCs,
also   plotted   in   Fig.~\ref{fig:cmd}.  We also fit the
[MgFe]$^\prime$ index for the NGC 5128 GCs with a unimodal
distribution and obtain a reduced $\chi^2$  value of 1.02, which is
significantly different than the bimodal fit.
The metallicity distribution  obtained from
the SSP models, [Z/H], shown in Fig.~\ref{fig:agemetafe}, for the clusters in  NGC 5128 is consistent with
a bimodal  distribution, but it is not strongly preferred  over other
possible distributions.  Yet, with  the shape of the metallicity index
[MgFe]$^\prime$ distribution,  we can state that  both the metallicity
(obtained   directly  from  high-S/N   spectroscopy)  and   the  color
distributions for  the same sample  of GCs, both appear bimodal.
\cite{yoon06} have  shown that a bimodal  metallicity distribution can
be  an  artifact  of  the  non-linearity in  the  transformation  from
metallicity to color \citep[see also][]{cantiello07}.
We show here that the GCs in NGC 5128 have a bimodal color {\it and} a
bimodal   [MgFe]$^\prime$   index,  which   is   independent  of   any
transformation between color and metallicity.  This finding is further
supported  by the  recent  observations of  \cite{spitler08} who  show
strong evidence  for color bimodality in  NGC 5128 GCs from  R band to
mid-IR  band observations,  which  are insensitive  to hot  horizontal
branch stars, and  thus a good indicator of  metallicity 
as well.

We calculate the iron abundance for  the GCs in NGC 5128 obtained from
the SSP models \citep[see equation  4 in][]{tmb03} and compare to that
obtained from  the color conversion  discussed above, [Fe/H]$_{C-T_1}$
for the GCs in NGC 5128, shown in Figure~\ref{fig:ZH_FeH}.  We perform
a  best  linear fit which has an rms of 0.36 (after  removal of  deviant  points  based on  the
Chauvenent Criteria \citep{parratt61}).  
The correlation does not match
a 1:1 relationship  as expected for direct comparison  between the two
values.   It appears  either  [Fe/H] determined  from  the models  is 
too  metal-rich, or  the [Fe/H]$_{C-T_1}$ values  obtained from
the color conversion are too  metal-poor.  This could be the result of
the small non-linearity in the color-to-metallicity conversion or a
progressive change in alpha abundance with metallicity.
 
In  Figure~\ref{fig:agemetafe_comp}, we  show [Z/H]  as a  function of
age.  The majority of GCs in NGC 5128 and the Milky Way share the same
relative ages in our study, but NGC 5128 contains a number
of intermediate-age and young GCs.  The spread in age that we find for
NGC 5128 GCs  is almost purely from the  metal-rich subpopulation.  We
find, specifically,  that 92$\%$ (23/25) of metal-poor  GCs and 56$\%$
(26/47) of metal-rich GCs in NGC 5128 have ages $> 8$ Gyr.  We may see
evidence for an age-metallicity relation,  however, no strong
conclusions can be drawn from our results because of
the large biases in our selected GC sample.

Massive elliptical galaxies, including  NGC 5128, show
a  super metal-rich  peak observed  in the [Z/H]  distribution of GCs.  This
metal-rich peak has been  predicted by the spatially resolved chemical
evolution  code of  elliptical galaxies  \citep{pipino04,pipino07} and
closely  matched  the  spectroscopically determined  metal-rich  [Z/H]
distribution  of  a   representative  sample  of  elliptical  galaxies
\citep{puzia06}.  Their models indicate  that this super metal-rich GC
peak found  in massive  elliptical galaxies could  be the result  of a
parent  galaxy  that  has  radially varying  photochemical  properties
resulting     from      the     outside-in     formation     mechanism
\citep{pipino04,pipino06}.   In  addition   to  the  radially  varying
metallicity of  the galaxy  (and of the  GCS), the  massive elliptical
would have undergone a violent merger history.
 
The photometric metallicity for the  stellar halo for NGC 5128 is also
shown  in  Fig.~\ref{fig:cmd} for  an  inner region  at  8  kpc and  a
combined outer  halo field of  21 and 31  kpc, at a maximum  extent of
$\sim 7 R_{\rm eff}$  of the galaxy \citep{harris00,harris02,rejkuba05}.
What is  strikingly evident  is that the stellar  halo and  GC populations
have very different metallicity  distributions in the regions studied.
The GCs plotted are all within $\sim 20$ kpc from the center of NGC
5128 and  we are  thus comparing  clusters  in relatively  the same  projected
region as the  inner stellar halo.  The
stellar halo  peaks in metallicity  near $-0.4$ dex, coinciding  with the
major high metallicity peak for GCs.  
However, $\sim35\%$ of GCs have [Z/H]$  < -1$, and only $< 10\%$ of the  stellar halo falls in
this metal-poor  regime, nearly identical  to the statistics  found by
\cite{beasley08}.

The  contrast  in metallicity  distributions  of  the  stellar and  GC
population requires explanation.  We must  be able to show in galaxy
formation models why there is  such a contrast between two populations
that are  likely to have formed during the  same events of  formation.
One possibility is  destruction efficiency differences between the two
populations, where the more metal-rich GCs are destroyed with a higher
fraction compared  to the metal-poor GCs. However,  the destruction of
GCs would only dominate in the  center regions of the galaxy and there
are plenty of metal-rich GCs in the outer halo as well. One could also
consider the destruction of metal-poor GCs being extremely high in the
inner halo only.   In this scenario, the metal-poor  GCs would have to
have  different structure than  the metal-rich  GCs which  would allow
them to be  destroyed in the inner regions.  This  is not supported by
the structure of  the metal-poor GCs in the  outer regions which would
have  survived, but  do  not show  any  major structural  differences \citep{gomez07}.

Another possibility is that the formation efficiency of metal-poor GCs
to metal-poor  stars is much  higher than the formation  efficiency of
metal-rich   GCs  to   metal-rich   stars  within   the  same   galaxy
\citep{harris02,peng08}.  We  could also be examining  a biased sample
of the  stellar population.  The inner  region of the  galaxy light is
dominated  by the  metal-rich stellar  spheroid, built  up  by merging
events.   In  this  scenario, it  may  be  possible  to detect  a  low
metallicity stellar population  that formed before or at  the onset of
hierarchical  merging,  within pregalactic  clouds  condensing in  the
outermost regions of the galaxy.  We would expect to find this stellar
halo population if we were to examine the stellar halo at much further
distances   from  the   center  of   the  galaxy.    Studies   of  M31
\citep{kalirai06}   and  NGC  3379   \citep{harris07}  have   shown  a
transition  from a metal-rich  to metal-poor  stellar population  at a
galactocentric  distance   of  $12  R_{\rm eff}$.   For   NGC  5128,  this
corresponds to a distance of 65 kpc along the isophotal major axis and
45 kpc along the isophotal minor axis, assuming an axial ratio of $b/a
= 0.77$.  If this transition were  detected in NGC 5128 also, it would
have strong  suggestions for galaxy  formation. We hope to  conduct an
outer-halo search for these low-metallicity stars with upcoming data.

\subsection{Alpha-to-Iron Abundance Ratios}
\label{sec:alpha_fe}

The  abundance  of  $\alpha$-elements  compared  to  that  of  Fe-type
elements  can  provide  information  on the  formation  timescale  and
chemical history of the gas  cloud from which the GCs form.  Supernova
type  II  events  produce  an overabundance  of  $\alpha$-elements  in
relation  to  Fe-peak  elements  and  occur over  a  short  progenitor
lifetime (a few 100 Myr).  Supernova type Ia events enrich the interstellar  medium
after $\sim1$ Gyr and produce Fe-peak  elements in preference to  
$\alpha$-elements.  When
both type II and Ia (or  only type Ia) are occurring the [$\alpha$/Fe]
ratio decreases.   The abundance ratio therefore tells  us whether the
GCs formed  over a long or short timescale \citep{tornambe86}  and it also tells  us the
chemical composition of the  cloud at the
time of GC formation.

Fig.~\ref{fig:agemetafe} shows distribution of  [$\alpha$/Fe] obtained  from the  SSP  models is
best  fit  by a  unimodal Gaussian  distribution.  
The  NGC  5128  GCs  have   on  average  a  lower
[$\alpha$/Fe] than the Milky Way GCs,  indicating  slower  formation  times  at  every  given
metallicity,  as  shown  in  Fig.~\ref{fig:agemetafe_comp}.   We  also
examine     age    as     a    function     of     [$\alpha$/Fe]    in
Fig.~\ref{fig:agemetafe_comp}  and  clearly  see  a  wider  spread  in
[$\alpha$/Fe] among  the older GCs in  NGC 5128 than  among those with
ages $<  8$ Gyr.  A  comparable spread is  also seen in the  Milky Way
GCs, however shifted to higher $\alpha$-element enhancement.  The low 
[$\alpha$/Fe] values in the NGC 5128 
GCs are seen to occur over their full age range.  The GCs in NGC 5128 could have
undergone prolonged formation timescales
compared to the GCs in other giant systems.
The GCs with older ages have a wider spread in [$\alpha$/Fe] extending to
higher values than GCs with younger ages in NGC 5128.  The older GCs
within NGC 5128 may have formed within a faster
burst or collapse, on average, than the younger GCs within NGC 5128.

\section{Discussion}
\label{sec:dis}

Thus far  the ages  of GCs determined  spectroscopically have  made it
difficult  to  support  one  clear  dominant  formation  scenario  for
galaxies.  The  general trend appears to  show the majority  of GCs in
galaxies  are old,  with spectroscopic  ages greater  than 8  Gyr from
early-type         galaxies          to         dwarf         galaxies
\citep{puzia08,puzia05,strader05,kisslerpatig98,forbes01}.     However,
some  galaxies also  have  noticeable components  of intermediate  and
young                 populations                of                GCs
\citep{mora08,puzia05,larsen03,strader03,goudfrooij01},  forming along
with perhaps the major starbursting activity in the galaxy.  The issue
is  complicated   further  by environmental   dependencies  of  GCs.
Early-type galaxies, such as giant ellipticals, show a higher relative
number  of GCs  per unit  luminosity, the specific
frequency  \citep{harris81},  than  spiral  galaxies,  which  in  turn
generally  have  a  higher  specific  frequency  than  dwarf  galaxies
\citep[see however][who show that GC fractions for low-mass systems are
strongly environmentally dependent]{peng08}.  Studies, therefore, tend
to  concentrate on  the most  massive galaxies  where there  is a higher
quantity of GCs  present, such as this study.   These studies are also
generally biased towards the brightest and more centrally concentrated
GCs in  the massive distant early-type galaxies  to minimize telescope
time and maximize the number of GCs in the sample.  This can lead to a
spatially  biased  samples  of  metal-rich and  younger GCs. 
In the case of NGC 5128, studies with larger, more representative
samples of GCs across its entire halo show there is a near 50-50 split
between metal-rich and metal-poor GCs \citep{woodley05,hhg04} with the
metal-rich  population  being  more  centrally concentrated  than  the
metal-poor \citep{beasley08,woodley07,woodley05,peng04}.

A previous study in  NGC 5128 conducted by \cite{peng04} used
UBV photometry of GCs obtained with the CTIO 4-meter telescope as well
as spectra from the HYDRA instrument to examine GC ages.  Although
their  photometric  data suffered from the age-metallicity  degeneracy for objects
older than 2 Gyr, they were  able to show from the photometric indices
and  the \cite{bc93}  models  that  the metal-poor  GCs  appear to  be
between  12-15  Gyr  old.   They   also  found  2  young  GCs,  GC0084
(HGHH-G279) and GC0103 (pff$\_$gc-029),  with ages $\leq 1$ Gyr,
the  latter of which  is associated  with a  young, blue  tidal stream
\citep{peng02}.   Using spectra obtained with  signal-to-noise  of at
least 40 per  resolution element calibrated to the  Lick index system,
\cite{peng04}  compared the  measured indices  to  \cite{tmb03} SSP
models.   Their   results  indicated   that  the 
metal-poor GCs in NGC 5128 have ages ranging between 8-15 Gyr, and the
metal-rich GCs  have a  large range extending  between 1-10  Gyr, with
$1/3$ of their sample greater than 8 Gyr.  They also indicate that all
the GCs studied have alpha enhancement similar to the Milky Way GCs.

A more  recent age, metallicity  and alpha-to-iron abundance  study of
the GCs in  NGC 5128 has been performed  by \cite{beasley08}.  For 147
GCs  extending out  to 40\arcmin,  they were  able to  obtain spectra
using the  2dF instrument  on the AAT  with signal-to-noise of  30 per
$\rm{\AA}$.  Their  spectra are  compared to two  sets of  SSP models,
\cite{lw05}  and \cite{tmb03} including  the higher-order  Balmer line
corrections of  \cite{tmk04}.  We cannot compare  our results directly
for  GCs in  common  with \cite{beasley08}  as  age, metallicity,  and
[$\alpha$/Fe]  model  derived results  were  not  tabulated.  They  do
present empirically derived metallicities  for a large fraction of GCs
in NGC 5128,  however, that are model independent.   These are derived
from  the dependency  of   known  metallicity  with   Lick  index
measurements  from  the  Milky  Way  GCs  \citep[please  see][for  more
details]{beasley08}.   We have 28 GCs in common with their study (with 
reasonable uncertainties of $\delta$[M/H]$<0.5$) and we plot the 
comparison in Figure~\ref{fig:mh_zh}. There is scatter about the 
1:1 relationship with an rms of $0.25$ dex.
We can  also  compare our  model derived  ages,
metallcities,  and [$\alpha$/Fe]  to their  graphical  results.  Their
results show  that $\sim85-90\%$  of the GC  sample is old,  with ages
similar  to  the  Milky  Way  GCs.  They  find  an  intermediate  aged
population of  4-8 Gyr  that constitutes $\leq  15\%$ of  their sample
dominated by  metal-rich GCs.  One  very young GC,  GC0084 (HGHH-G279)
with an  age of $\sim  1-2$ Gyr was  found.  The GCs have  lower alpha
abundances than the Milky Way GCs at a given metallicity.

We find similar  results to both  previous spectroscopic
studies in  NGC 5128,  although we do  not find  either as high a  fraction of
metal-rich  GCs  with young  ages  as do  \cite{peng04}  or  as low  a
fraction of young  GCs as \cite{beasley08}.  From our  sample, we find
13/72 (18$\%$) of the clusters have ages less  than 5 Gyr, and none
less  than 3.5  Gyr, including  GC0084 for  which we  obtained  a more
intermediate age of 6.2$\pm0.4$ Gyr.   All studies do show that the
majority of GCs  in NGC 5128 have ages greater than  8 Gyr, similar to
those in the Milky  Way.   An important finding, strongly  reinforcing
that of \cite{beasley08}, is that a  majority of {\it both}
metal-poor ($92\%$) and metal-rich ($56\%$) GCs have old ages.  We  do not find a
strong indication  that metal-rich clusters formed  exclusively at slightly
later times than the metal-poor GCs as suggested by the multiphase in situ
scenario \citep{forbes97}.  However, the range of [$\alpha$/Fe] for the old GCs
is large,  indicating perhaps  a range  in formation
timescale and/or strong chemical variance during the primordial phase.  Also the SSP models themselves
have  limited ability  to distinguish  between  GCs of  old ages  (see
Fig.~\ref{fig:diagnostic_hydrogen}).   We also  do not  find  that all
metal-rich GCs are  old as would be expected  in an accretion scenario.
In this scenario, the metal-rich  GCs are  formed in  a massive  seed  galaxy, and
metal-poor    GCs    are    accreted   through    metal-poor    dwarfs
\citep{cote99,cote00,cote02}.  But our study indicates a population of
intermediate ($14\%$) and younger ($18\%$) GCs, the  large majority of
which are metal-rich.  These younger clusters do not appear to have formed
primordially and it seems more probable to suggest they built up by merging events later
in the evolutionary history of the galaxy.

The supersolar  [$\alpha$/Fe] for GCs in NGC 5128 indicates  a rapid formation
of  both  metal-rich  and   metal-poor  GCs, with a mean value of
$0.14\pm0.04$.  However,  the  formation
timescales  of these GCs  are on  average slower  than those  in other
elliptical galaxies and more along  the lines of spiral and lenticular
galaxies \citep{puzia06}.   In  the sparse environment
of NGC 5128, the formation of  the more metal-rich GCs could be extended over
a longer  timescale, supported by both  the range of ages  for the red
GCs and the low [$\alpha$/Fe] values compared to other giant galaxies.

Different formation scenarios  predict different absolute and relative
ages  of metal-rich  and metal-poor  GCs.  The  hierarchical  model of
accretion and  merging of smaller subunits to  build larger structures
over time suggests that both metal-rich  and metal-poor GCs with ages $\geq
9$ Gyr might be found in a giant galaxy.   The semi-analytic models 
of \cite{beasley02}  showed in the
hierarchical formation  scenario of  galaxies that the  metal-rich GCs
were younger  than the metal-poor GCs, with  mean ages of 9  and 12 Gyr,
respectively.   However,  extended  metal-rich  GC
formation can take  place  in  the  galaxy merging  model
\citep{schweizer87,ashmanzepf92} forming GCs down to the
present redshift. 

On the formation history of GCs, our results indicate that the
majority of today's GCS in NGC 5128, both metal-rich and metal-poor,
have formed early and are coeval within the uncertainties of our age
determinations, likely forming with the bulk of the galaxy field
stellar populations.  However, this last point is entirely based on
integrated-light properties of the diffuse galaxy light as current
work with color magnitude diagrams does not go deep enough to reach the
main sequence turn-off or sub-giant branch of an intermediate-age
sub-population.  In any case, we see trends towards younger GC ages
with increasing metallicity which is supported by the hierarchical
merging scenario.

A formation scenario by \cite{strader05}, combining
the    in    situ    \citep{forbes97}    and    accretion    scenarios
\citep{cote99,cote00,cote02}, is also consistent
with our  findings.  The metal-poor  GCs could have formed  in small
halos in the early universe until their formation was truncated by the
reionization process \citep{santos03}.   The field stars and metal-rich  GCs could then
form in massive early-type galaxies.  The  old ages obtained  for the stellar
population in the NGC 5128 stellar halo \citep{rejkuba05,harris08} support
this scenario as well.   In this case, the  youngest GCs
that are  found in NGC 5128 could be produced in  more recent merging
events.

\section{Conclusions}
\label{sec:conclusions}

Using the Gemini  South 8-meter telescope with GMOS, we obtained
spectra of GCs in NGC 5128 to obtain radial velocities, ages,
metallicities, and alpha-to-iron abundance ratios.  The main results
from this study are:

\begin{itemize}
\item We discovered 35
new  GCs from radial  velocity measurements  as  well as
remeasuring radial velocities for  101 previously confirmed GCs.

\item Our quantitative age estimates  suggest $68\%$ of our GCs have  old ages ($>8$ Gyr),
$14\%$ have  intermediate ages ($5-8$  Gyr), and $18\%$  have young ages ($< 5$
Gyr), suggesting several, or at least extended, formation epochs.  

\item Separating the GCs  into metal-rich and
metal-poor  subpopulations by  the model-derived  [Z/H]  quantity, we
find that 92$\%$ of  metal-poor ([Z/H]$<-1$) and 56$\%$ of metal-rich
([Z/H]$\geq -1$)  GCs   in  NGC  5128   have  ages  $>  8$   Gyr.   

\item We  also find both  the Washington  color and  spectroscopic synthetic
metallicity  index  [MgFe]$^\prime$   show  a bimodal
distribution for  the GCs, indicating  that there
are  two  distinct  metallicity   subpopulations.

\item The [$\alpha$/Fe] for  the GCs  are super solar, but not as
  large as in the Milky Way.  We suggest that the GCs in NGC 5128
  formed on slower timescales than GCs in the Milky Way and other
  massive elliptical galaxies \citep{puzia06}.   
\end{itemize}

Our results support a scenario with a rapid, early formation of the
bulk of the GCS and significant subsequent major accretion and/or GC-
and star-forming events in more recent times.
We note that  our relatively high
fraction of  metal-rich and  young GCs is  likely biased  because we probed
the inner regions of  NGC  5128 and observed mainly  the
brightest GCs.  We  consider our results, therefore, to  be a
fractional upper  limit on the number  of metal-rich and  young GCs in
NGC 5128.

\acknowledgements  Based  on   observations  obtained  at  the  Gemini
Observatory, which is operated  by the Association of Universities for
Research  in Astronomy, Inc.,  with a  cooperative agreement  with the
NSF,  on  behalf  of  the  Gemini partnership:  the  National  Science
Foundation  (United  States), the  Science  and Technology  Facilities
Council  (United  Kingdom), the  National  Research Council  (Canada),
CONICYT   (Chile),  the   Australian  Research   Council  (Australia),
Minist{\'e}rio da Ci{\^e}ncia e  Tecnologia (Brazil) and Ministerio de
Ciencia, Tecnolog{\'i}a e  Innovai{\'o}n Productiva (Argentina).  This
research  used the facilities  of the  Canadian Astronomy  Data Centre
operated by the  National Research Council of Canada  with the support
of  the  Canadian  Space  Agency.   KAW,  WEH,  and  GLHH  acknowledge
financial  support  from NSERC.   THP  acknowledges  support from  the
National Research Council of Canada in form of the Plaskett Fellowship
at   the  Herzberg   Institute  of   Astrophysics.    D.G.  gratefully
acknowledges support from the  Chilean {\sl Centro de Astrof\'\i sica}
FONDAP  No. 15010003  and from  the  Chilean Centro  de Excelencia  en
Astrof\'\i sica y Tecnolog\'\i  as Afines (CATA).  KAW appreciates the
help  with Gemini GMOS  data reduction  provided by  Rodrigo Carrasco,
Bryan Miller,  and Gelys  Trancho, and the  hospitality of  all Gemini
South  astronomers  and  staff  during  a stay  at  the  Gemini  South
facilities in 2006.




\clearpage

\begin{deluxetable*}{llllc}
\tablecolumns{5}
\tabletypesize{\small}
\tablecaption{Gemini-South GMOS Observation Data Summary\label{tab:gmos}}
\tablewidth{0pt}
\tablehead{
\colhead{Field} & \colhead{Field Center}&
\colhead{Time} & \colhead{Year} & \colhead{N Objects}  \\
\colhead{ } &\colhead{(J2000)}
&\colhead{(hr)} & \colhead{ } & \colhead{Observed}  \\
}\startdata
1 & 13 25 42.5 -42 58 05.9 & 3.75 & 2005 & 23\\        
2 & 13 24 56.4 -42 58 04.8 & 3    & 2005 & 19\\
3 & 13 24 57.6 -43 04 55.2 & 3.5  & 2005 & 23\\        
4 & 13 25 45.6 -43 06 00.0 & 3.5  & 2005 & 18\\
5 & 13 25 54.0 -42 58 21.8 & 3    & 2007 & 25\\        
6 & 13 25 20.0 -42 57 23.0 & 3    & 2007 & 22\\
7 & 13 24 55.0 -43 00 32.4 & 3    & 2007 & 25\\        
8 & 13 25 16.4 -43 05 43.4 & 3    & 2007 & 21\\
\enddata
\end{deluxetable*}

\begin{deluxetable*}{lllllrll}
\tablecolumns{8}
\tabletypesize{\small}
\tablecaption{New Radial Velocity Measurements of Known Globular Clusters\label{tab:rv_known}}
\tablewidth{0pt}
\tablehead{
\colhead{GC Name} & {Old GC Name} &\colhead{RA}     &\colhead{Decl.}   & \colhead{Mask \& Slit} & \colhead{S/N{\tablenotemark{a}}} & \colhead{v$_{r,GMOS}$} & \colhead{v$_{r}$} \\
\colhead{ }       & \colhead{ }   &\colhead{(J2000)}&\colhead{(J2000)} & \colhead{ }            & \colhead{($\rm{\AA}^{-1}$) }            & \colhead{(km s$^{-1}$)}& \colhead{(km s$^{-1}$)}  \\
}
\startdata
GC0046 & pff$\_$gc-015& 13 24 41.20 &-43 01 45.6 &m=7 s=17 &  54 &508$\pm$31&518$\pm$18\\
GC0049 & pff$\_$gc-017& 13 24 43.63 &-42 58 16.4 &m=2 s=01 &  54 &442$\pm$53&717$\pm$36\\
GC0050 & WHH-5        & 13 24 44.58 &-43 02 47.3 &m=3 s=03 &  47 &716$\pm$32&718$\pm$16\\
       &              &             &            &m=7 s=22 &  37 &735$\pm$26& \\
GC0052 & AAT109711    & 13 24 45.35 &-42 59 33.5 &m=2 s=02 &  66 &167$\pm$73&210$\pm$33\\
       &              &             &            &m=7 s=09 &  60 &186$\pm$48& \\         
GC0053 & AAT109788    & 13 24 45.78 &-43 02 24.5 &m=3 s=02 &  27 &471$\pm$47&504$\pm$17\\
       &              &             &            &m=7 s=23 &  20 &508$\pm$26& \\
\enddata	
\tablenotetext{a}{Approximate S/N is measured per $\rm{\AA}$ between 4700$\rm{\AA}$ - 4830$\rm{\AA}$.}
\tablenotetext{b}{GCs from \cite{woodley09}.}
\end{deluxetable*}	 

\begin{deluxetable*}{llllrlllllll}
\tablecolumns{12}
\tabletypesize{\small}
\tablecaption{Radial Velocity Measurements of New Globular Clusters\label{tab:rv_new}}
\tablewidth{0pt}
\tablehead{
\colhead{GC Name} & \colhead{RA}    &\colhead{Decl.}   & \colhead{Mask \& Slit} &\colhead{S/N{\tablenotemark{a}}}   &\colhead{v$_{r,GMOS}$}& \colhead{C} & \colhead{$\sigma$ C} & \colhead{M} & \colhead{$\sigma$ M} & \colhead{T$_1$} & \colhead{$\sigma$ T$_1$} \\
\colhead{ }       &\colhead{(J2000)}&\colhead{(J2000)} & \colhead{ }            &\colhead{($\rm{\AA}^{-1}$)}           & \colhead{(km s$^{-1}$)} &\colhead{(mag)}&\colhead{(mag)}&\colhead{(mag)}&\colhead{(mag)}&\colhead{(mag)}&\colhead{(mag)}    \\
}
\startdata
GC0445 &13 24 40.45 &-43 01 56.9& m=7 s=24&  16 & 504$\pm$45&21.86 &0.03 &20.93 &0.02 &20.04 &0.03\\
GC0446 &13 24 50.43 &-43 04 51.0& m=3 s=05&  30 & 358$\pm$27&21.27 &0.01 &20.08 &0.01 &19.09 &0.01\\
GC0447 &13 24 50.90 &-42 59 59.0& m=7 s=11&  07 & 607$\pm$77&22.42 &0.03 &21.90 &0.02 &20.73 &0.02\\
GC0448 &13 24 53.64 &-42 57 59.3& m=7 s=03&  15 & 623$\pm$53&22.12 &0.03 &21.16 &0.02 &20.22 &0.02\\
GC0449 &13 24 53.96 &-43 00 43.4& m=7 s=16&  18 & 332$\pm$42&21.79 &0.04 &20.63 &0.03 &19.71 &0.04\\
GC0450 &13 24 54.97 &-43 02 52.4& m=7 s=21&  25 & 415$\pm$92&21.37 &0.05 &21.03 &0.04 &19.08 &0.02\\
GC0451 &13 24 55.42 &-42 59 49.0& m=2 s=10&  41 & 675$\pm$42{\tablenotemark{b}}&20.89 &0.03 &20.14 &0.03 &19.69 &0.03\\ 
       &            &           & m=7 s=07&  36 &           &      &     &      &     &      &    \\
\enddata
\tablenotetext{a}{Approximate S/N is measured per $\rm{\AA}$ between 4700$\rm{\AA}$ - 4830$\rm{\AA}$.}
\tablenotetext{b}{The quoted radial velocity is a weighted average.
  The individual measurements were 655$\pm$69 for m=2,s=10 and
  687$\pm$52 for m=7,s=07.}
\tablenotetext{c}{The quoted radial velocity is a weighted average.
  The individual measurements were 410$\pm$46 for m=7,s=05 and
  398$\pm$34 for m=6,s=22.}		 
\end{deluxetable*}

\clearpage
\tabletypesize{\scriptsize}
\setlength{\tabcolsep}{0.02in} 
\LongTables 
\begin{landscape}
\begin{deluxetable}{lrrrrrrrrrrrrrrrrrrrr}
\tablecolumns{21}
\tablecaption{Lick Indices for Globular Clusters in NGC 5128\label{tab:index}}
\tablewidth{0pt}
\tablehead{
\colhead{GC Name}&\colhead{H$\delta_A$} &\colhead{H$\delta_F$}&\colhead{CN$_1$} &\colhead{CN$_2$} &\colhead{Ca4227} &\colhead{G4300}&\colhead{H$\gamma_A$} &\colhead{H$\gamma_F$}&\colhead{Fe4383} &\colhead{Ca4455}&\colhead{Fe4531} &\colhead{C$_2$4668}&\colhead{H$\beta$} &\colhead{Fe5015}&\colhead{Mg$_1$} &\colhead{Mg$_2$} &\colhead{Mg$_b$} &\colhead{Fe5270}&\colhead{Fe5335} &\colhead{Fe5406} \\
\colhead{ }      &\colhead{($\rm{\AA}$)}     &\colhead{($\rm{\AA}$)}    &\colhead{(mag)}&\colhead{(mag)}&\colhead{($\rm{\AA}$)}  &\colhead{($\rm{\AA}$)}&\colhead{($\rm{\AA}$)}     &\colhead{($\rm{\AA}$)}    &\colhead{($\rm{\AA}$)}  &\colhead{($\rm{\AA}$)} &\colhead{($\rm{\AA}$)}  &\colhead{($\rm{\AA}$)} &\colhead{($\rm{\AA}$)}  &\colhead{($\rm{\AA}$)} &\colhead{(mag)}&\colhead{(mag)}&\colhead{($\rm{\AA}$)}&\colhead{($\rm{\AA}$)} &\colhead{($\rm{\AA}$)}  &\colhead{($\rm{\AA}$)}     \\
}
\startdata
GC0046  &  3.1664 &  2.9822 & -0.0504 & -0.0782 &  0.0855 &  3.8816 &  0.9354 &  1.9789 &  1.2878 &  0.7032 &  1.5413 &  0.1498 &  1.9535 &  1.8411 &  0.0263 &  0.0911 &  1.1370 &  1.2358 &  1.6591 &  0.6536\\
GC0049  &  3.4197 &  2.2441 & -0.0133 & -0.0481 &  0.1772 &  2.9800 &  1.3203 &  0.8744 &  0.2306 &  0.6245 &  1.0969 &  0.8011 &  2.4616 &  2.3498 &  0.0396 &  0.0949 &  1.5542 &  1.2490 &  1.2960 &  0.4232\\
GC0050  & -2.0442 &  1.7229 & -0.0041 &  0.0192 &  0.3911 &  5.8169 & -2.9499 & -1.3527 &  1.4816 &  0.1630 &  2.3403 &  2.5039 &  1.6482 &  5.9435 &  0.0714 &  0.0701 & -0.3681 &  2.8971 &  1.9032 &  0.9223\\
GC0052  &  2.7035 &  2.9233 & -0.0706 & -0.0160 &  0.1576 &  1.1184 &  2.4258 &  2.3386 &  0.7019 &  0.4223 &  0.1181 & -0.5694 &  2.5340 &  0.3784 &  0.0200 &  0.0576 &  0.8554 &  0.8646 &  0.9145 &  0.2960\\
GC0053  &  0.0554 &  1.3056 &  0.0470 &  0.0600 &  1.7039 &  5.6664 & -4.0525 & -2.2466 &  3.4825 &  0.7872 &  2.8172 &  4.2024 &  1.4969 & -2.2065 &    -    &    -    &    -    &    -    &    -    &    -   \\
GC0056  & -1.6691 &  0.4412 &  0.1364 &  0.2364 &  1.7137 &  6.1523 & -4.7779 & -1.4771 &  5.3029 &  0.1963 &  1.6534 &  5.8177 &  1.6702 &  4.0582 &  0.1296 &  0.3104 &  4.7575 &  3.4308 &  2.4358 &  2.2057\\
GC0057  &  1.9973 &  2.1368 & -0.0272 & -0.0015 &  0.3901 &  2.6582 & -0.0862 &  1.1814 &  2.4523 &  0.6786 &  1.0395 &  1.2480 &  2.1312 &  1.9875 &  0.0356 &  0.1034 &  1.1820 &  1.6324 &  1.5620 &  0.9284\\
GC0062  &  2.1028 &  2.1492 & -0.0234 &  0.0226 &  0.1673 &  2.8996 & -0.0080 &  1.6349 &  2.0537 &  0.5363 &  1.7744 &  1.4755 &  1.9964 &  2.2511 &  0.0176 &  0.0822 &  1.0767 &  0.9118 &  1.3295 &  0.6253\\
\enddata     
\end{deluxetable}
\clearpage
\end{landscape}
                                                           
\clearpage
\LongTables 
\begin{landscape}                                                               
\begin{deluxetable}{lrrrrrrrrrrrrrrrrrrrr}
\tablecolumns{21}
\tabletypesize{\small}
\tablecaption{Lick Indices Uncertainties for Globular Clusters in NGC 5128\label{tab:index_uncert}}
\tablewidth{0pt}
\tablehead{
\colhead{GC Name} &\colhead{H$\delta_A$} &\colhead{H$\delta_F$}&\colhead{CN$_1$} &\colhead{CN$_2$} &\colhead{Ca4227} &\colhead{G4300}&\colhead{H$\gamma_A$} &\colhead{H$\gamma_F$}&\colhead{Fe4383} &\colhead{Ca4455}&\colhead{Fe4531} &\colhead{C$_4$668}&\colhead{H$\beta$} &\colhead{Fe5015}&\colhead{Mg$_1$} &\colhead{Mg$_2$} &\colhead{Mg$_b$} &\colhead{Fe5270}&\colhead{Fe5335} &\colhead{Fe5406} \\
\colhead{ }       &\colhead{($\rm{\AA}$)}     &\colhead{($\rm{\AA}$)}    &\colhead{(mag)}&\colhead{(mag)}&\colhead{($\rm{\AA}$)}  &\colhead{($\rm{\AA}$)}&\colhead{($\rm{\AA}$)}     &\colhead{($\rm{\AA}$)}    &\colhead{($\rm{\AA}$)}  &\colhead{($\rm{\AA}$)} &\colhead{($\rm{\AA}$)}  &\colhead{($\rm{\AA}$)} &\colhead{($\rm{\AA}$)}  &\colhead{($\rm{\AA}$)} &\colhead{(mag)}&\colhead{(mag)}&\colhead{($\rm{\AA}$)}&\colhead{($\rm{\AA}$)} &\colhead{($\rm{\AA}$)}  &\colhead{($\rm{\AA}$)}     \\
}
\startdata
GC0046  & 0.0315  &  0.0180  &  0.0009  &  0.0012  &  0.0185  &  0.0617  &  0.0436  &  0.0240  &  0.1045  &  0.0283   & 0.0514  &  0.0667  &  0.0282  &  0.0682  &  0.0006  &  0.0006  &  0.0613  &  0.0894  &  0.0420  &  0.0289\\                         
GC0049  & 0.3988  &  0.1915  &  0.0080  &  0.0104  &  0.1605  &  0.4021  &  0.2501  &  0.2194  &  0.5541  &  0.2638   & 0.6326  &  0.5825  &  0.2701  &  0.5488  &  0.0066  &  0.0070  &  0.2338  &  0.2638  &  0.2725  &  0.2492\\                         
GC0050  & 0.0354  &  0.0253  &  0.0020  &  0.0024  &  0.0311  &  0.0678  &  0.0515  &  0.0232  &  0.0812  &  0.0339   & 0.0473  &  0.0724  &  0.0496  &  0.0489  &  0.0005  &  0.0007  &  0.0446  &  0.0268  &  0.0290  &  0.0271\\                         
GC0052  & 0.1512  &  0.1165  &  0.0062  &  0.0084  &  0.1218  &  0.3424  &  0.1615  &  0.1273  &  0.2711  &  0.1386   & 0.4907  &  0.6585  &  0.3217  &  0.5554  &  0.0048  &  0.0051  &  0.1610  &  0.2058  &  0.3869  &  0.2480\\                         
GC0053  & 0.2361  &  0.1861  &  0.0066  &  0.0087  &  0.1154  &  0.1796  &  0.3988  &  0.1371  &  0.3084  &  0.1169   & 0.1893  &  0.5849  &  0.1345  &  0.5756  &    -     &    -     &    -     &    -     &    -     &    -   \\                         
GC0056  & 0.2413  &  0.1032  &  0.0068  &  0.0040  &  0.0511  &  0.0761  &  0.1034  &  0.0777  &  0.1357  &  0.0547   & 0.0989  &  0.2055  &  0.0582  &  0.1830  &  0.0021  &  0.0020  &  0.0585  &  0.0507  &  0.0805  &  0.0742\\ 
GC0057  & 0.2027  &  0.1361  &  0.0059  &  0.0067  &  0.1264  &  0.1885  &  0.2396  &  0.1768  &  0.3815  &  0.1981   & 0.3174  &  0.7393  &  0.1897  &  0.3978  &  0.0052  &  0.0057  &  0.2627  &  0.2080  &  0.1744  &  0.2031\\ 
GC0062  & 0.2027  &  0.1361  &  0.0116  &  0.0142  &  0.1463  &  0.3201  &  0.3590  &  0.1370  &  0.4307  &  0.2213   & 0.4136  &  0.6322  &  0.1506  &  0.7987  &  0.0035  &  0.0040  &  0.1805  &  0.4216  &  0.3824  &  0.1885\\
\enddata     
\end{deluxetable}
\clearpage
\end{landscape}
 
\tabletypesize{\small}
\begin{deluxetable*}{lrrr}
\tablecolumns{4}
\tablecaption{Ages, Metallicities, and [$\alpha$/Fe]
  of Globular Clusters in NGC 5128\label{tab:agemetafe}}
\tablewidth{0pt}
\tablehead{
\colhead{GC Name} & \colhead{Age}    &\colhead{[Z/H]}   & \colhead{[$\alpha$/Fe]} \\
\colhead{ }       &\colhead{(Gyr)}   &\colhead{(dex)}   &\colhead{(dex)}   \\
}
\startdata
GC0046 & 8.4$^{+3.4}_{-0.3}$ &  -1.05$^{+0.06}_{-0.15}$ &  -0.10$^{+0.03}_{-0.02}$\\
GC0049 &10.8$^{+4.1}_{-4.2}$ &  -0.89$^{+0.16}_{-0.24}$ &   0.14$^{+0.13}_{-0.12}$\\
GC0050 & 6.8$^{+5.1}_{-4.4}$ &  -0.29$^{+0.56}_{-0.20}$ &  -0.99$^{+0.01}_{-0.01}$\\
GC0052{\tablenotemark{a}} &11.5$^{+2.2}_{-2.4}$ &  -1.67$^{+0.16}_{-0.10}$ &   0.02$^{+0.29}_{-0.27}$\\
GC0057{\tablenotemark{b}} &12.0$^{+1.2}_{-3.1}$ &  -1.04$^{+0.19}_{-0.14}$ &  -0.11$^{+0.08}_{-0.08}$\\
GC0062 &11.3$^{+0.6}_{-1.0}$ &  -1.22$^{+0.14}_{-0.03}$ &   0.15$^{+0.24}_{-0.22}$\\
GC0065 & 9.3$^{+3.9}_{-2.2}$ &  -1.01$^{+0.16}_{-0.24}$ &   0.01$^{+0.19}_{-0.18}$\\
GC0068 &10.5$^{+0.7}_{-2.0}$ &  -0.56$^{+0.05}_{-0.02}$ &   0.40$^{+0.09}_{-0.09}$\\
\enddata     
\tablenotetext{a}{From a second spectrum with lower S/N we found age$=9.2^{+0.1}_{-0.1}$ Gyr, [Z/H]$=-1.51^{+0.01}_{-0.01}$ and [$\alpha$/Fe]$=-0.23^{+0.06}_{-0.06}$.}
\tablenotetext{b}{From a second spectrum with lower S/N we found age$=8.8^{+0.1}_{-0.9}$ Gyr, [Z/H]$=-0.78^{+0.03}_{-0.01}$ and [$\alpha$/Fe]$=0.11^{+0.02}_{-0.03}$.}
\tablenotetext{c}{From a second spectrum with lower S/N we found age$=8.2^{+1.1}_{-0.2}$ Gyr, [Z/H]$=-1.43^{+0.03}_{-0.02}$ and [$\alpha$/Fe]$=-0.62^{+0.05}_{-0.05}$.}
\tablenotetext{d}{From a second spectrum with lower S/N we found age$=9.3^{+3.9}_{-4.8}$ Gyr, [Z/H]$=-1.08^{+0.92}_{-0.09}$ and [$\alpha$/Fe]$=-0.50^{+0.06}_{-0.07}$.}
\end{deluxetable*}                  
 
\clearpage
\begin{landscape}
\tabletypesize{\small}
\begin{deluxetable}{llcccccccccc}
\tablecolumns{12}
\tablecaption{RMIX Fits for Figure~\ref{fig:agemetafe}\label{tab:rmixfits}}
\tablewidth{0pt}
\tablehead{
\colhead{Distribution} & \colhead{Modality}  &\colhead{NGC 5128}   &\colhead{}  &\colhead{}  &\colhead{}   & \colhead{} &\colhead{Milky Way}  &\colhead{}   &\colhead{} &\colhead{}   & \colhead{} \\
\colhead{}  &\colhead{}  &\colhead{Proportion}   &\colhead{Mean}
&\colhead{Sigma}&\colhead{$\chi^2_{red}$}&\colhead{p-value}&\colhead{Proportion}   &\colhead{Mean}  &\colhead{Sigma} &\colhead{$\chi^2_{red}$}&\colhead{p-value}  \\
}
\startdata
Age             & unimodal & 1.0           & 8.58$\pm0.33$ & 2.78$\pm0.23$ &2.40 &0.003& 1.0           & 11.33$\pm0.23$ & 1.50$\pm0.17$& 2.22 & 0.018   \\
                & bimodal  & 0.29$\pm0.07$ & 4.85$\pm0.47$ & 1.32$\pm0.35$ &1.12 &0.346& 0.52$\pm0.18$ & 11.19$\pm0.47$ & 2.03$\pm0.42$&  0.89& 0.522  \\
                &          & 0.71$\pm0.07$ &10.12$\pm0.28$ &  1.44$\pm0.22$&     &     & 0.48$\pm0.18$ & 11.47$\pm0.20$ & 0.38$\pm0.16$&      &        \\ 
                & trimodal & 0.32$\pm0.07$ & 5.02$\pm0.46$ & 1.41$\pm0.36$ &0.74 &0.617&      -         &       -         &     -      &      &        \\
                &          & 0.25$\pm0.09$ & 8.76$\pm0.26$ & 0.56$\pm0.26$ &     &     &      -         &       -         &     -      &      &        \\
                &          & 0.43$\pm0.08$ &11.11$\pm0.24$ & 0.75$\pm0.02$ &     &     &      -         &       -         &     -      &      &        \\
Metallicity     & unimodal & 1.0           &-0.68$\pm0.07$ & 0.55$\pm0.05$ &1.29 &0.200& 1.0            & -0.95$\pm0.08$& 0.53$\pm0.06$& 1.28 & 0.197  \\
                & bimodal  & 0.37$\pm0.10$ &-1.27$\pm0.11$ & 0.29$\pm0.08$ &1.43 &0.158& 0.58$\pm0.09$ & -1.35$\pm0.07$ & 0.28$\pm0.06$& 1.13 & 0.343  \\
\enddata     
\end{deluxetable} 
\clearpage
\end{landscape}

                  
\clearpage        
\begin{figure}
\plotone{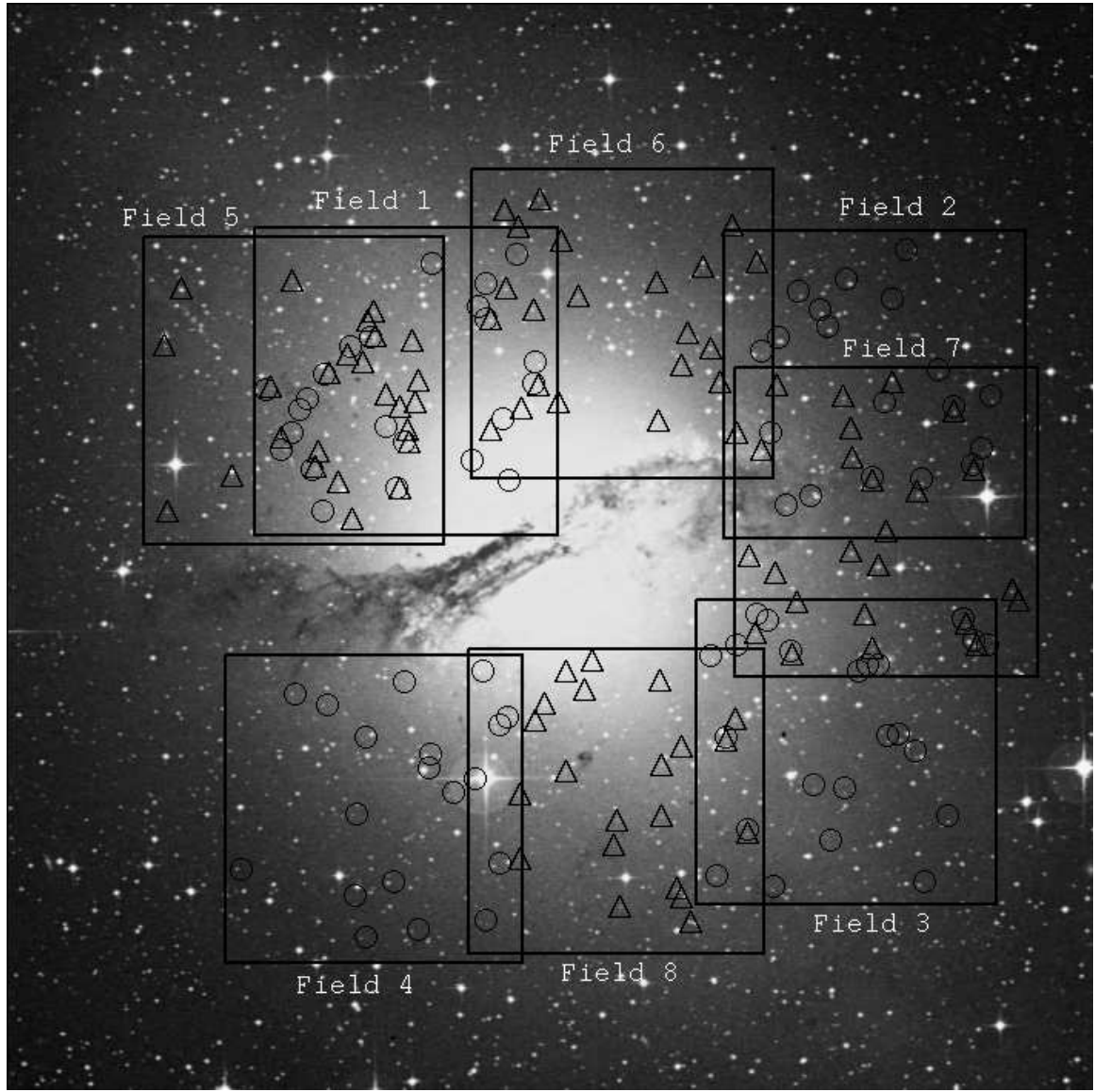}
\caption{The positions of GMOS fields 1-4 GMOS from 2005 and fields 5-8 from 2007.  The squares indicate a 5.5\arcmin\ region
and the circles (triangles) indicate where the observed objects are located in
each field from 2005 (2007).
The fields are overplotted on a DSS image of NGC 5128,
clearly showing the field's proximity to the center of the galaxy.} 
\label{fig:gmospositionsg}
\end{figure}
       
\begin{figure}
\plotone{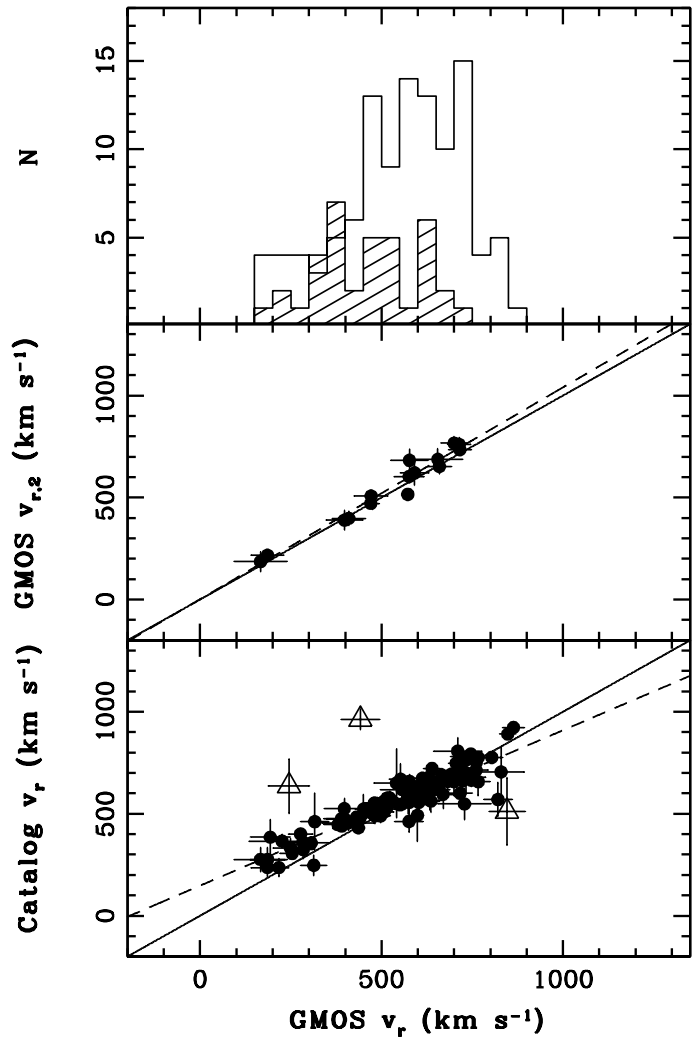}
\caption{{\it  Top:}  The projected  radial  velocities measured  with
  Gemini/GMOS for the previously known GCs in Table~\ref{tab:rv_known}
  (open    histogram)    and    the    newly   discovered    GCs    in
  Table~\ref{tab:rv_new}   (hatched   histogram).   {\it  Middle:}   A
  comparison of radial velocities of GCs measured twice with GMOS. The
  lines are a 1:1 (solid) and a least square fit with slope = 1.04 and
  y-intercept =  1.071 with an rms  error of 34 km  s$^{-1}$ about the
  line. {\it  Bottom:} The  radial velocity of  the GCs  measured with
  GMOS compared  to the  cataloged velocity from  \cite{woodley07} and
  from M. G{\'o}mez \& K. A. Woodley (2009, in preparation).  The lines are a  1:1 (solid) and a least squares
  fit with slope = 0.76 and  y-intercept = 147.5 with and rms error of
  72  km s$^{-1}$  about  the  line.  The  open  triangles are  GC0049
  (measured  once before  from \cite{peng04b})  and GC0188  and GC0226
  (both  measured once  before in  \cite{beasley08}, removed  from the
  fit.}
\label{fig:vel}
\end{figure}

\begin{figure}
\plotone{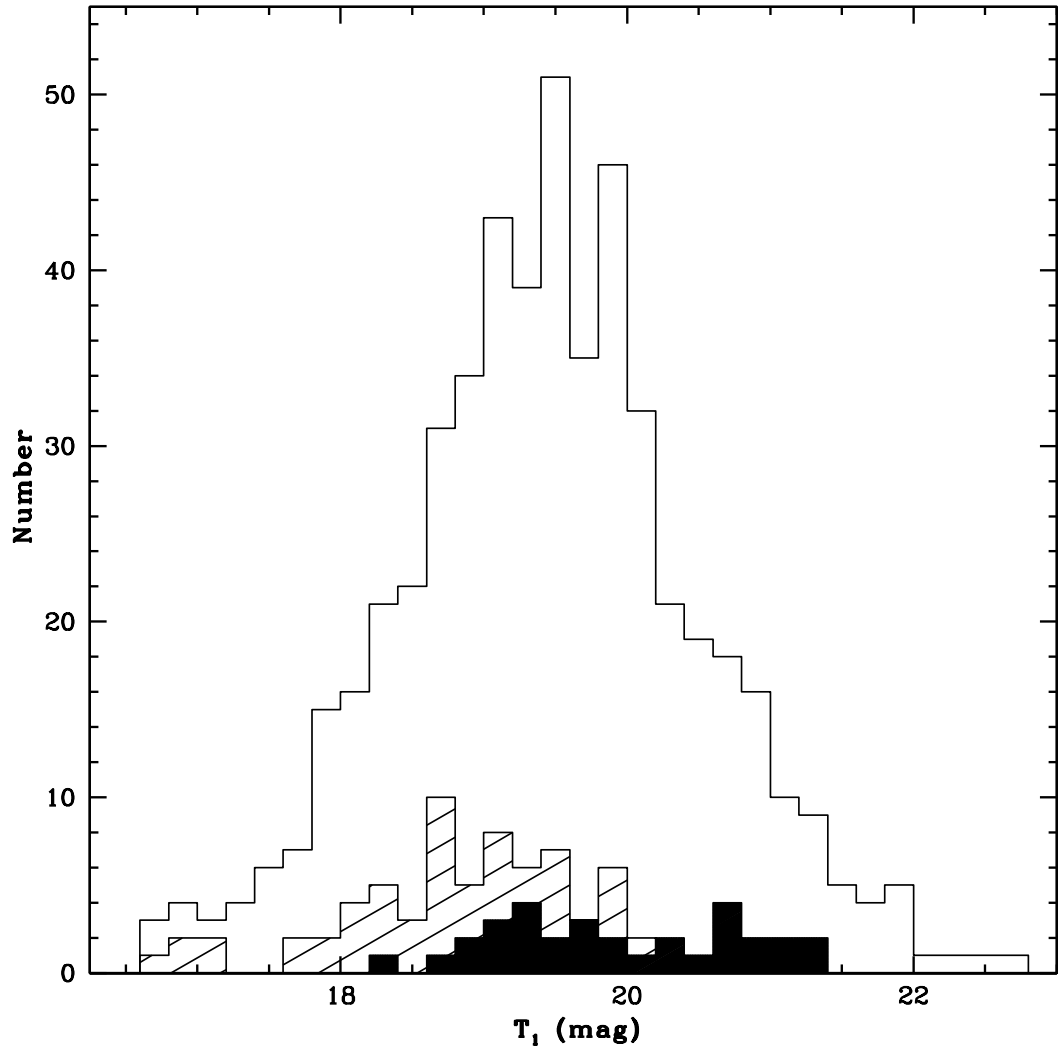}
\caption{The luminosity functions for the entire GCS of NGC 5128, consisting of 524 GCs with T$_1$ measurements ({\it open histogram}), 33 of the newly confirmed GCs from this study that have T$_1$ magnitudes ({\it solid histogram}), and 69 of the targetted 72 GCs with available T$_1$ magnitudes with ages, metallicities, and [$\alpha$/Fe] from this study ({\it hatched histogram}).} 
\label{fig:lum_N}
\end{figure}

\begin{figure}
\centering
\begin{tabular}{cccc}
\epsfig{file=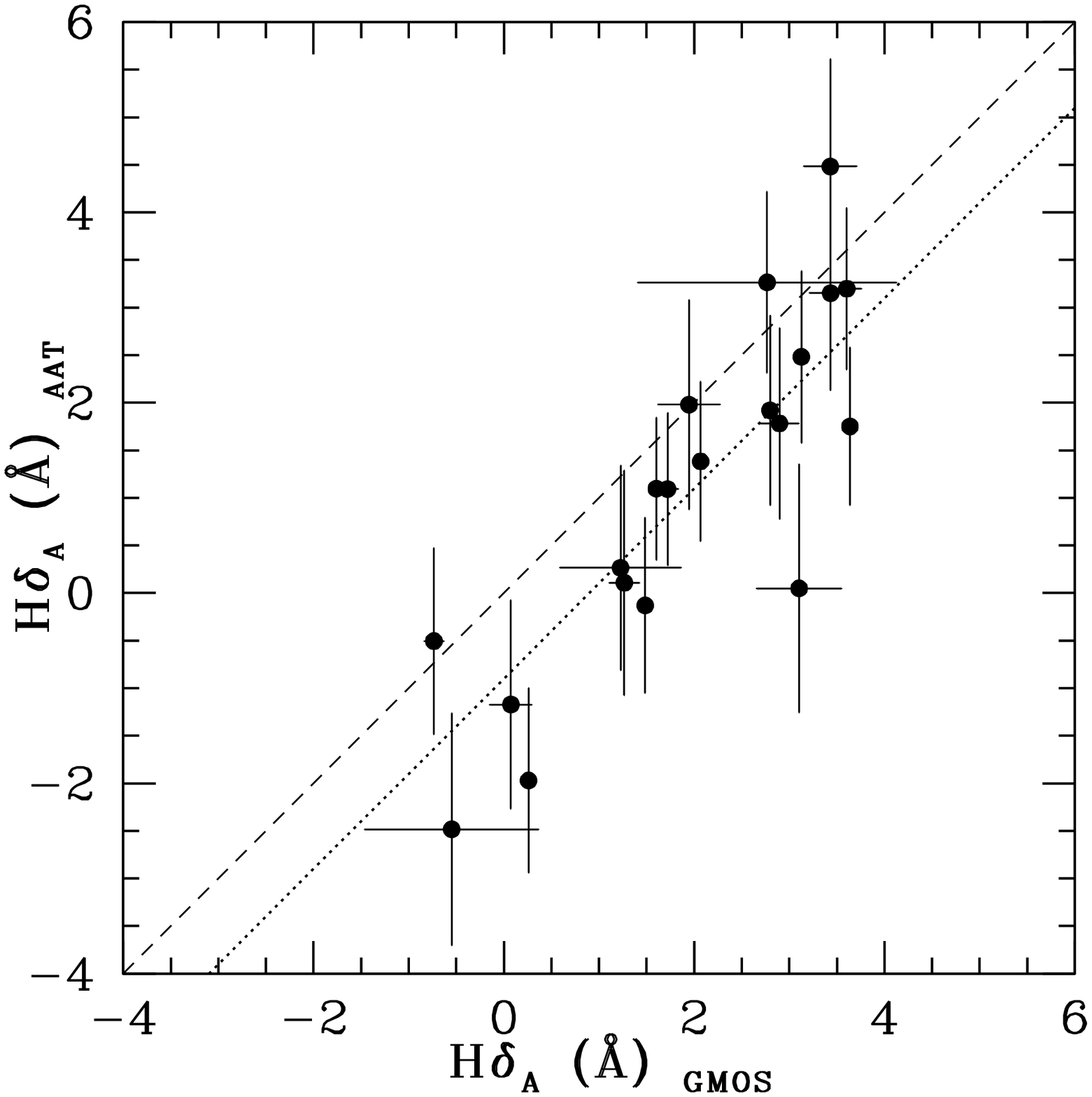,width=0.2\linewidth,clip=} &
\epsfig{file=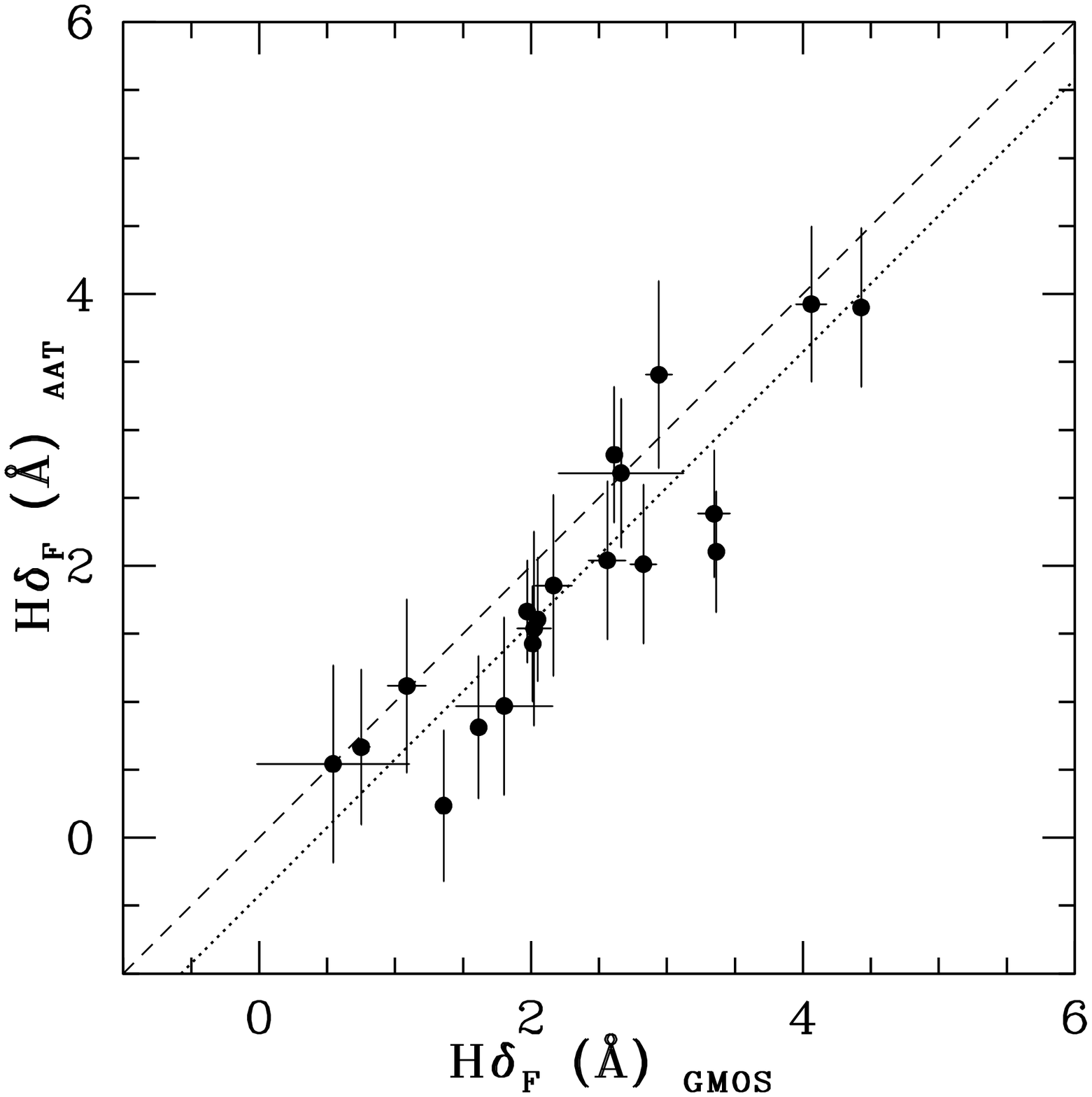,width=0.2\linewidth,clip=} &
\epsfig{file=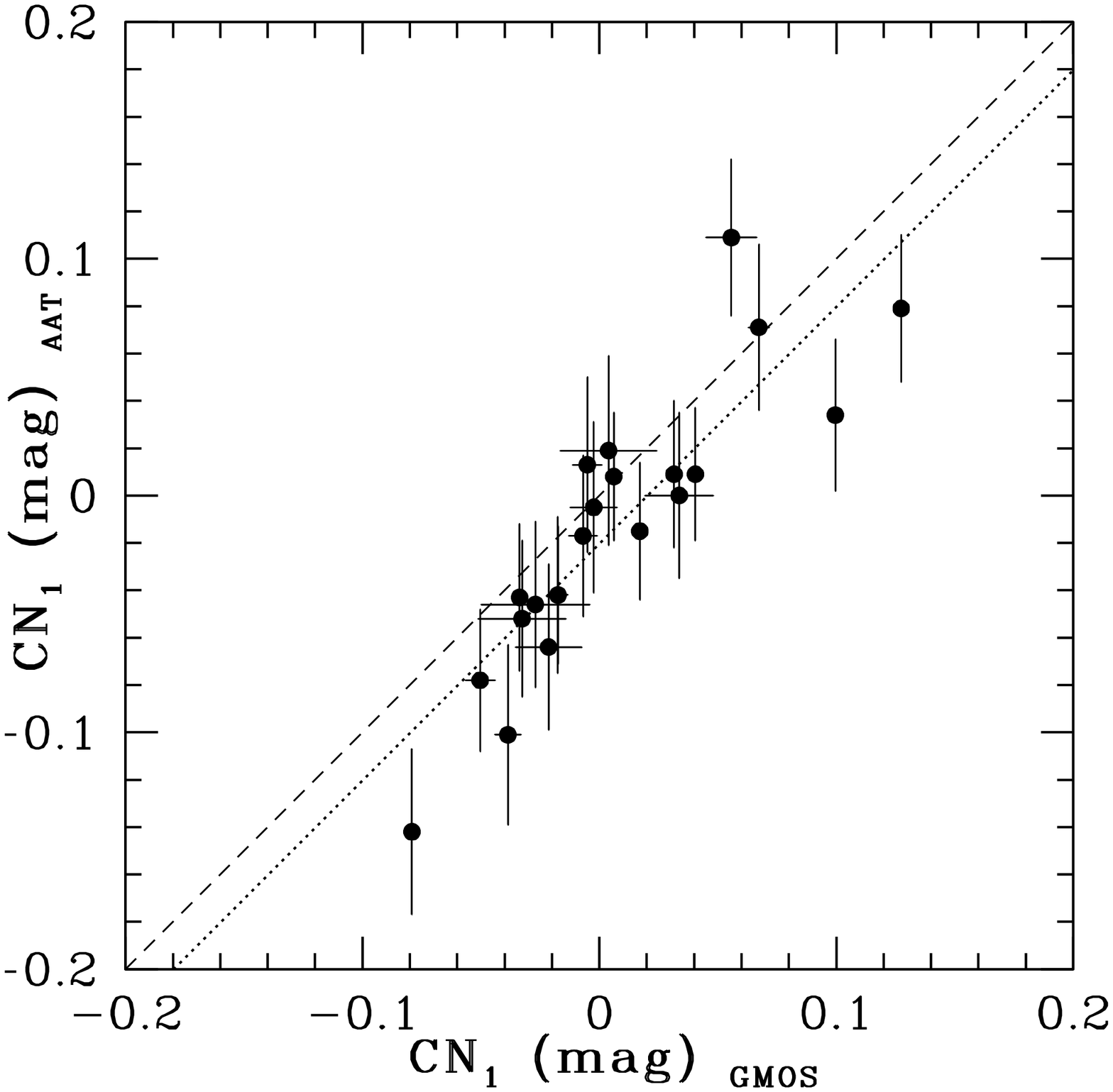,width=0.2\linewidth,clip=} &
\epsfig{file=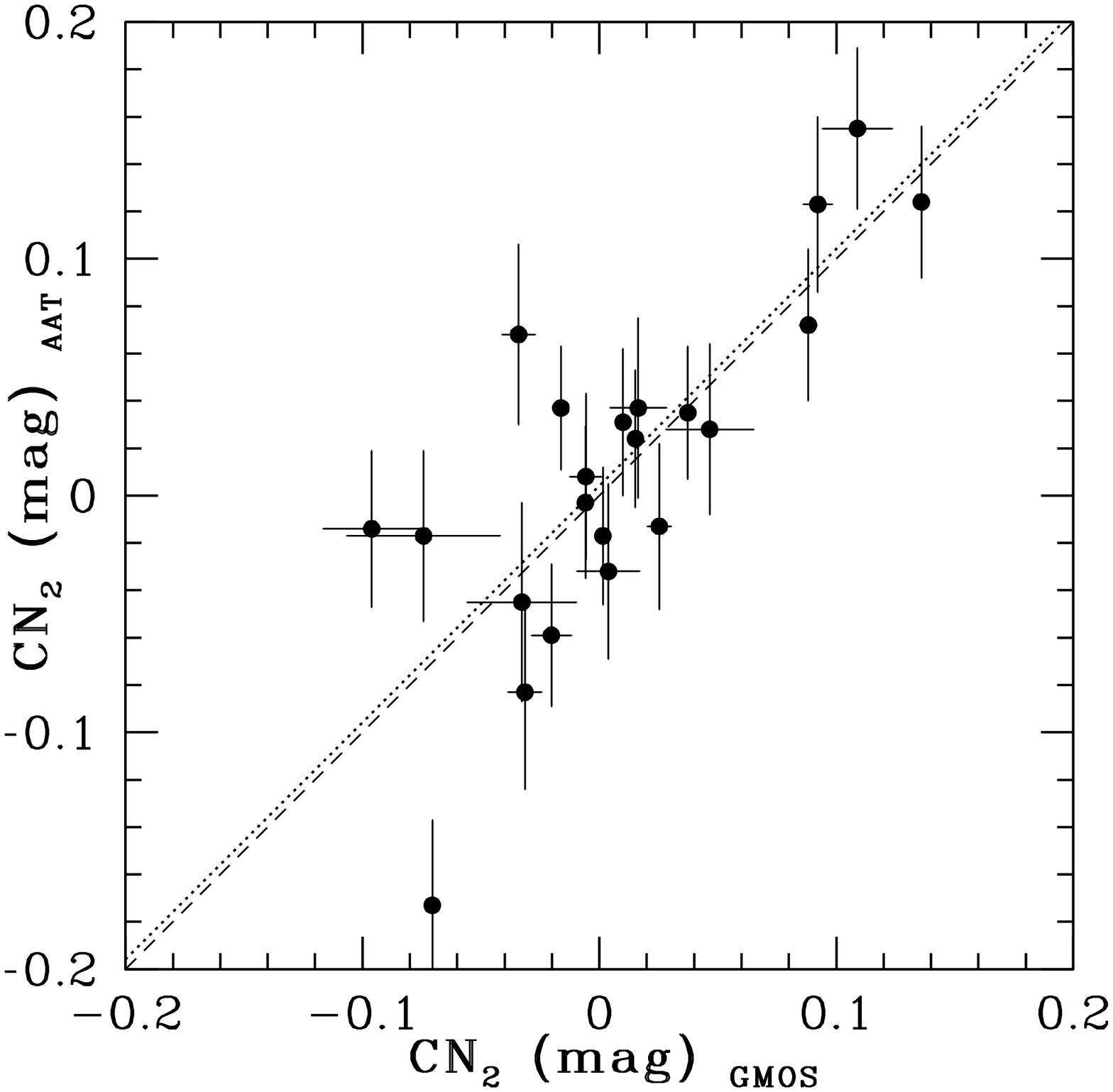,width=0.2\linewidth,clip=} \\
\epsfig{file=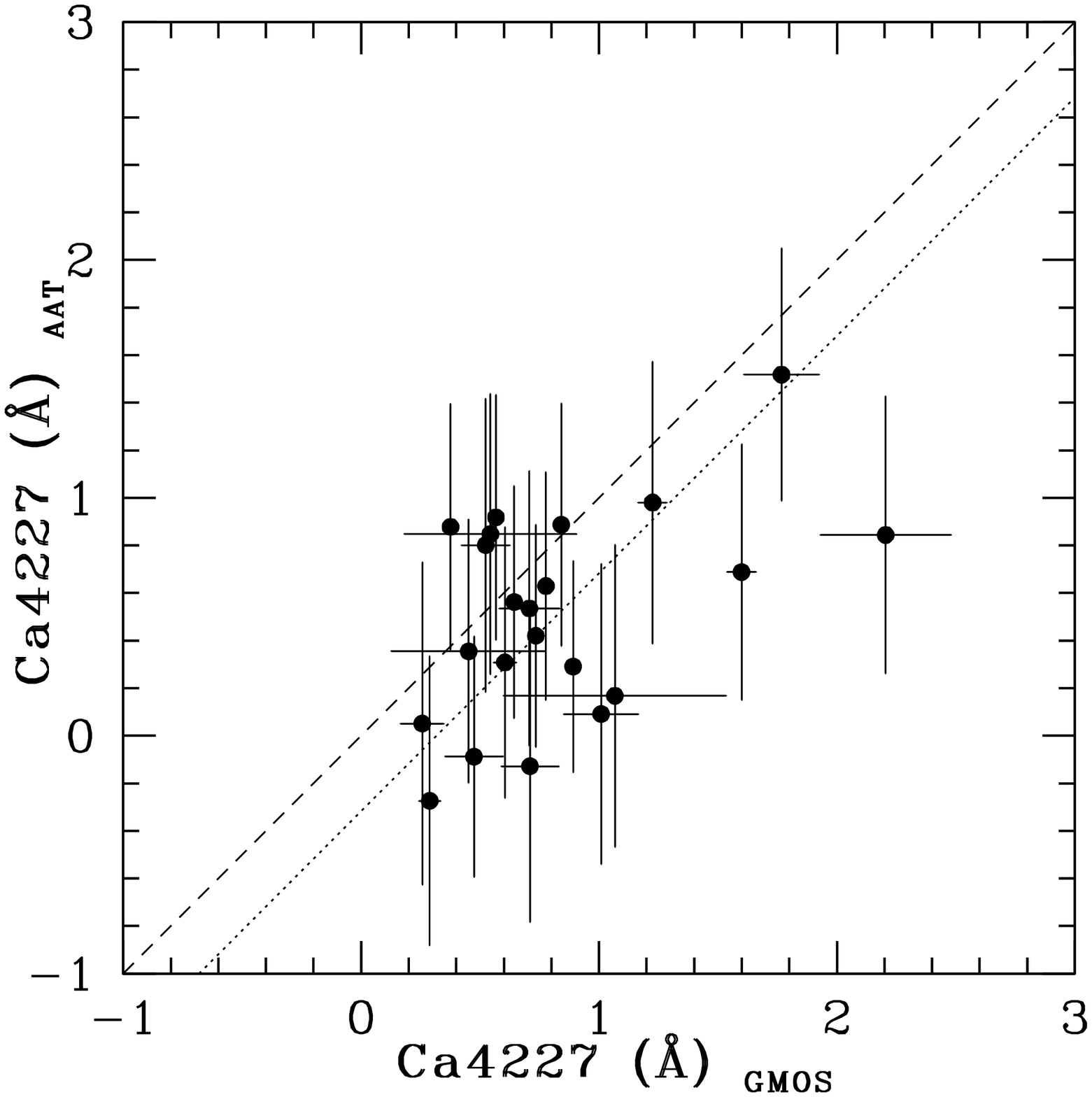,width=0.2\linewidth,clip=} &
\epsfig{file=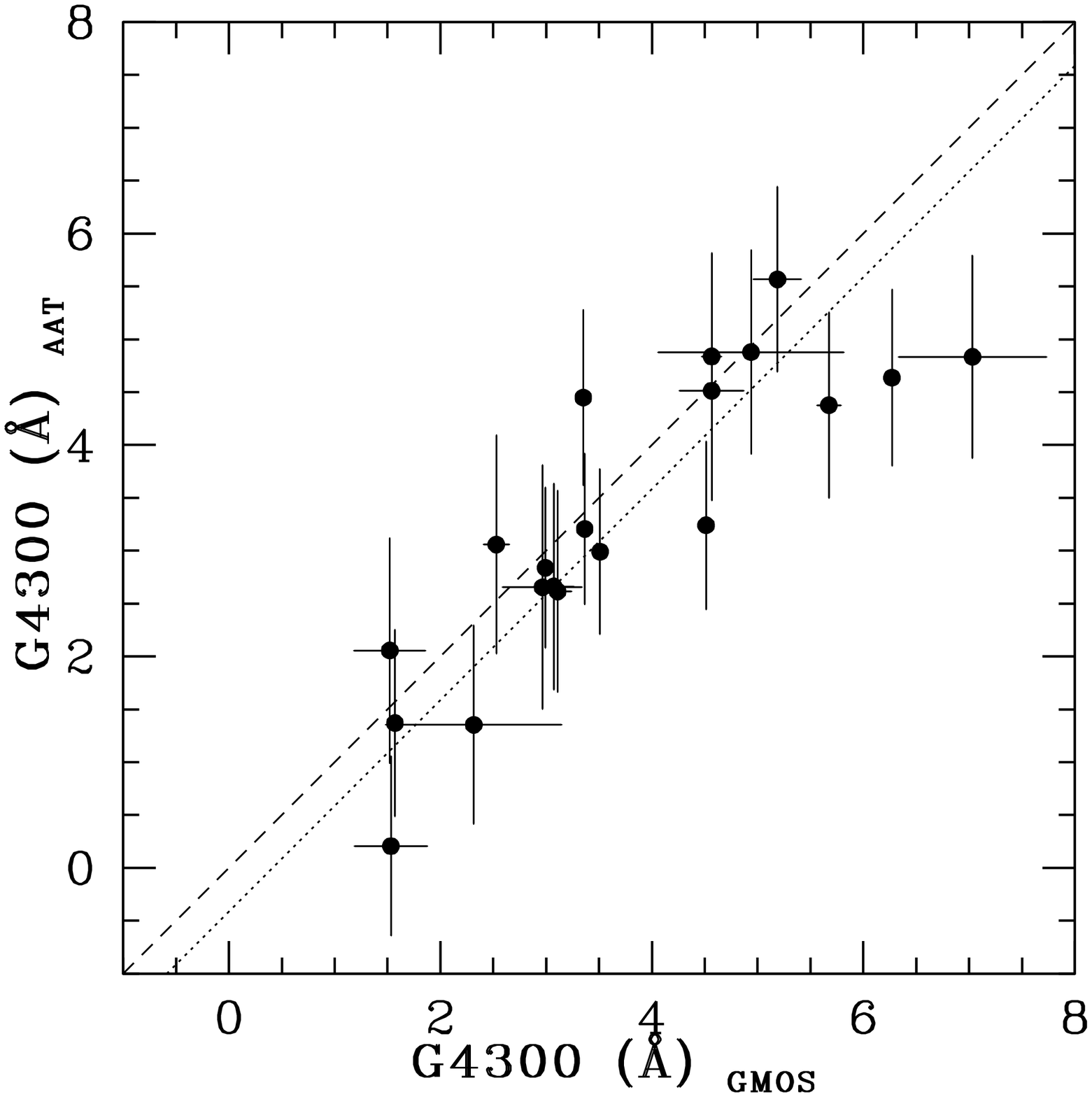,width=0.2\linewidth,clip=} &
\epsfig{file=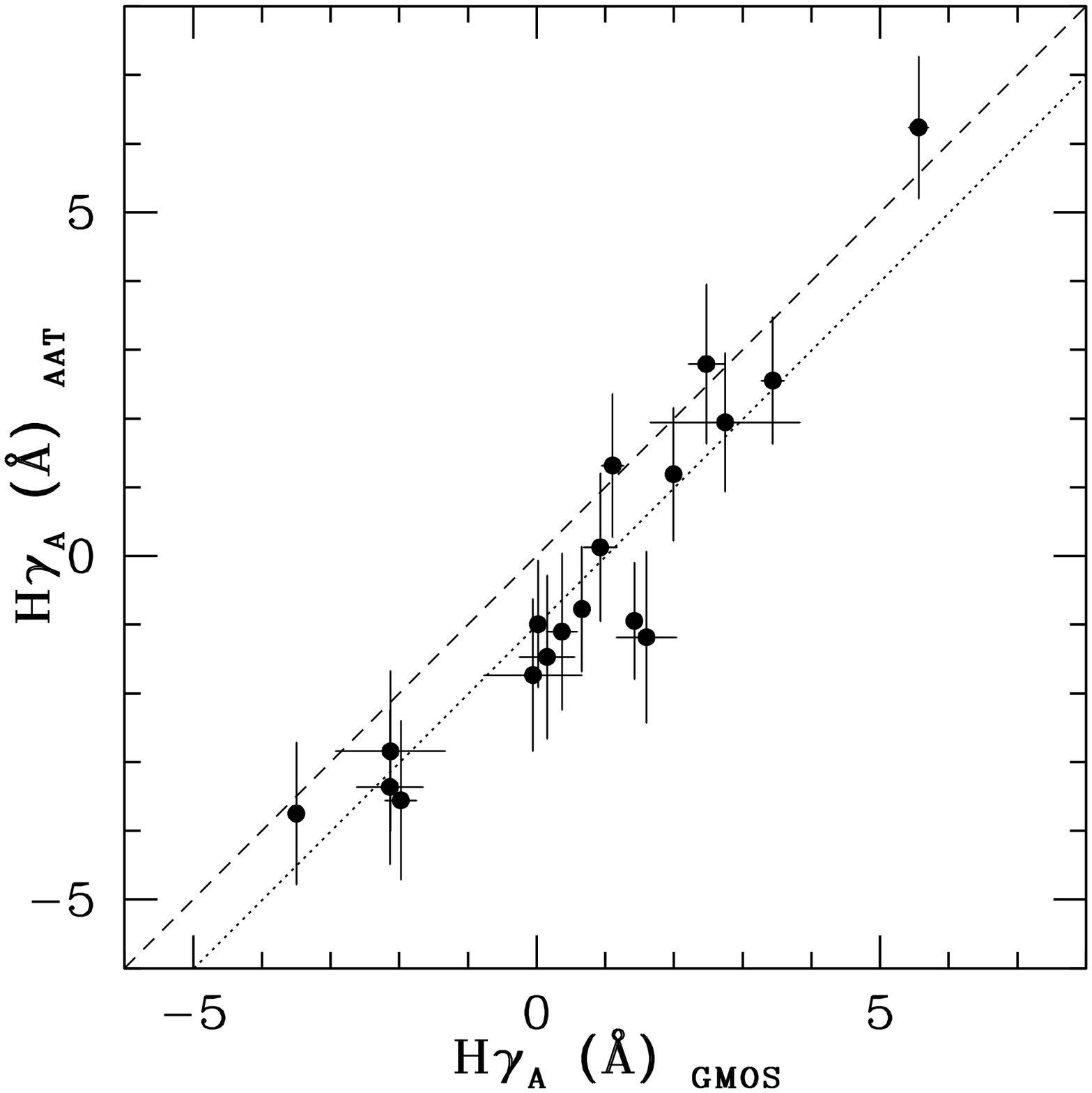,width=0.2\linewidth,clip=} &
\epsfig{file=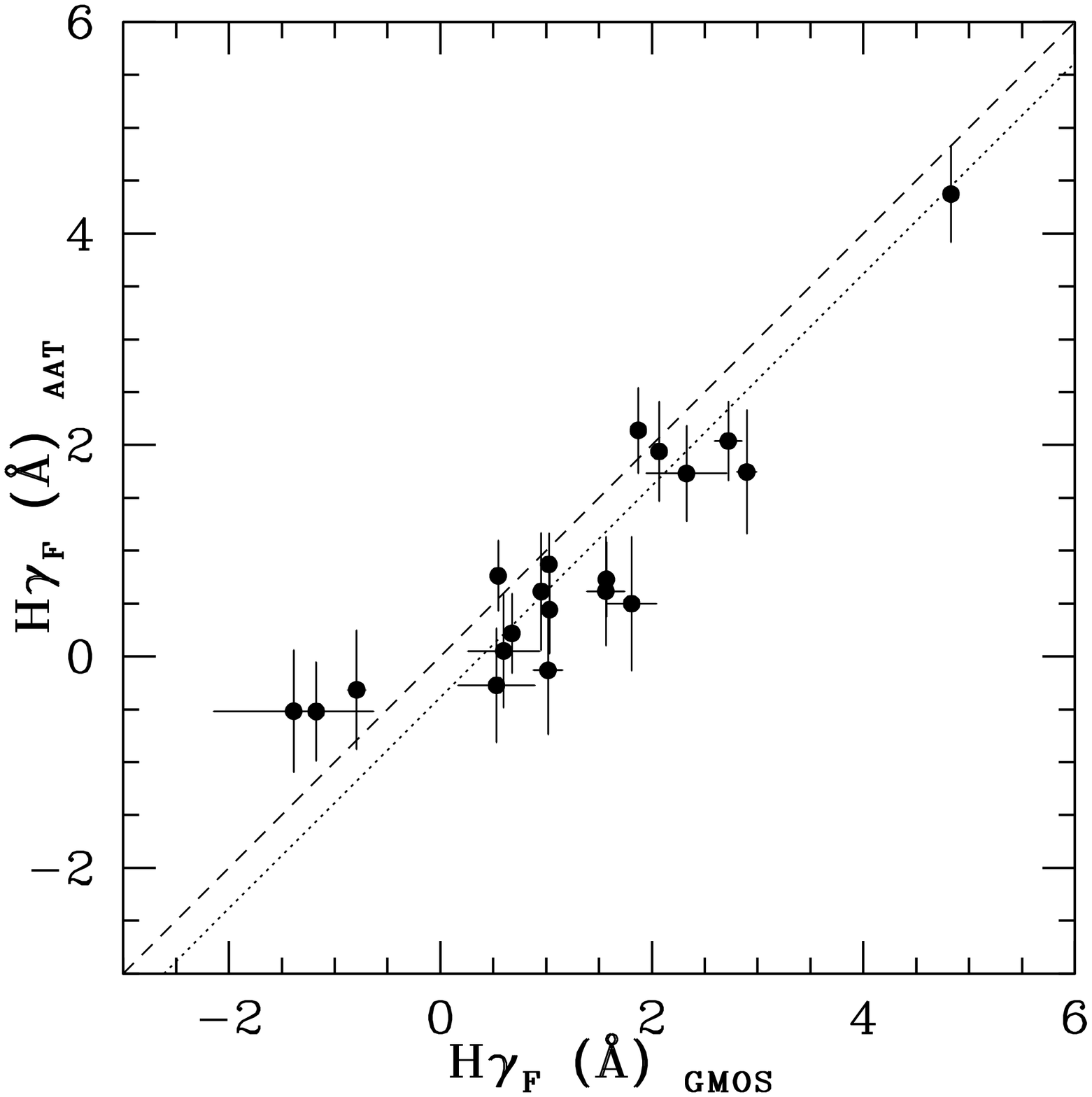,width=0.2\linewidth,clip=} \\
\epsfig{file=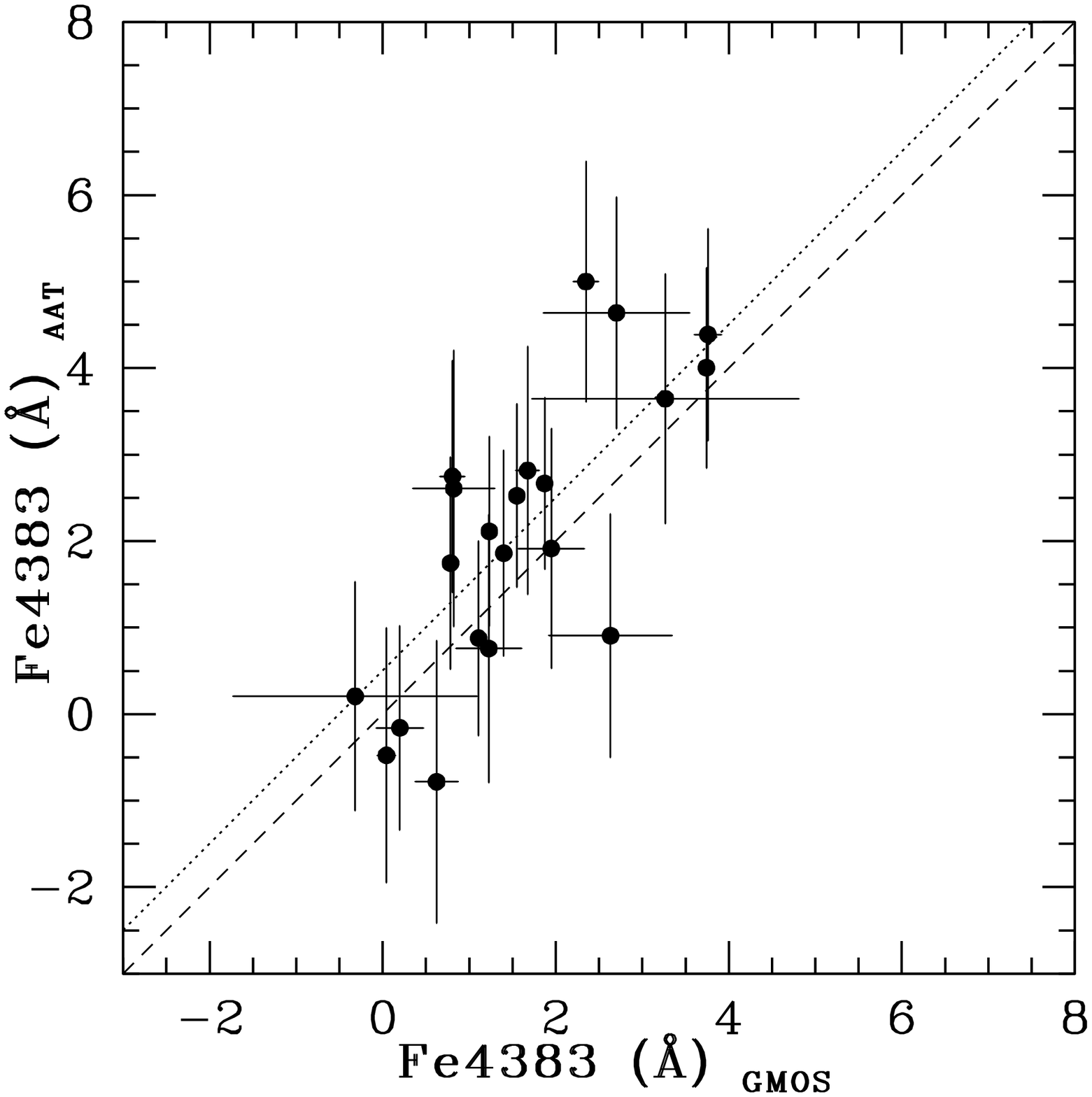,width=0.2\linewidth,clip=} &
\epsfig{file=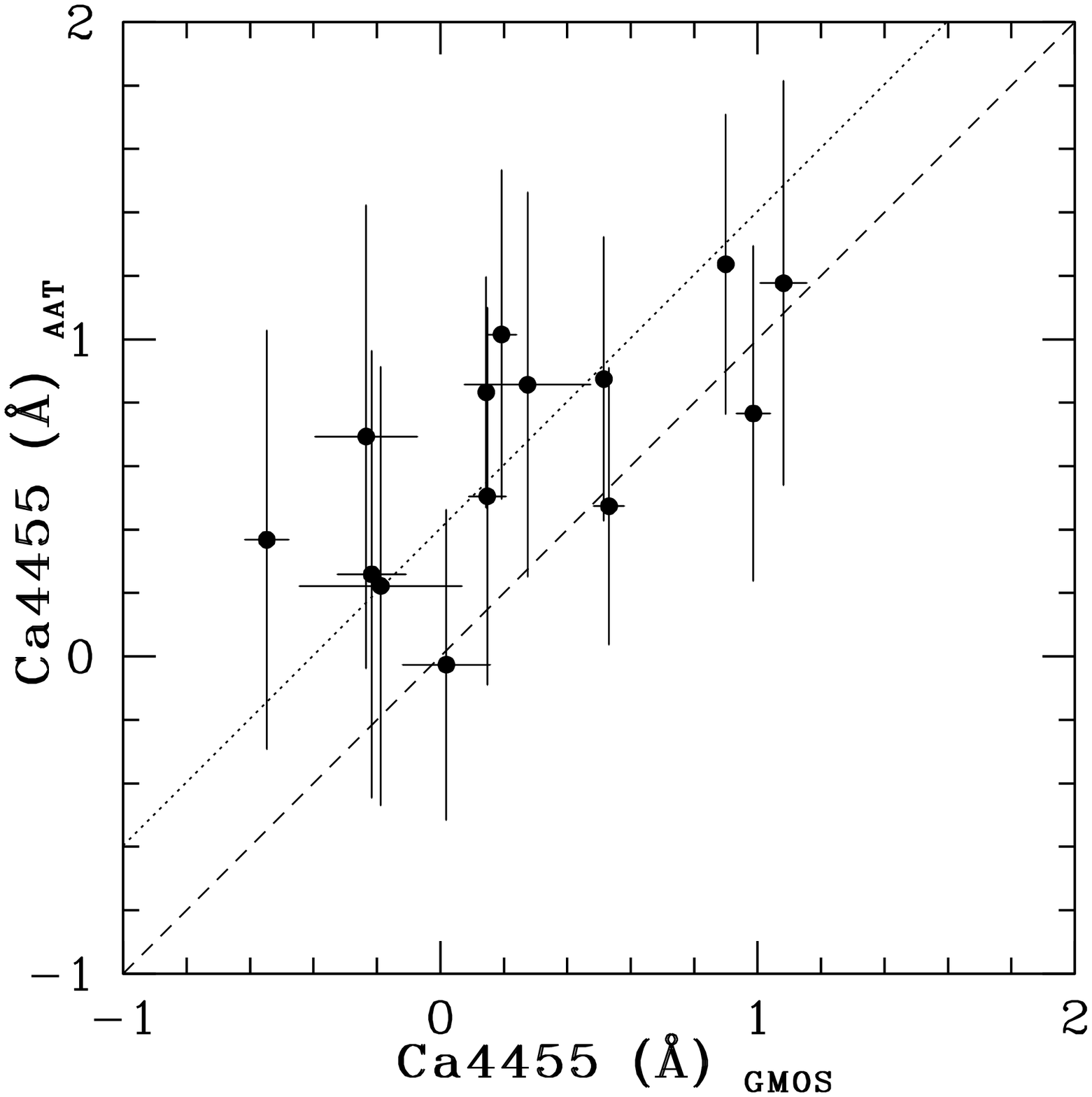,width=0.2\linewidth,clip=} &
\epsfig{file=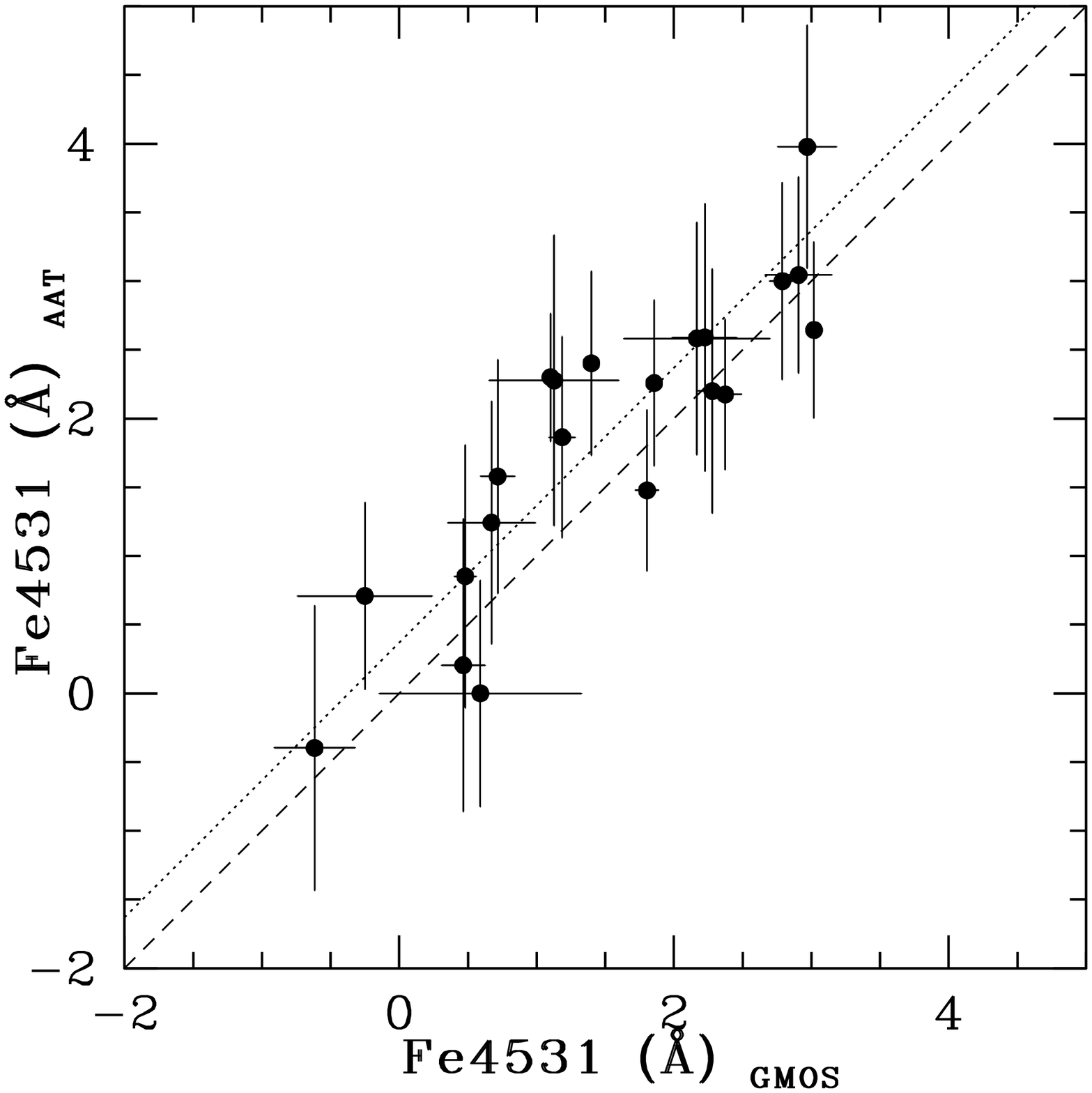,width=0.2\linewidth,clip=} &
\epsfig{file=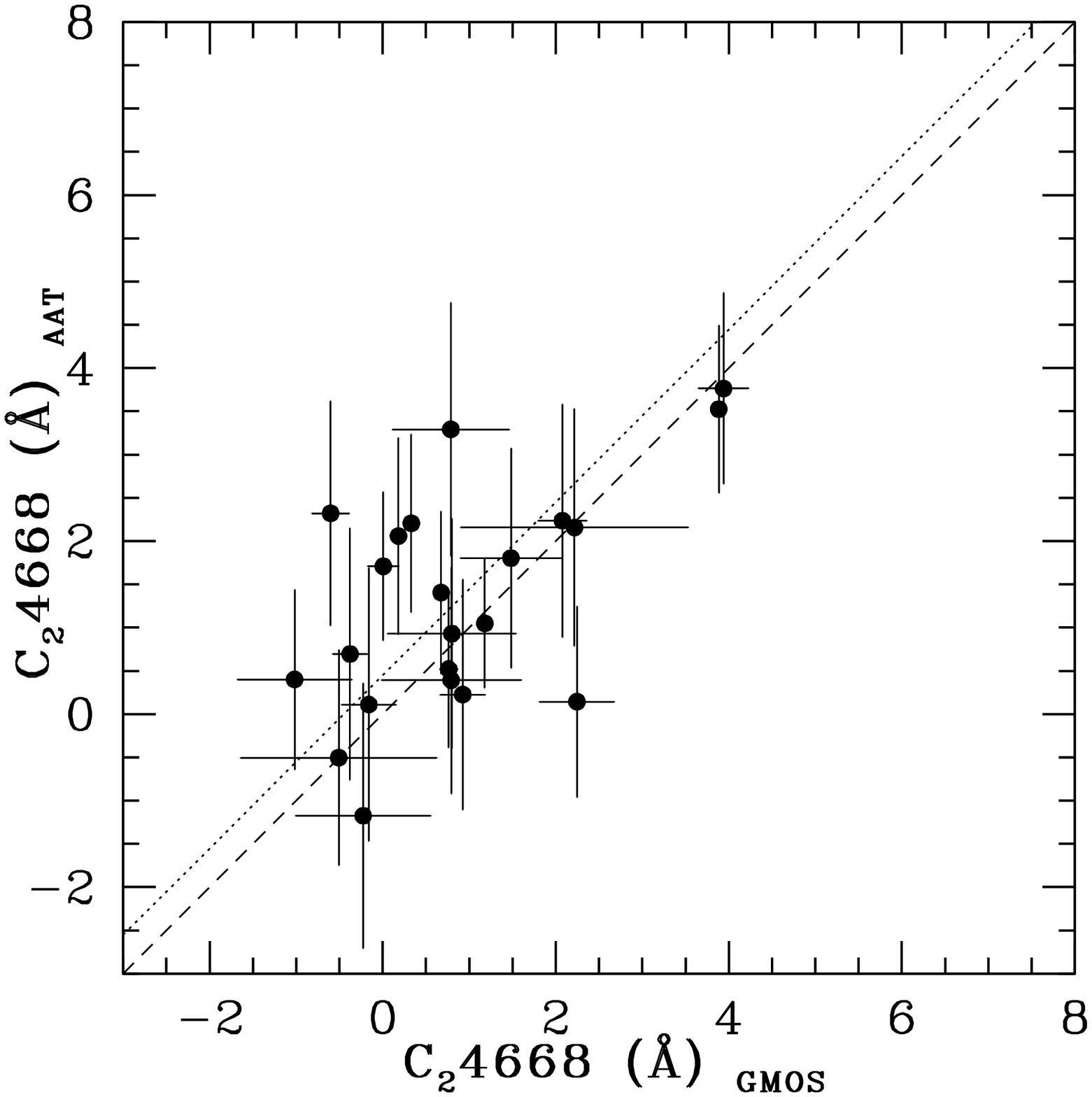,width=0.2\linewidth,clip=} \\
\epsfig{file=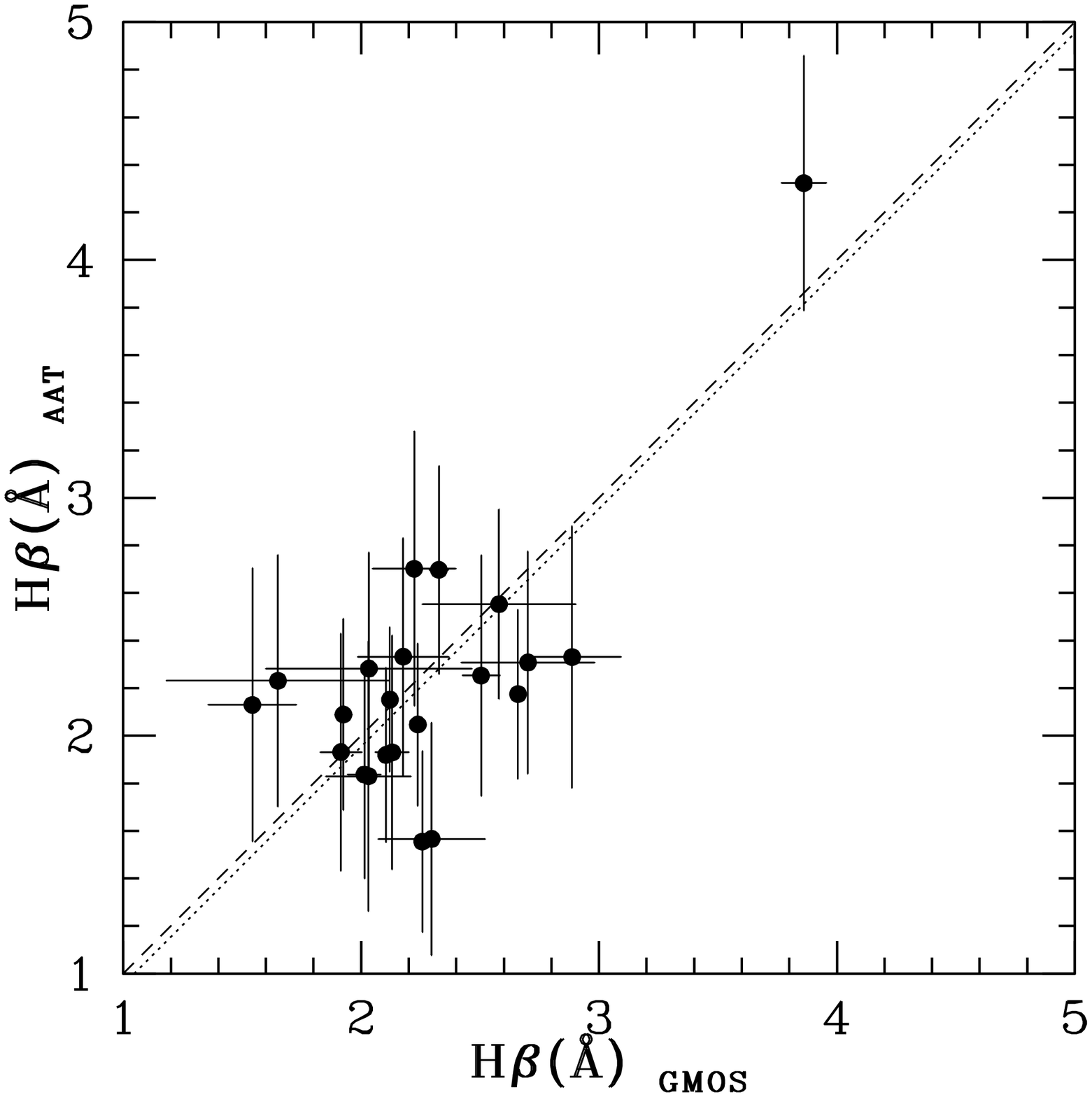,width=0.2\linewidth,clip=} &
\epsfig{file=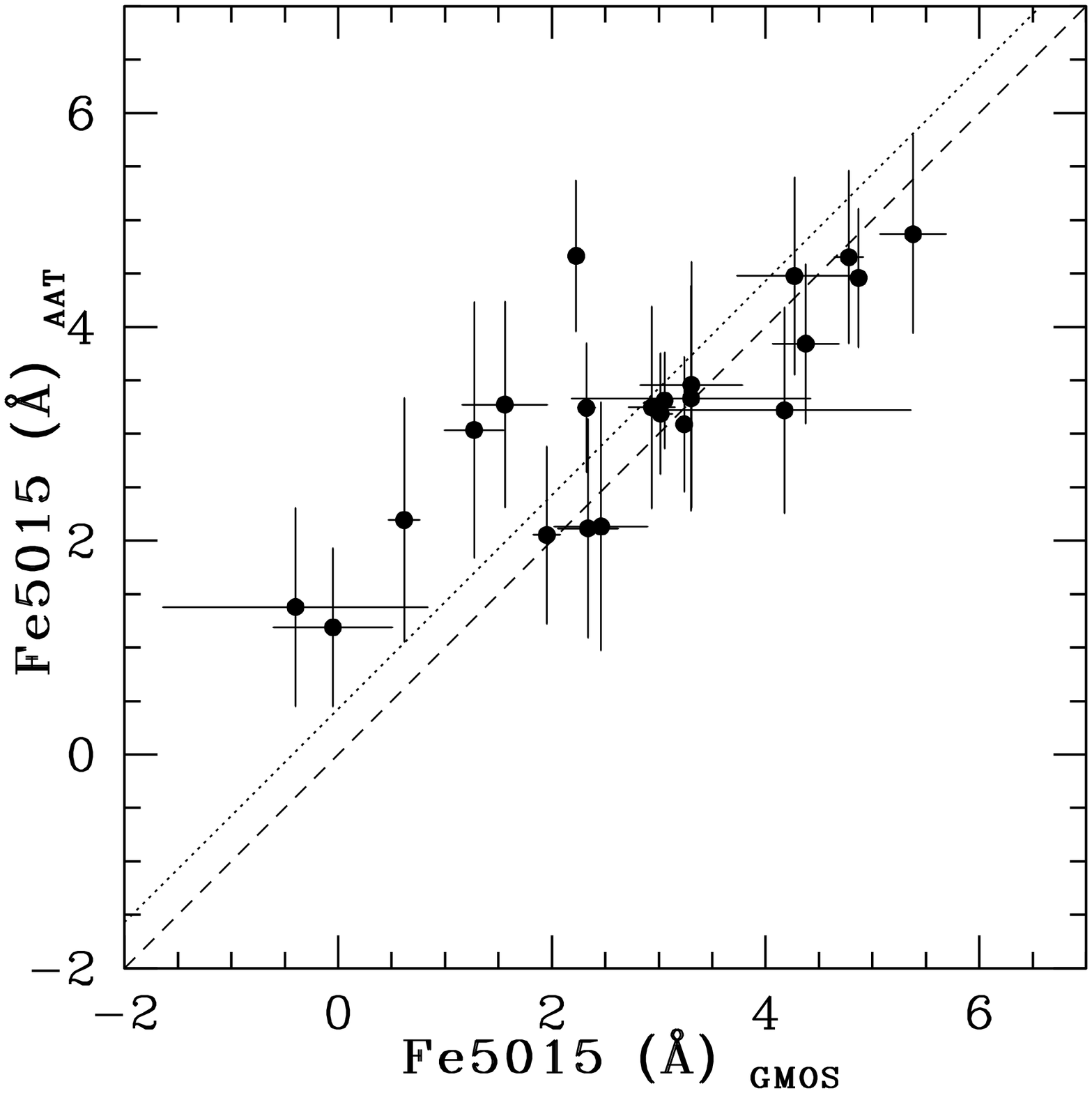,width=0.2\linewidth,clip=}&
\epsfig{file=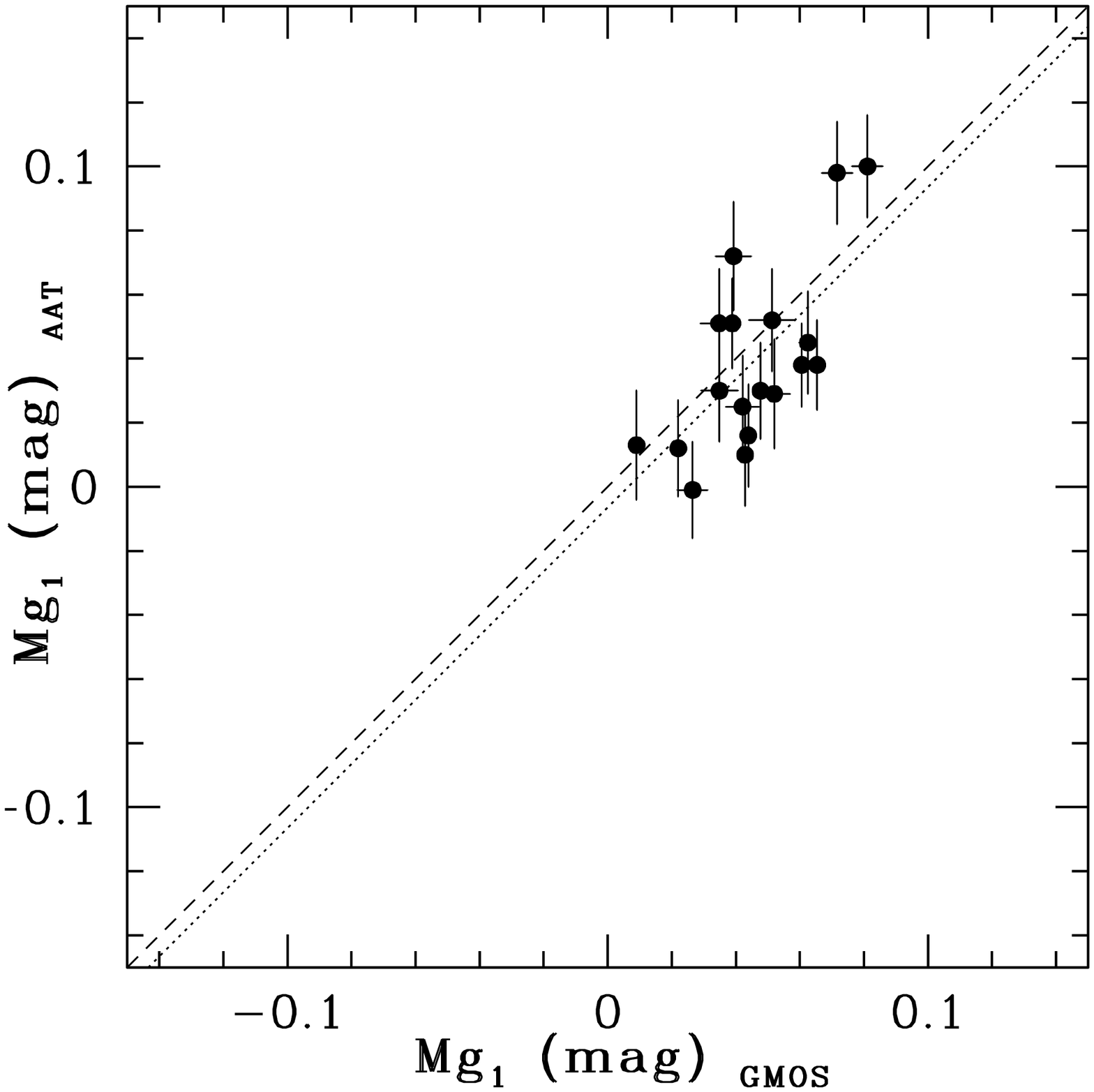,width=0.2\linewidth,clip=}&
\epsfig{file=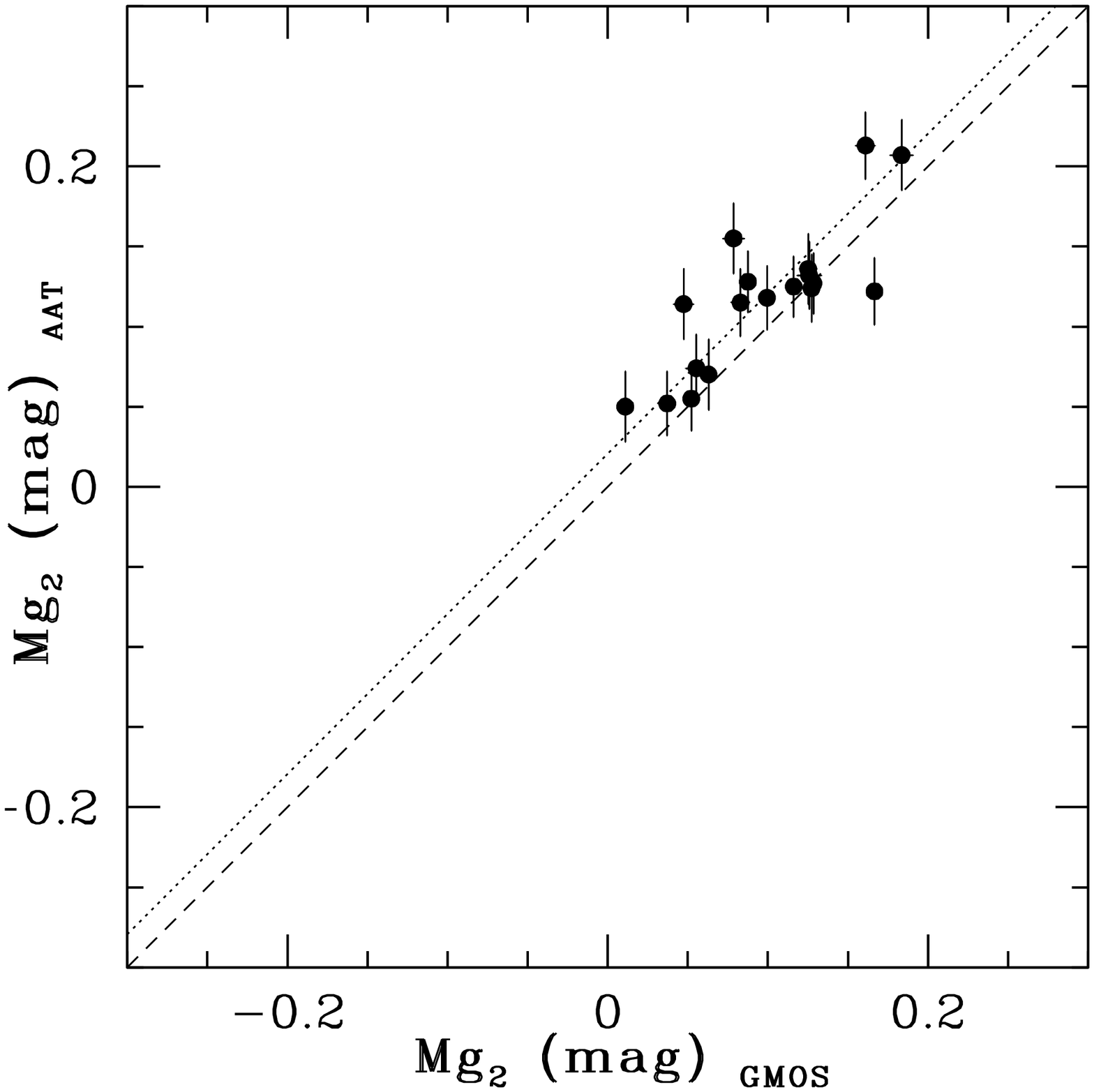,width=0.2\linewidth,clip=} \\
\epsfig{file=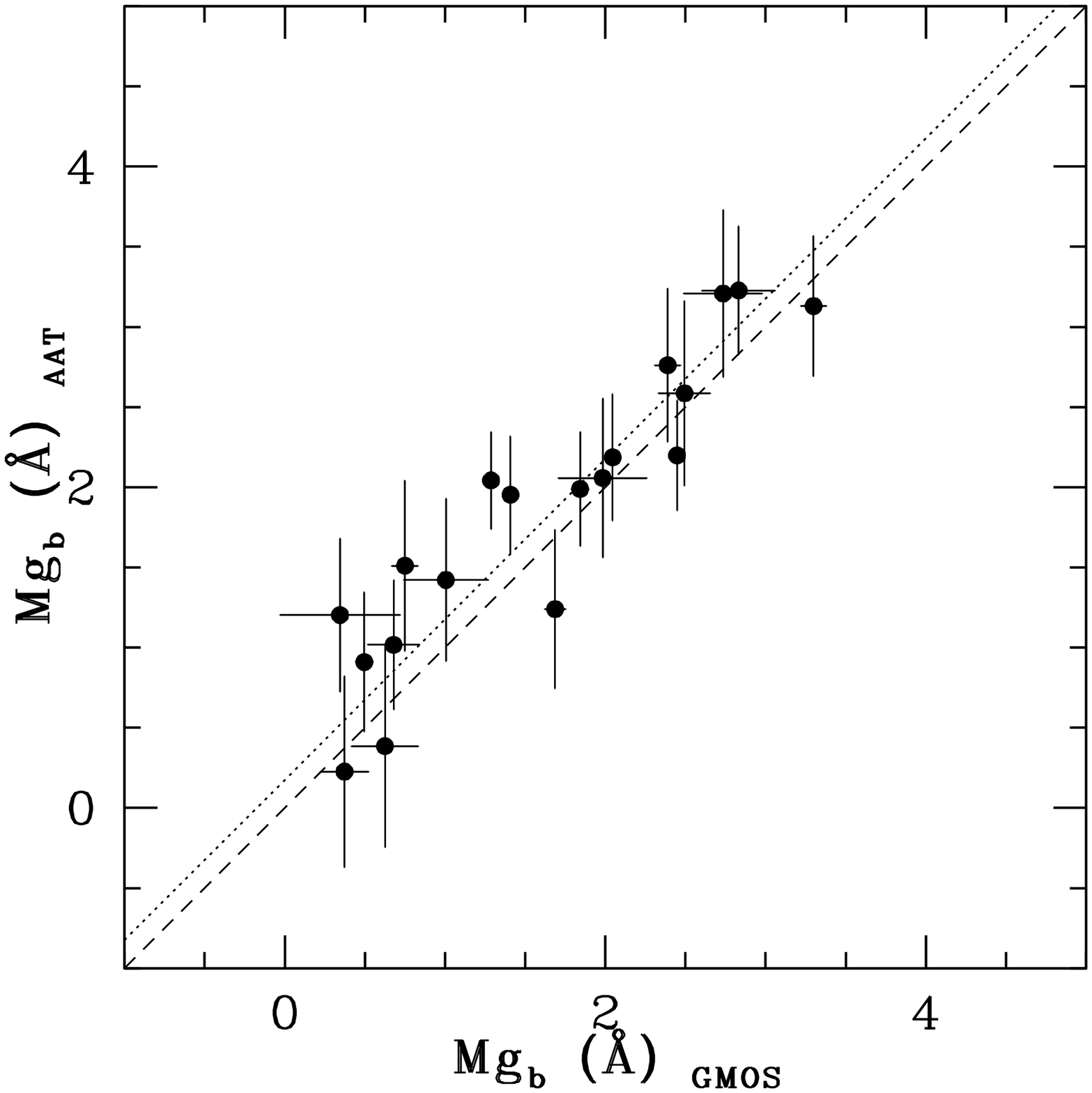,width=0.2\linewidth,clip=} &
\epsfig{file=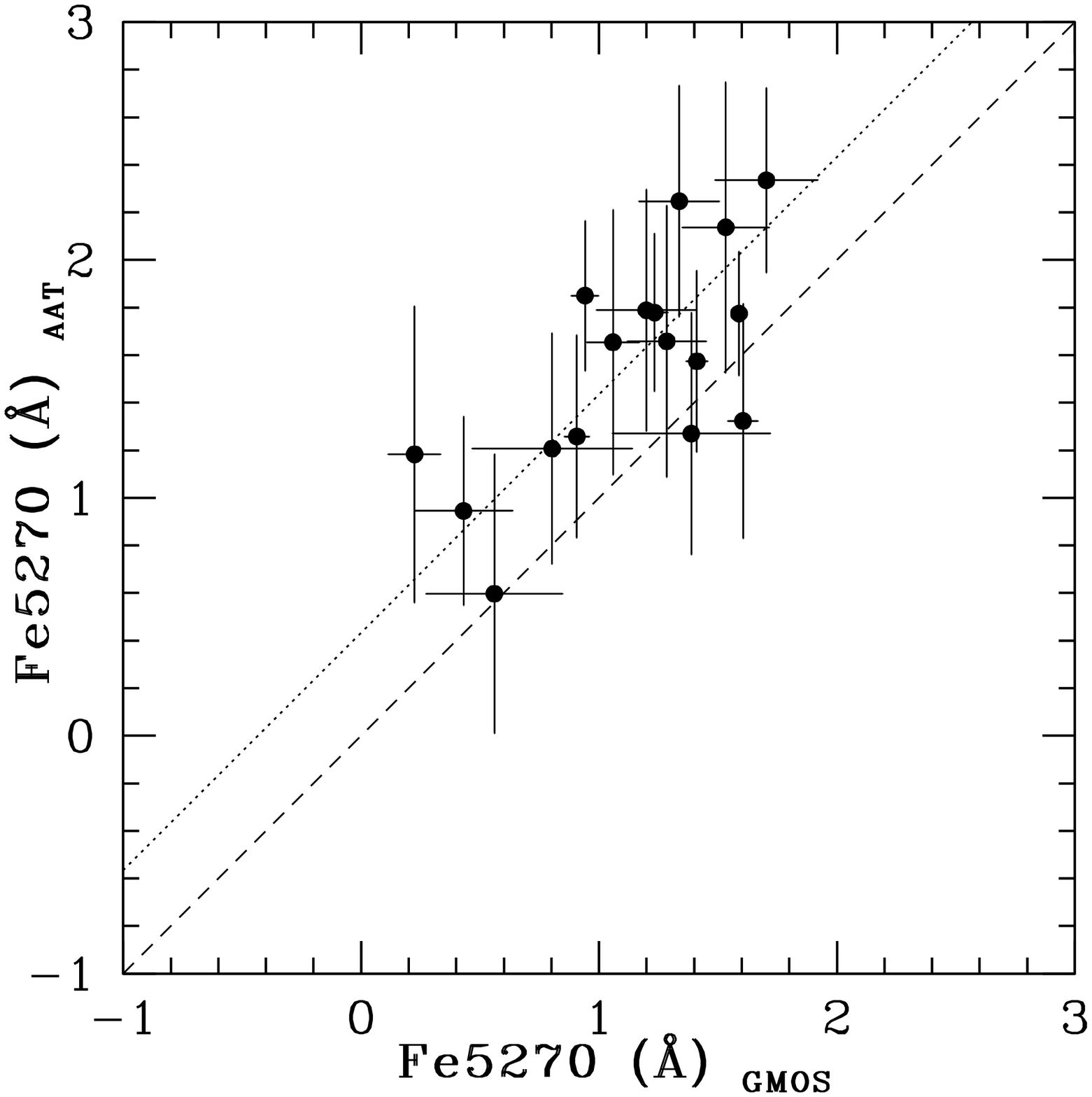,width=0.2\linewidth,clip=} &
\epsfig{file=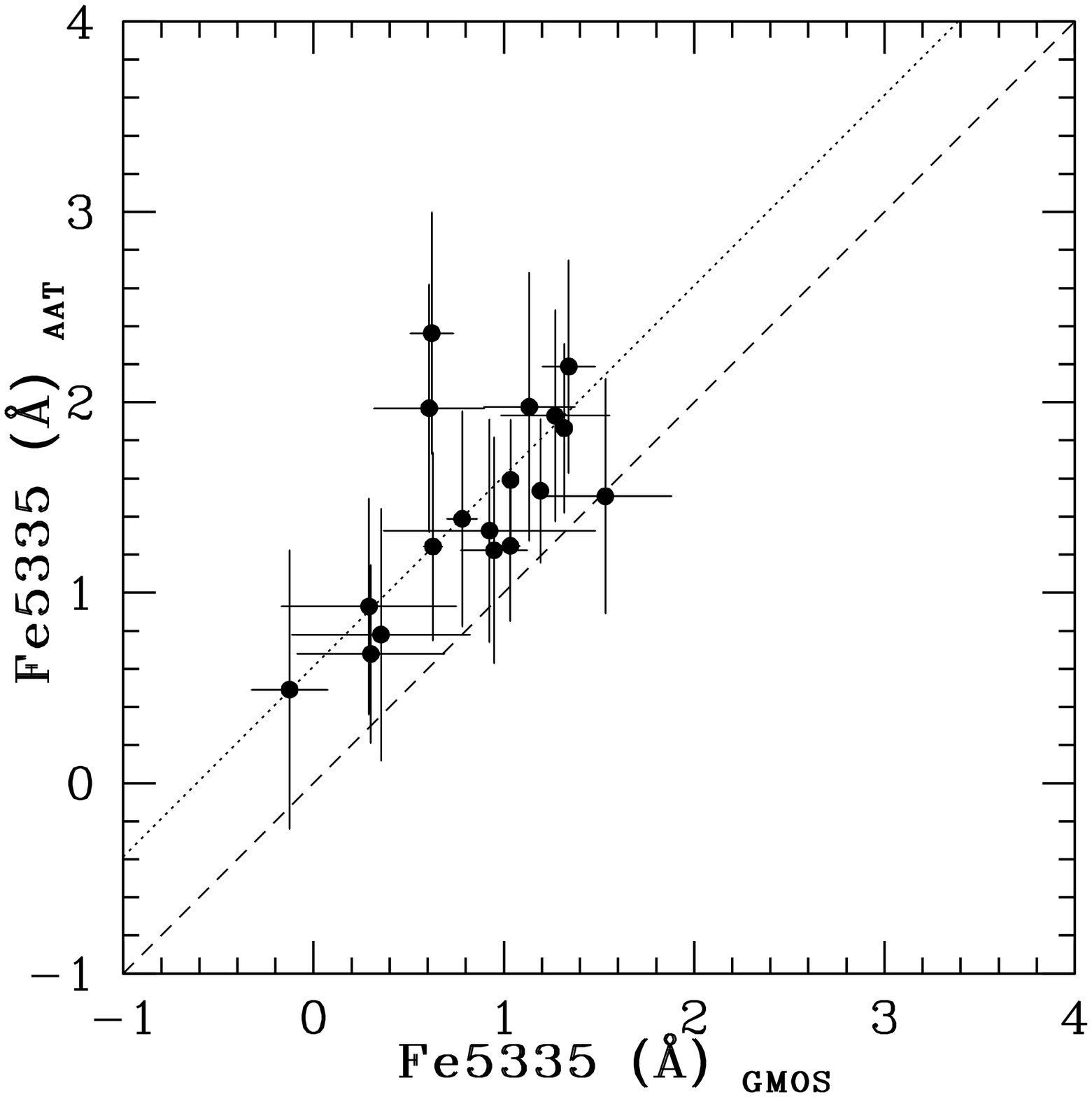,width=0.2\linewidth,clip=} &
\epsfig{file=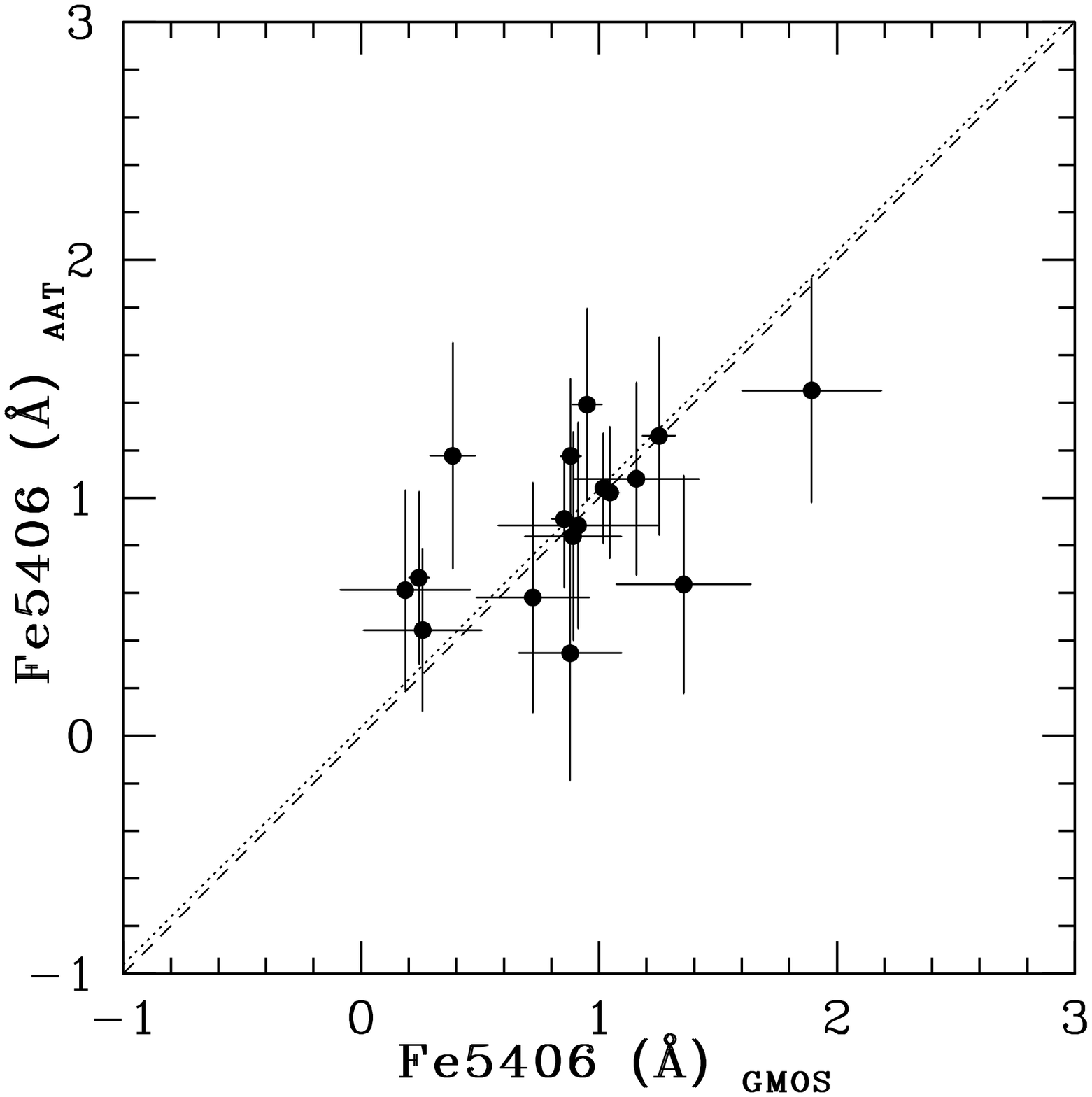,width=0.2\linewidth,clip=} \\

\end{tabular}
\caption{The calibration of index measurement from the common GMOS
spectra of 2005 with those of the AAT study by \cite{beasley08},
which have been calibrated to the Lick index system. All globular
clusters within $2\sigma$ of the mean shift between the two
  systems were kept for the fit.  The dashed line is the one-to-one
  line and the dotted line is the mean shift with a slope of one.  Within the scatter of the two
  datasets, the mean shift is a valid correction to the Lick index
  system.} 
\label{fig:cal05}
\end{figure}

\begin{figure}
\centering
\begin{tabular}{cccc}
\epsfig{file=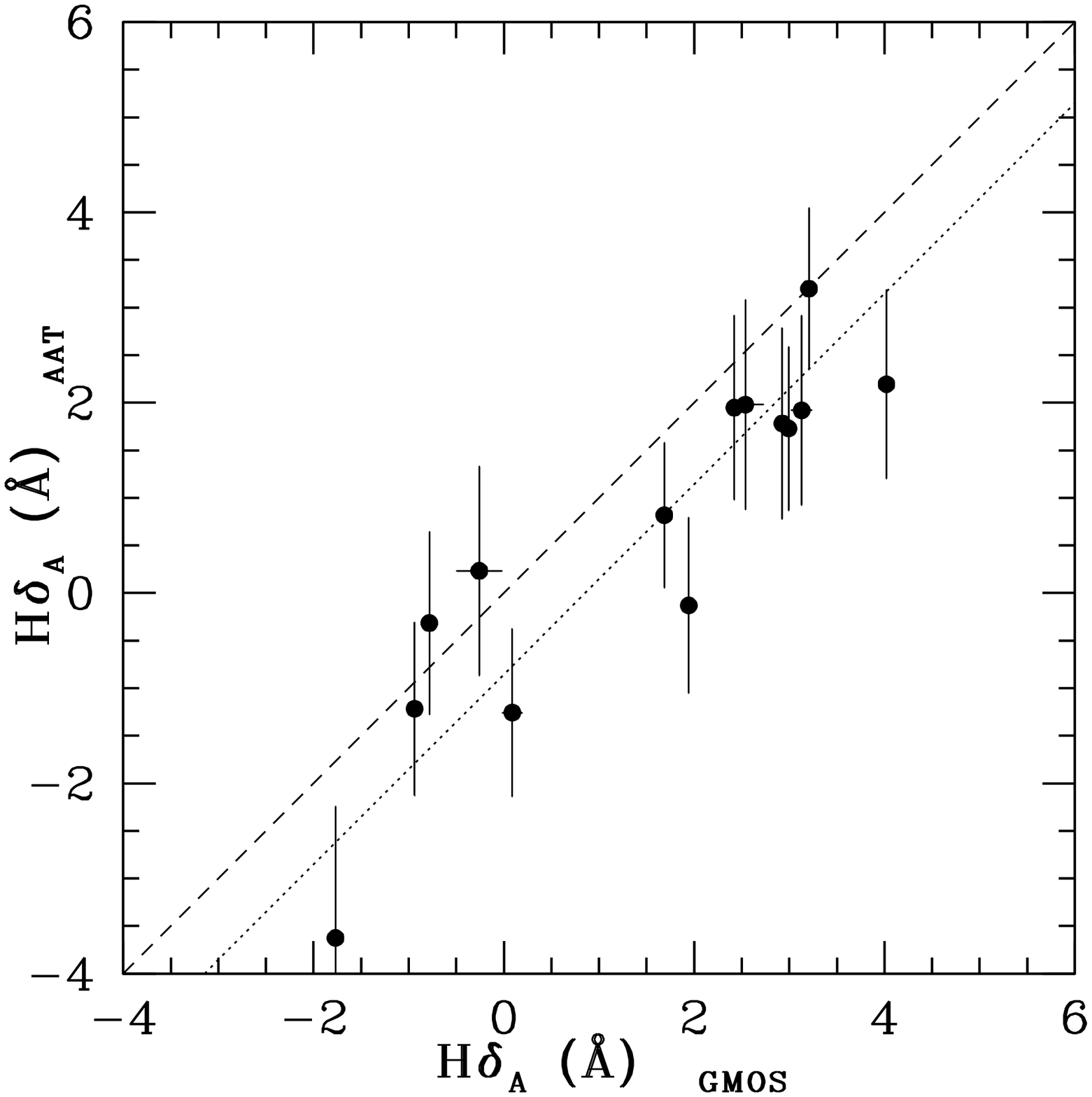,width=0.2\linewidth,clip=} &
\epsfig{file=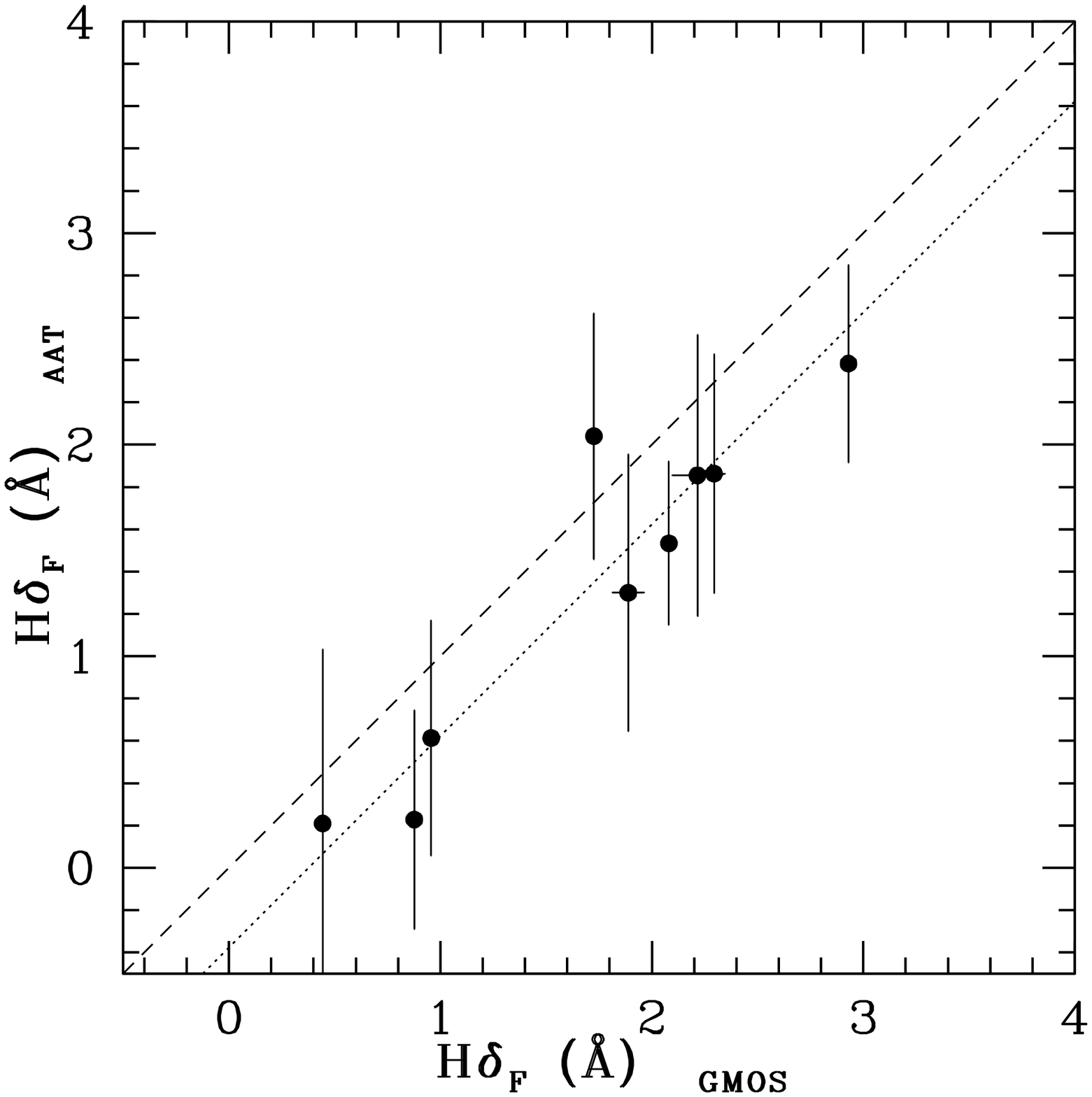,width=0.2\linewidth,clip=} &
\epsfig{file=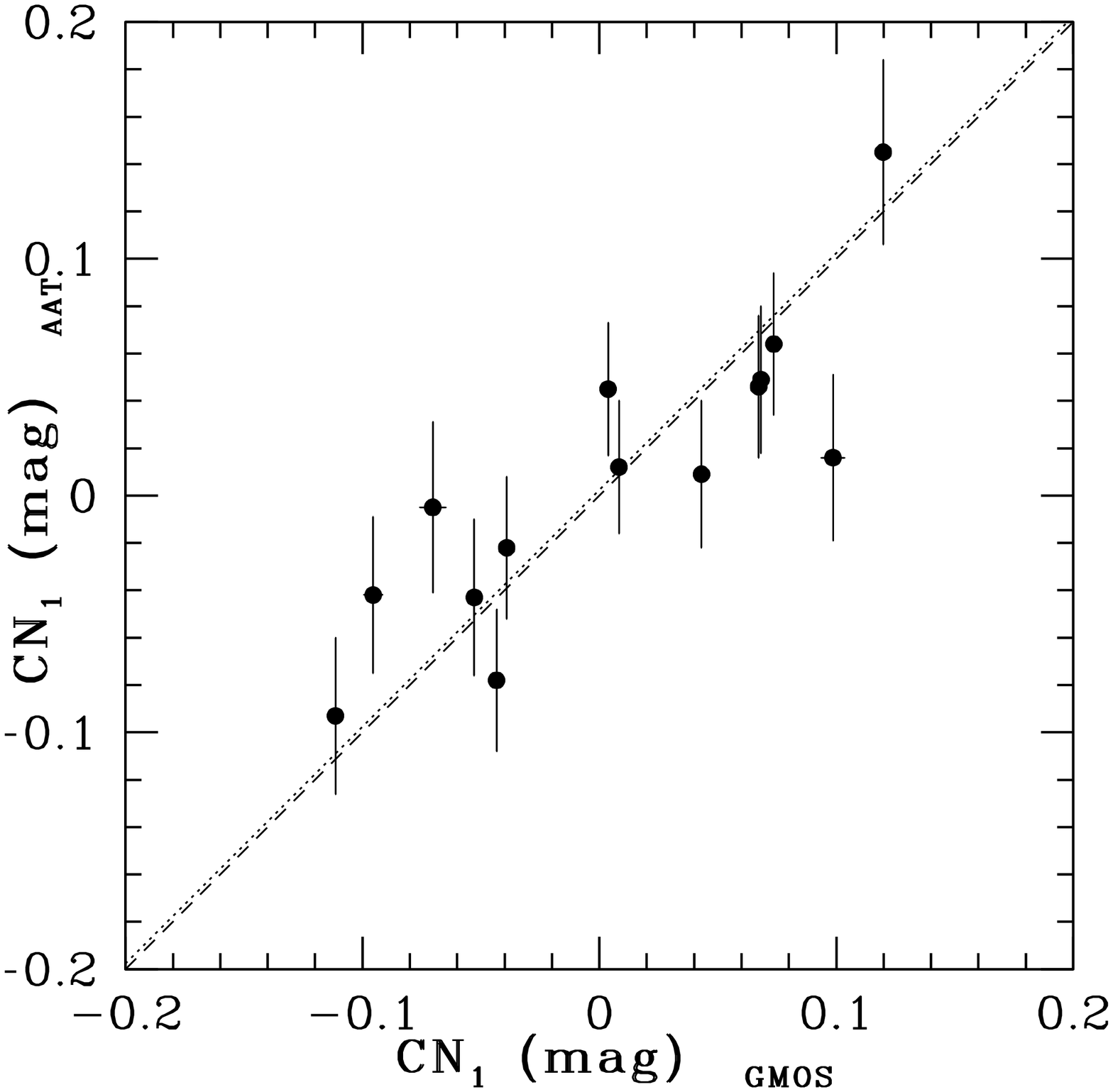,width=0.2\linewidth,clip=} &
\epsfig{file=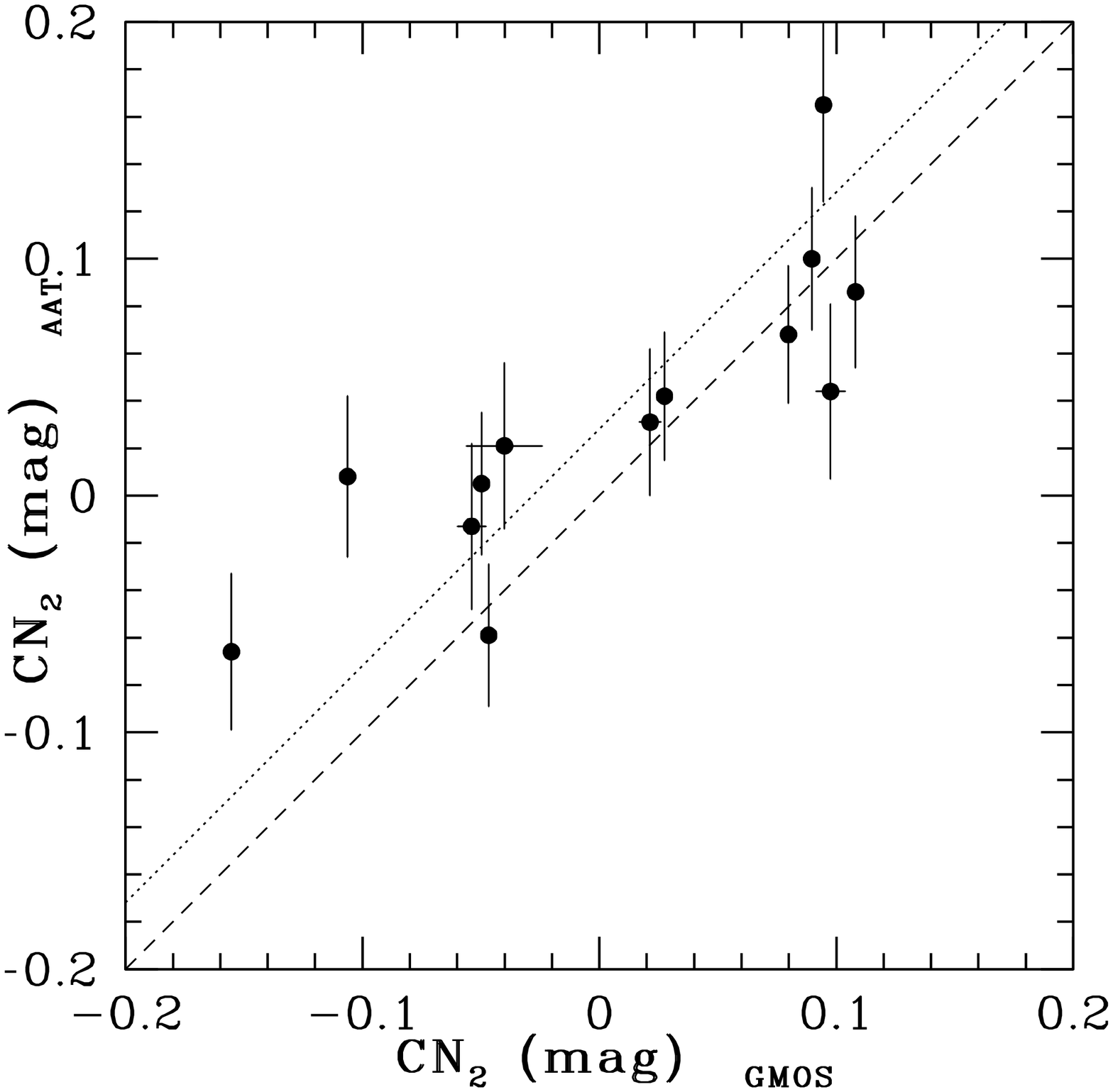,width=0.2\linewidth,clip=} \\
\epsfig{file=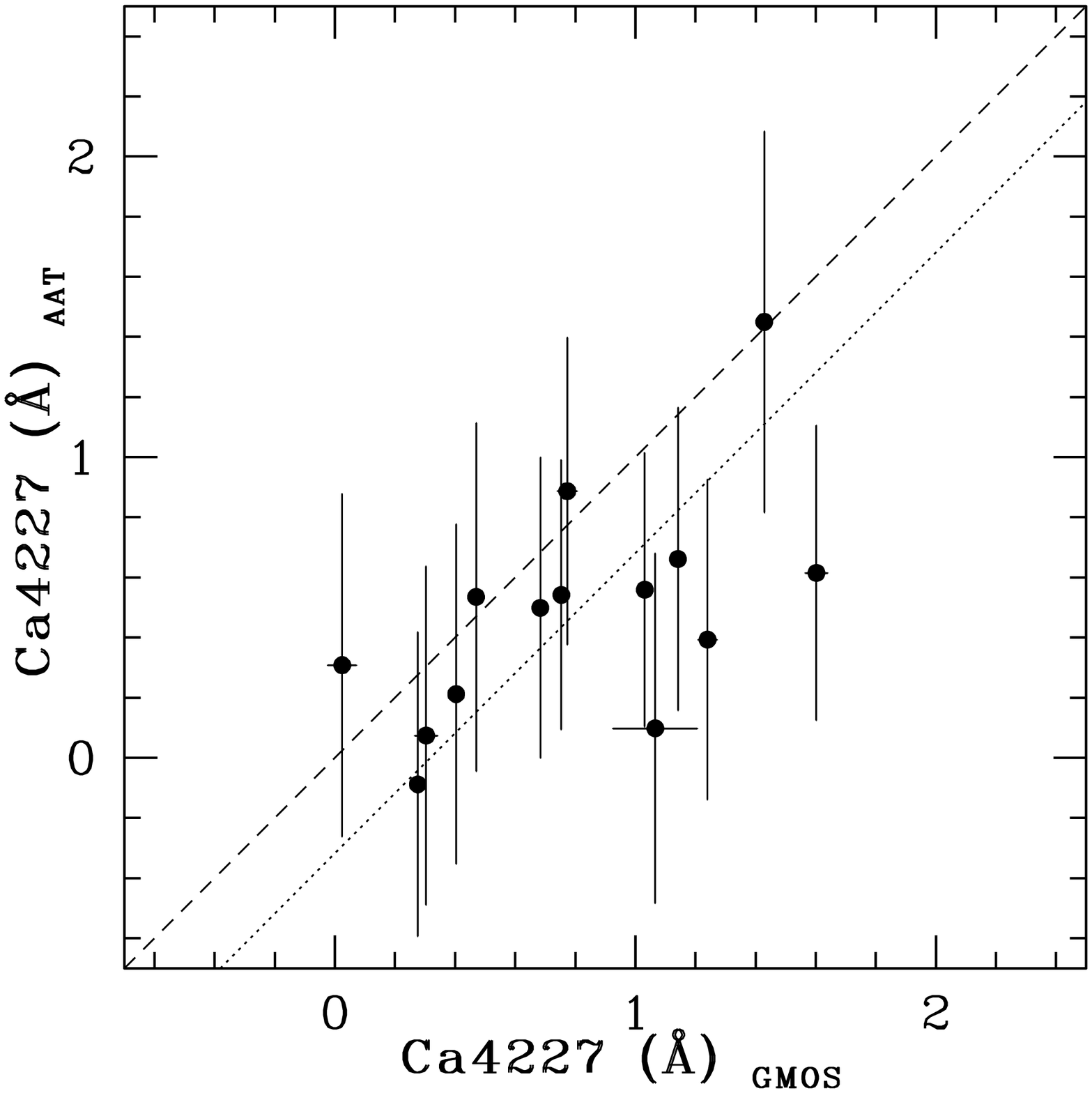,width=0.2\linewidth,clip=} &
\epsfig{file=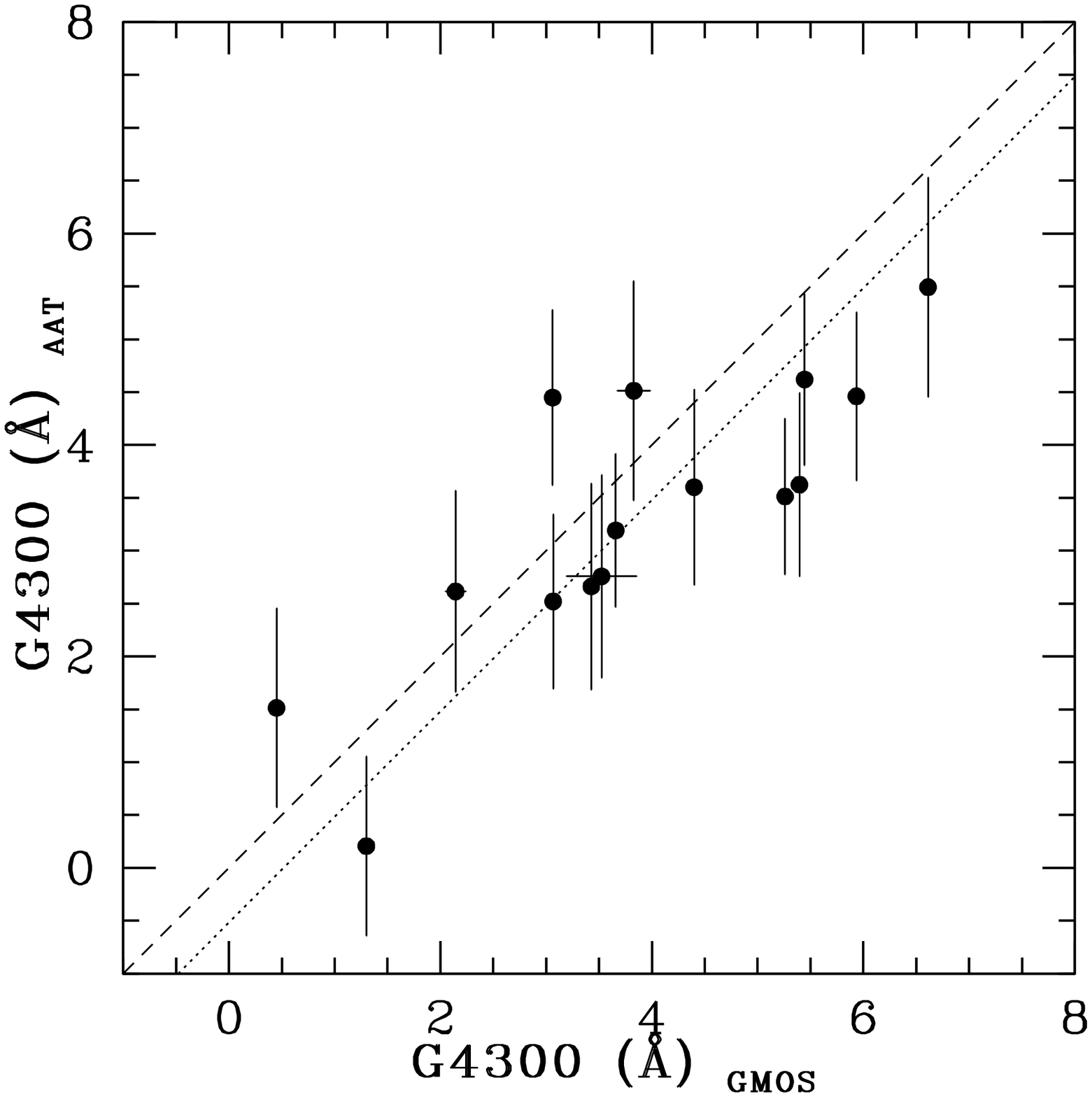,width=0.2\linewidth,clip=} &
\epsfig{file=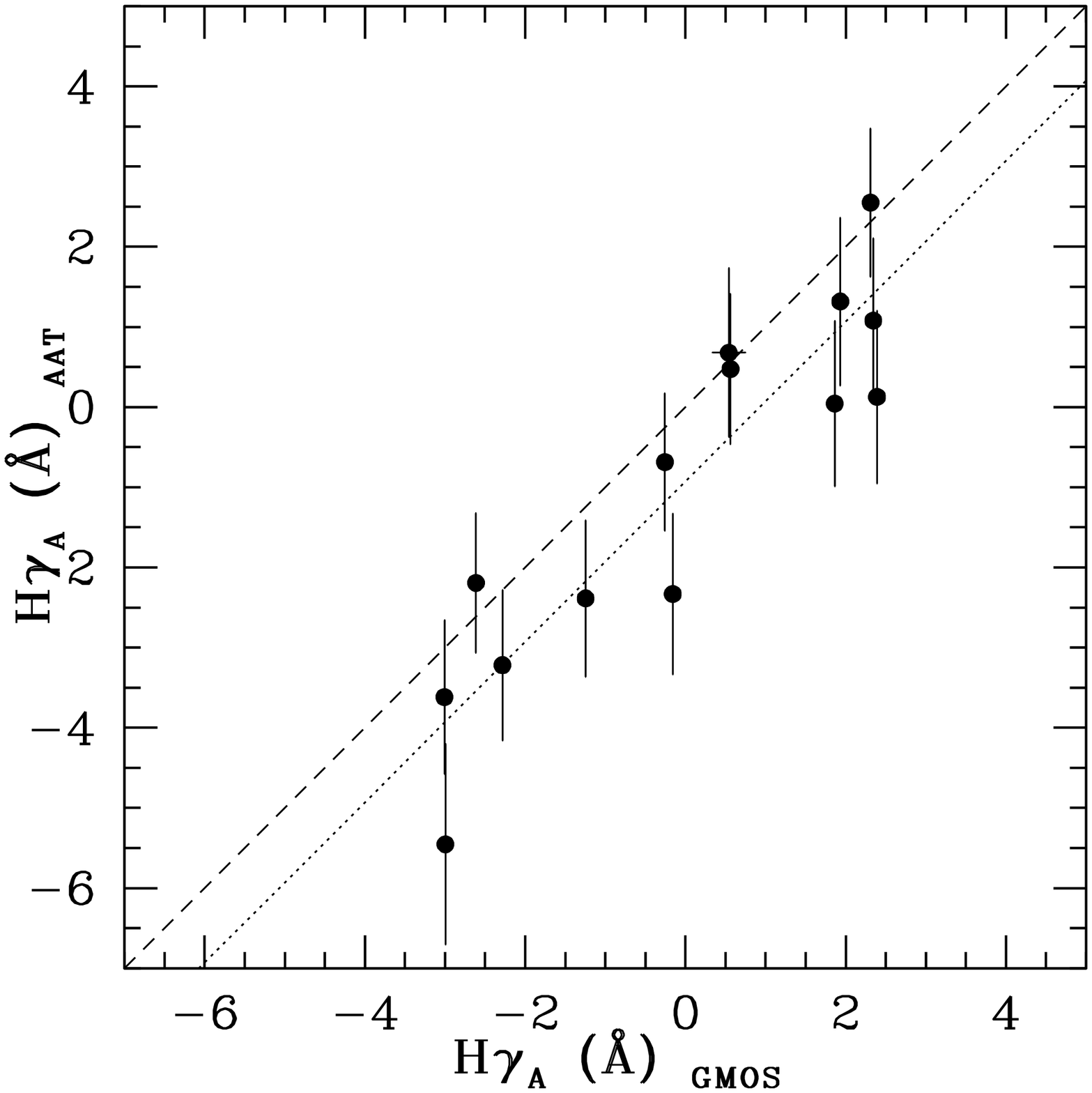,width=0.2\linewidth,clip=} &
\epsfig{file=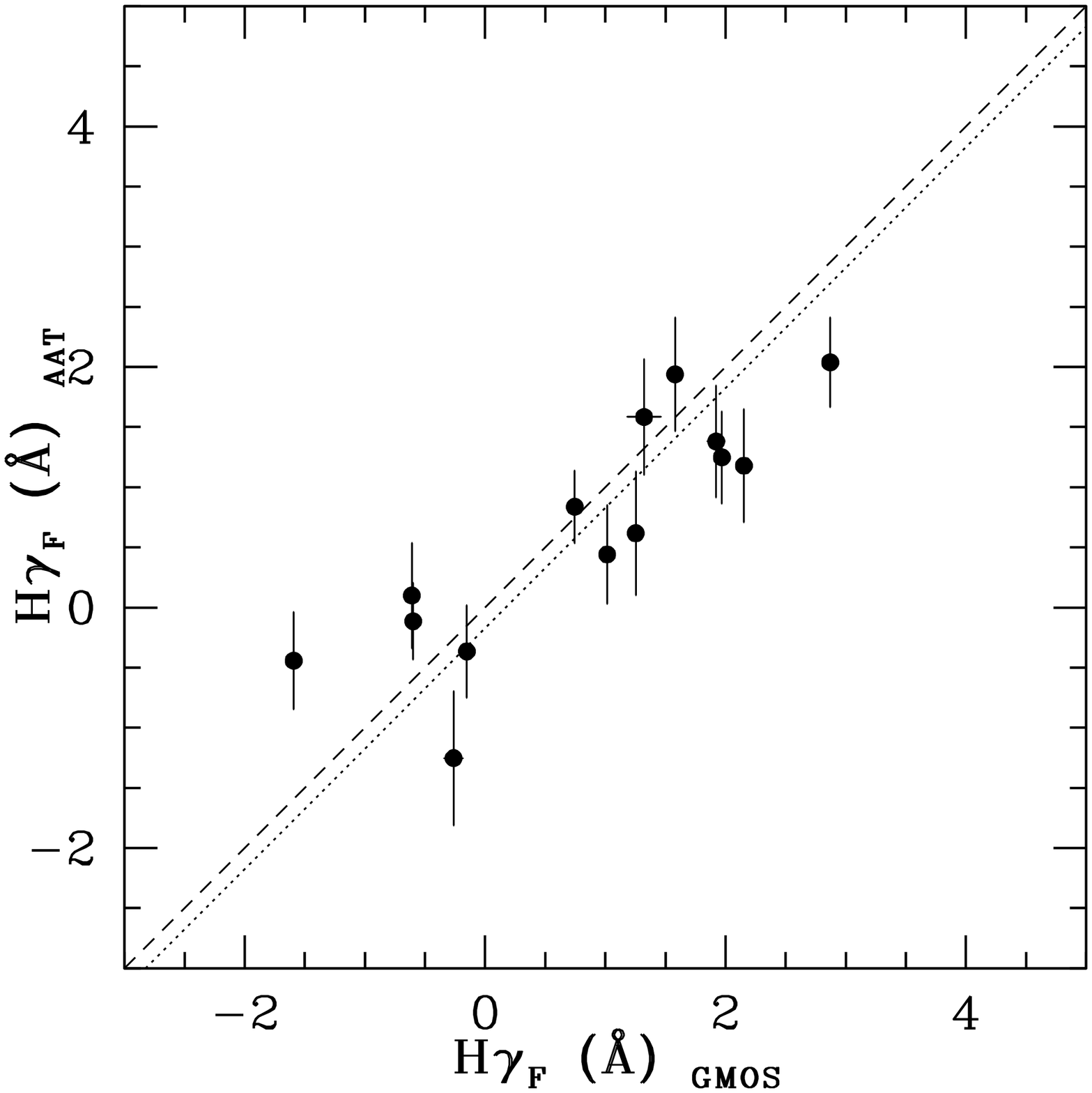,width=0.2\linewidth,clip=} \\
\epsfig{file=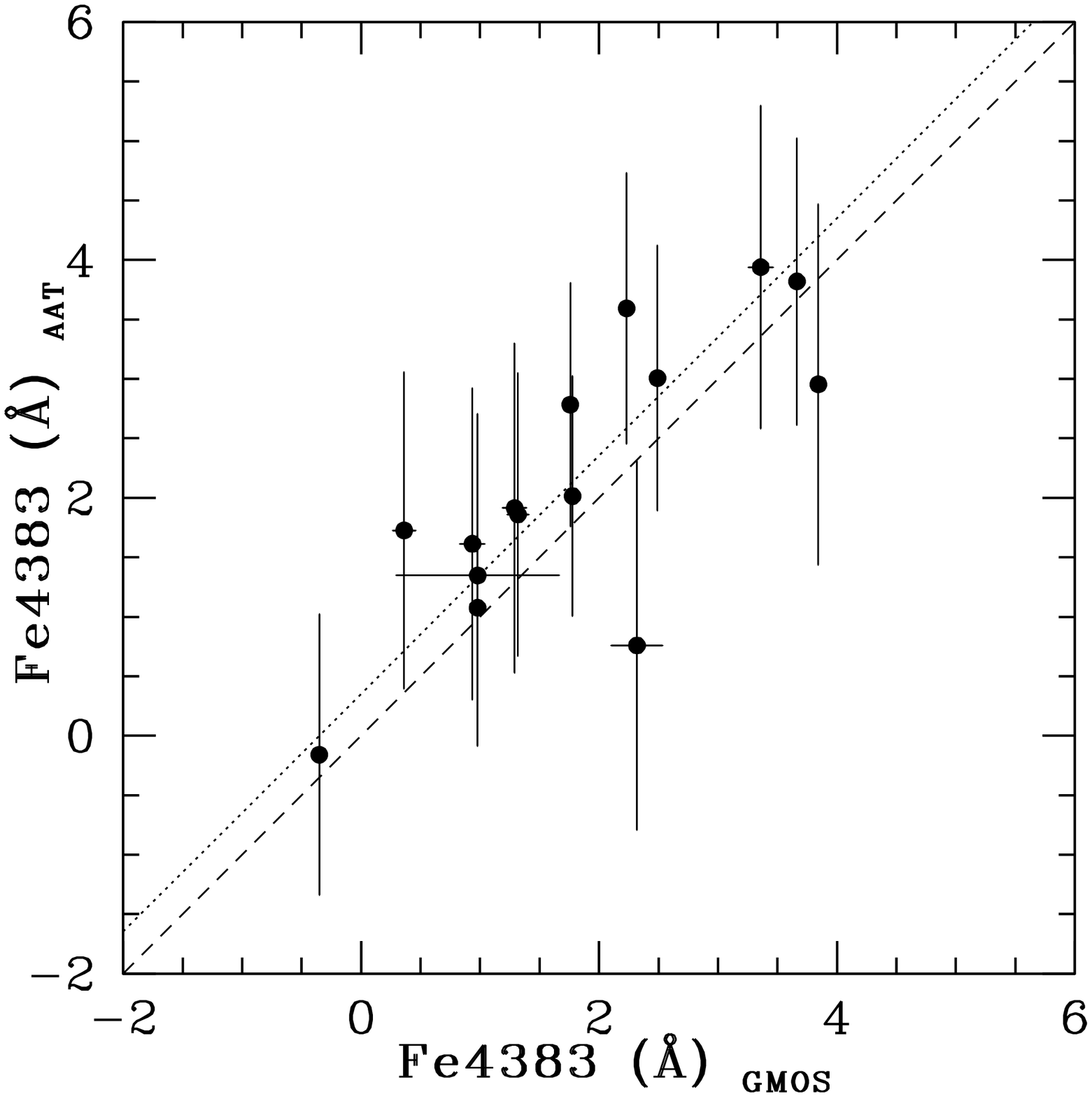,width=0.2\linewidth,clip=} &
\epsfig{file=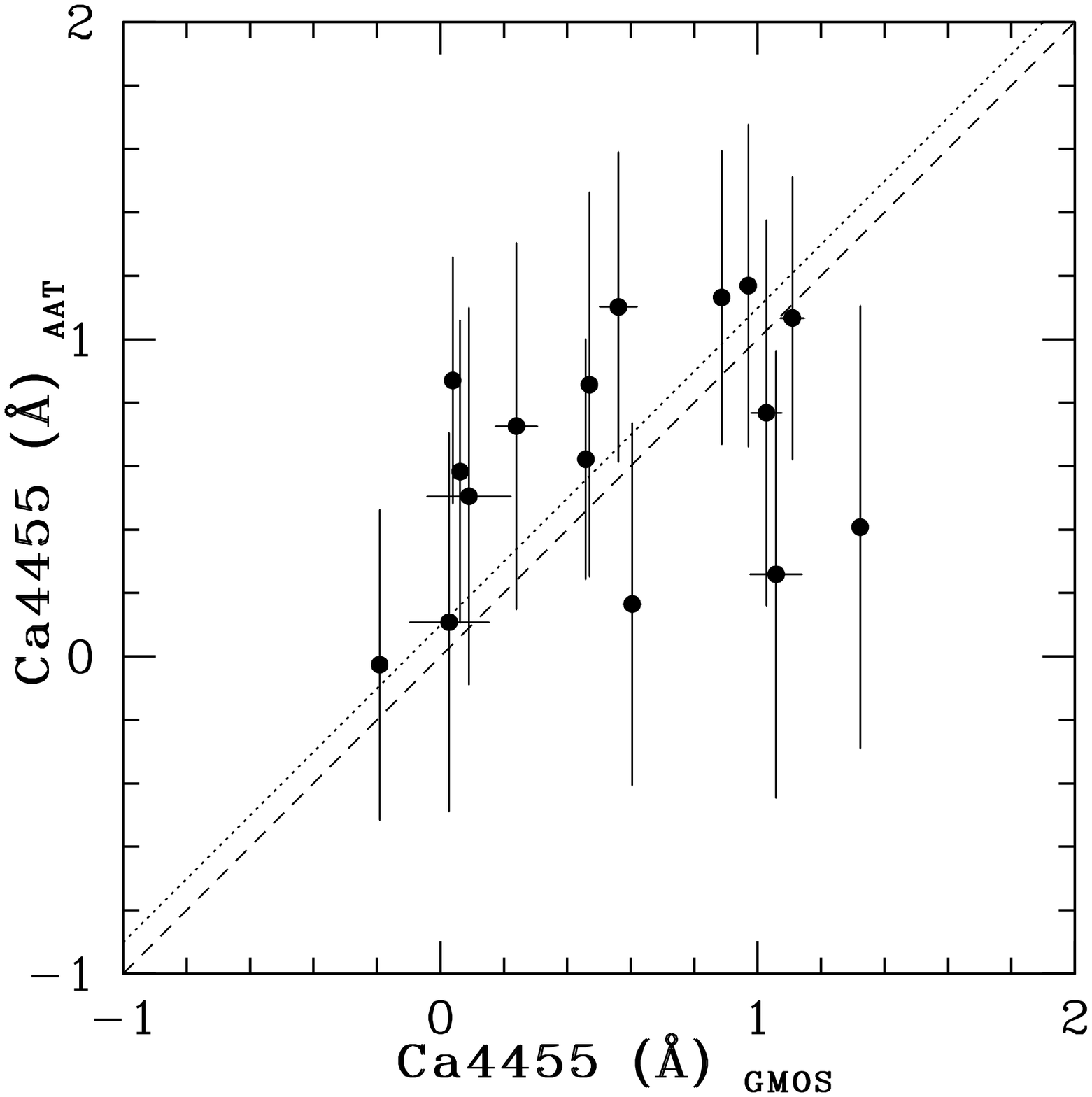,width=0.2\linewidth,clip=} &
\epsfig{file=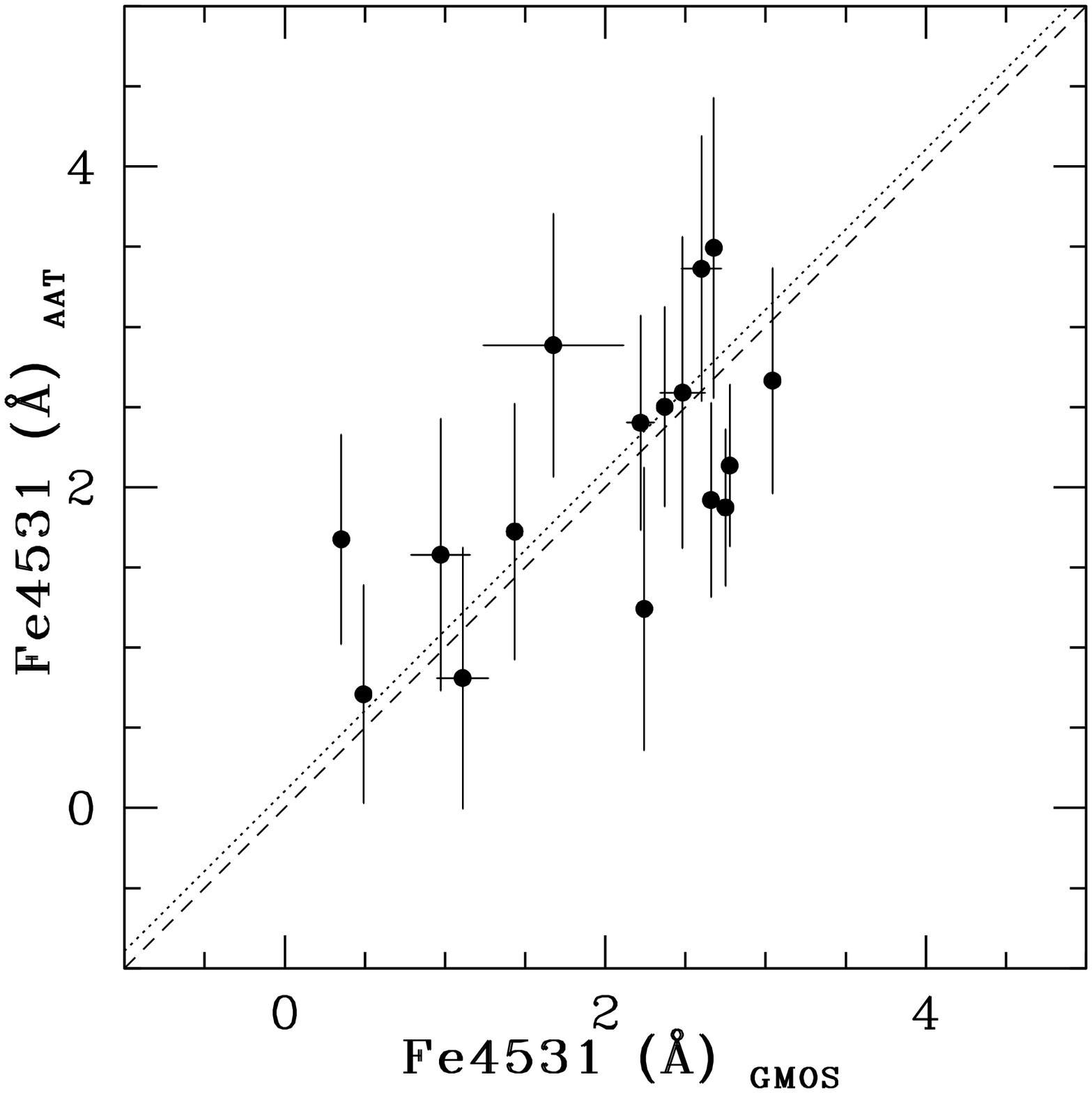,width=0.2\linewidth,clip=} &
\epsfig{file=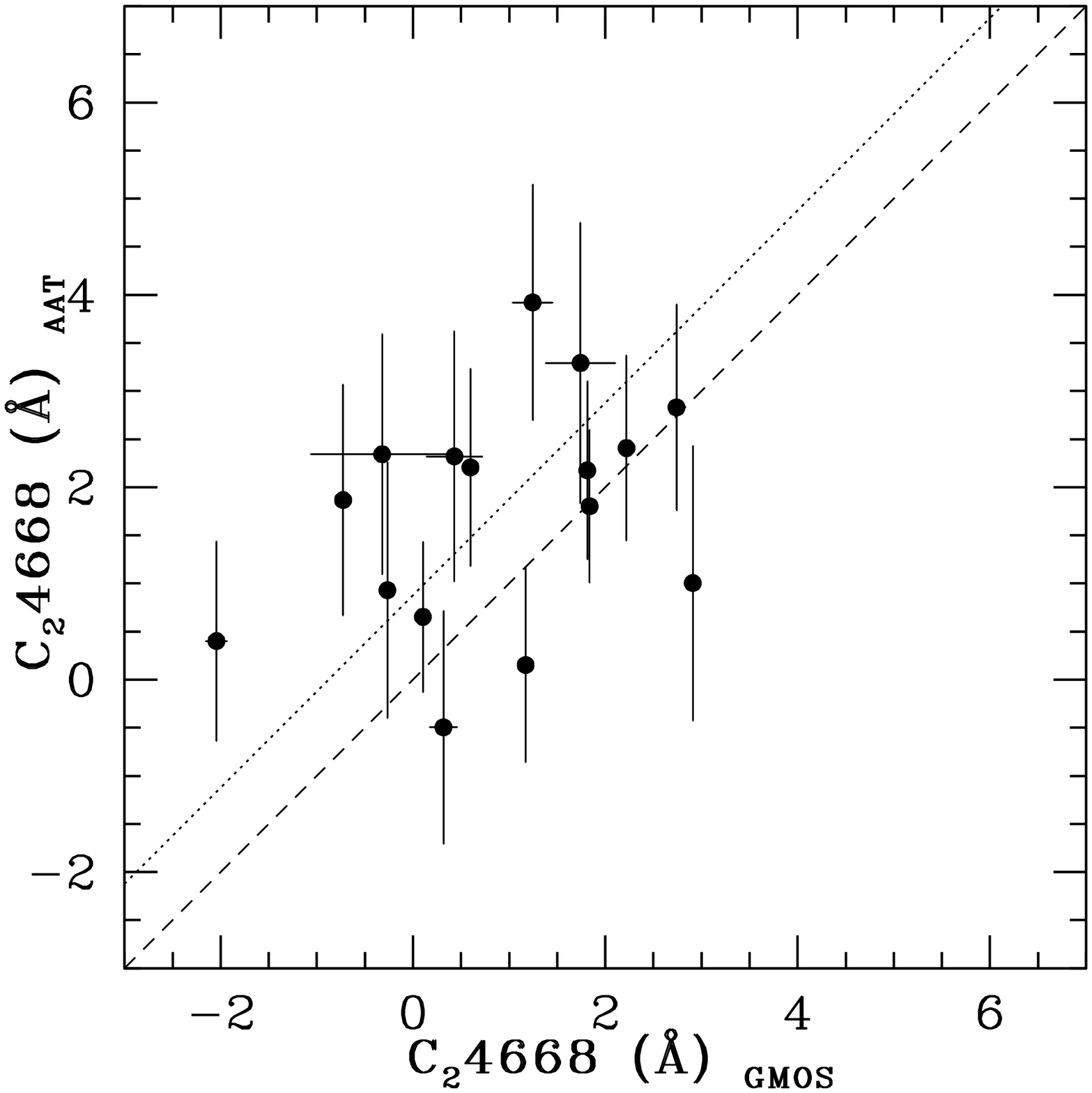,width=0.2\linewidth,clip=} \\
\epsfig{file=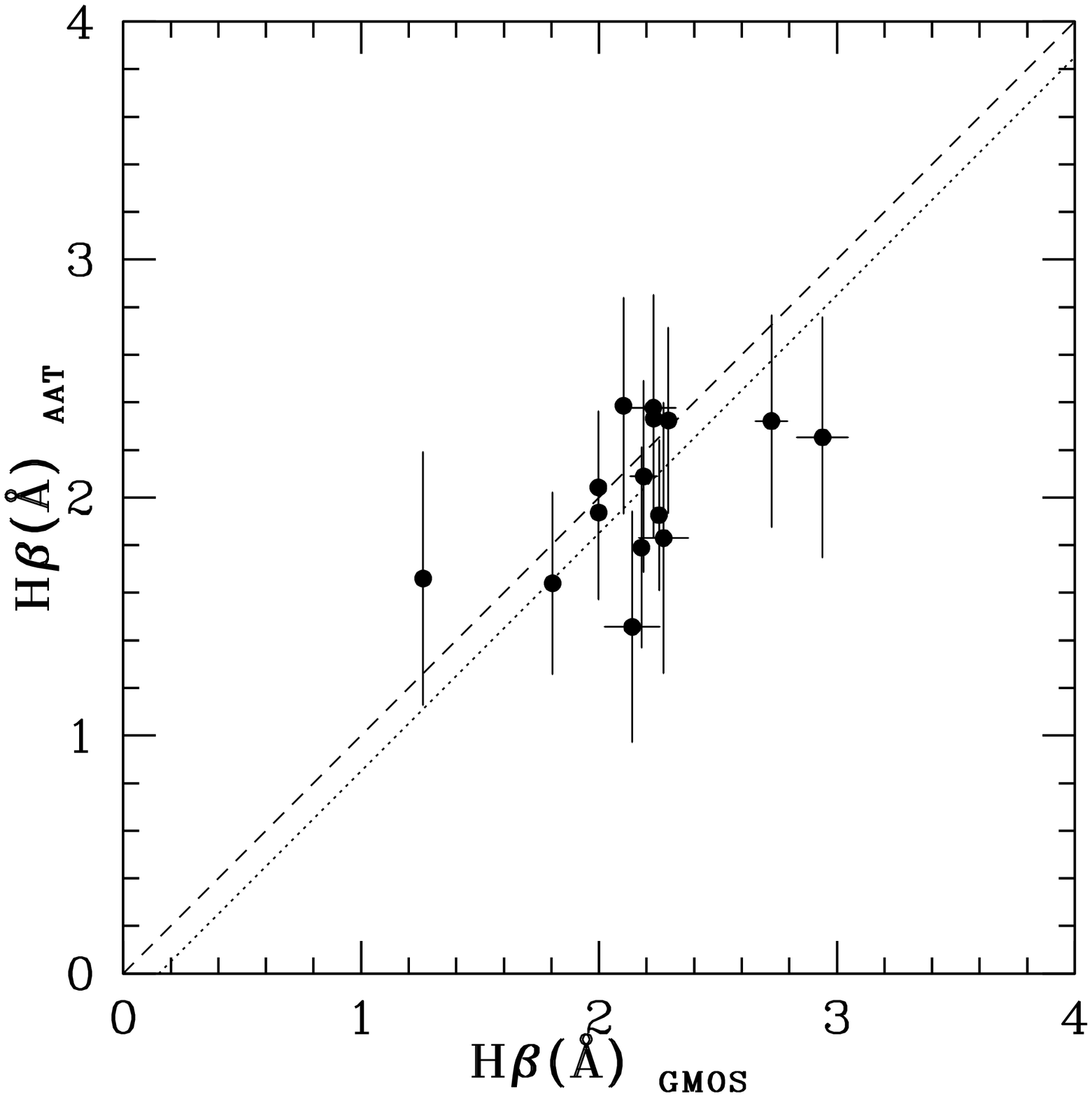,width=0.2\linewidth,clip=} &
\epsfig{file=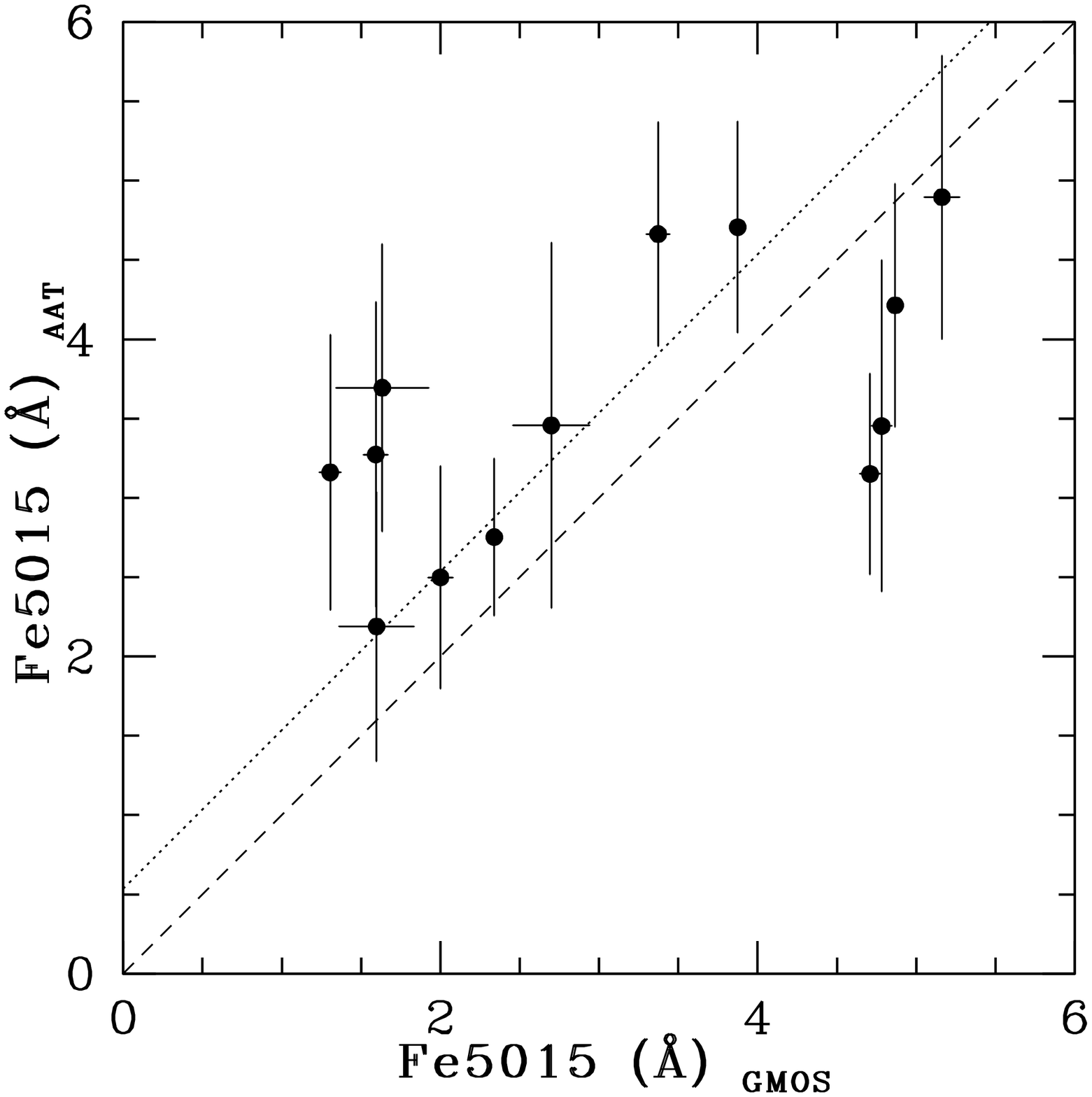,width=0.2\linewidth,clip=}&
\epsfig{file=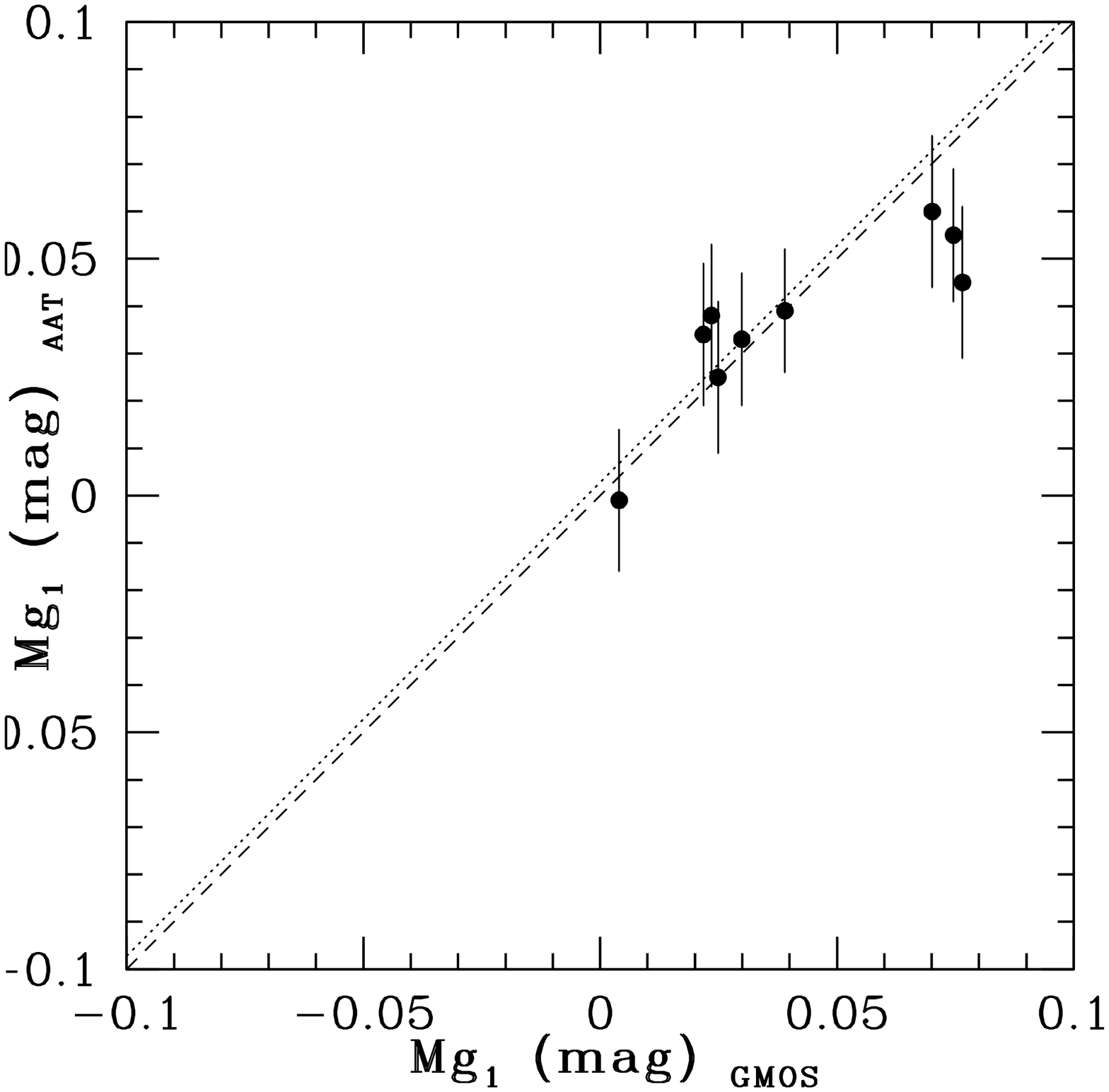,width=0.2\linewidth,clip=}&
\epsfig{file=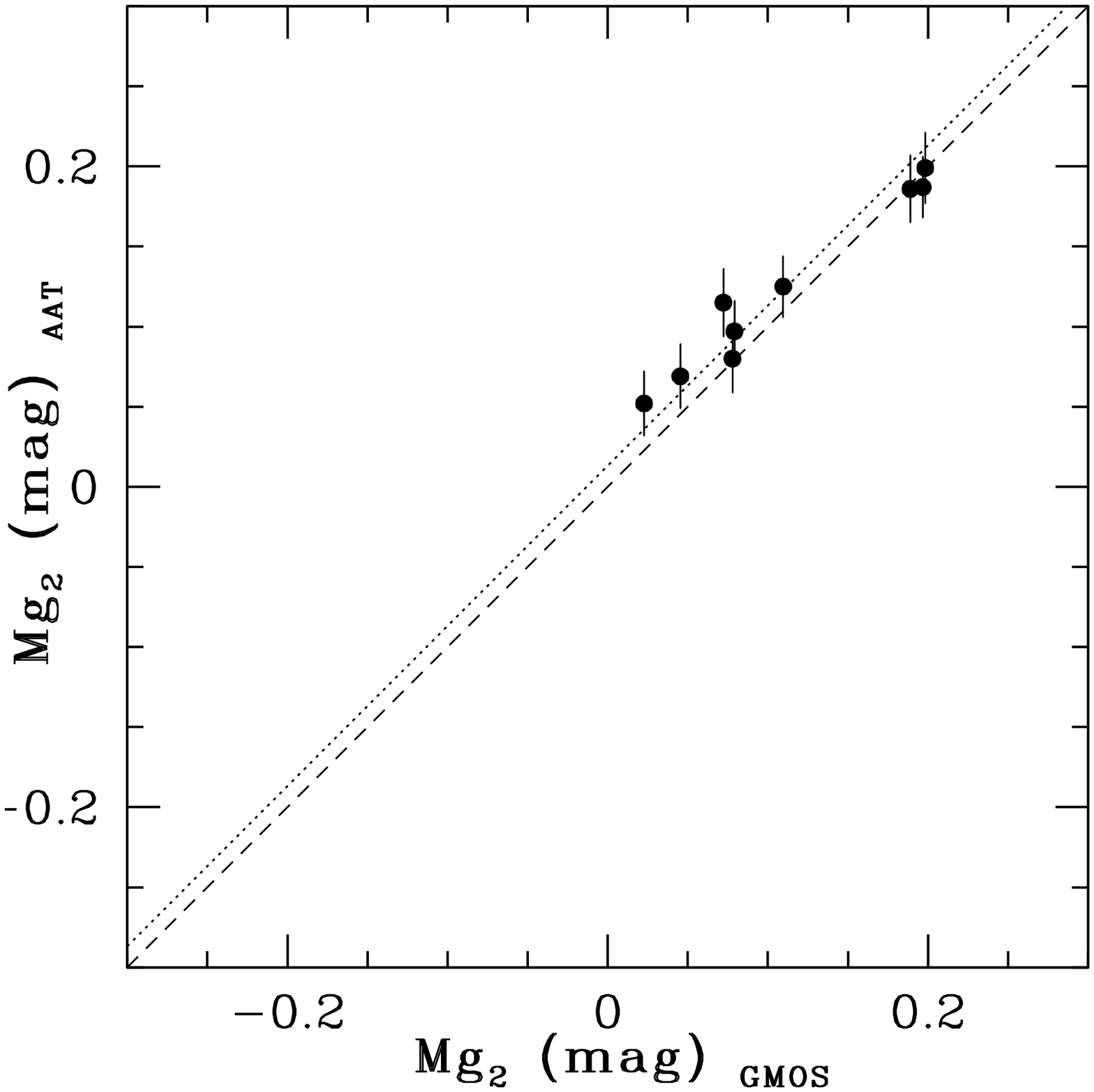,width=0.2\linewidth,clip=} \\
\epsfig{file=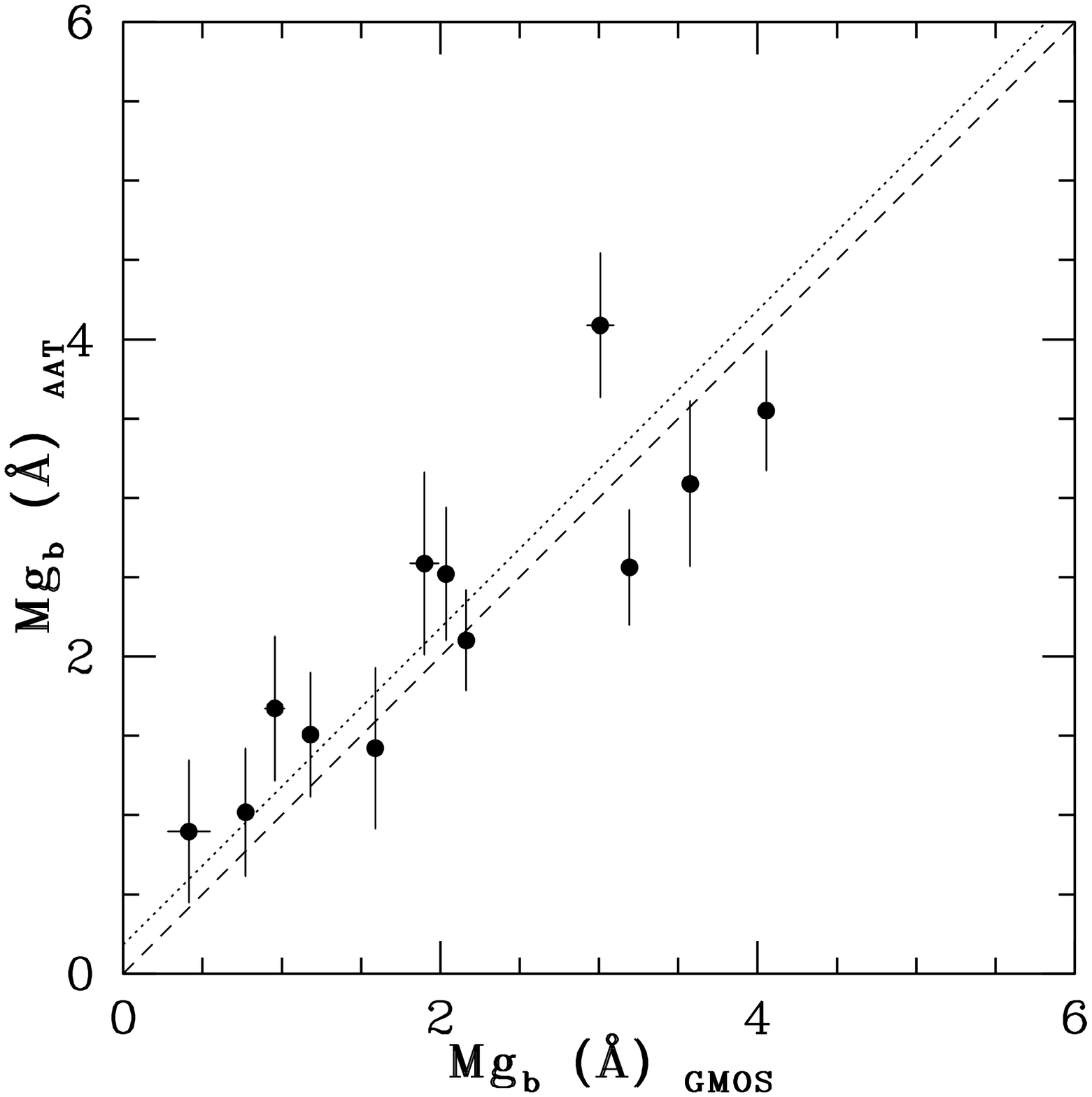,width=0.2\linewidth,clip=} &
\epsfig{file=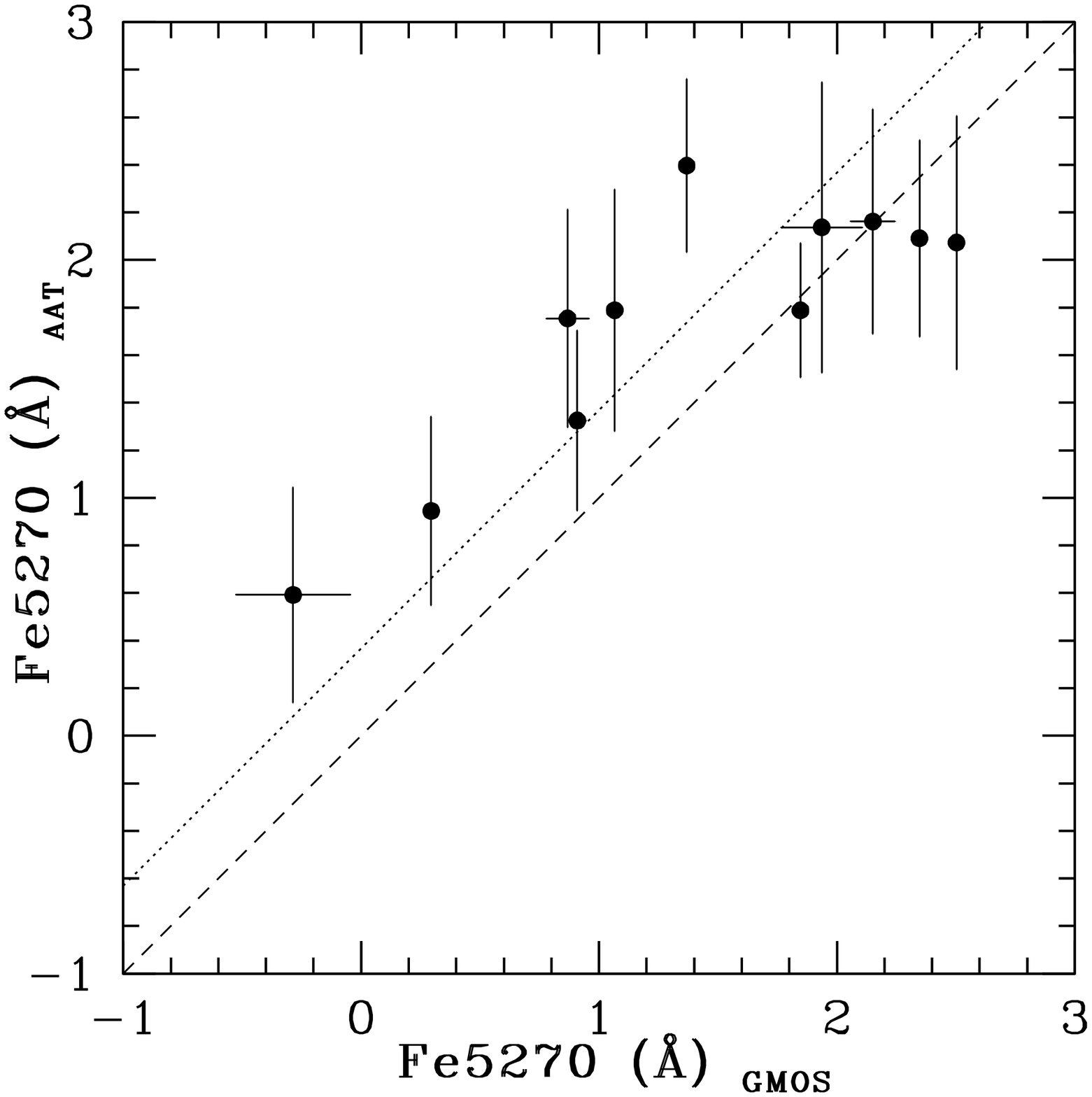,width=0.2\linewidth,clip=} &
\epsfig{file=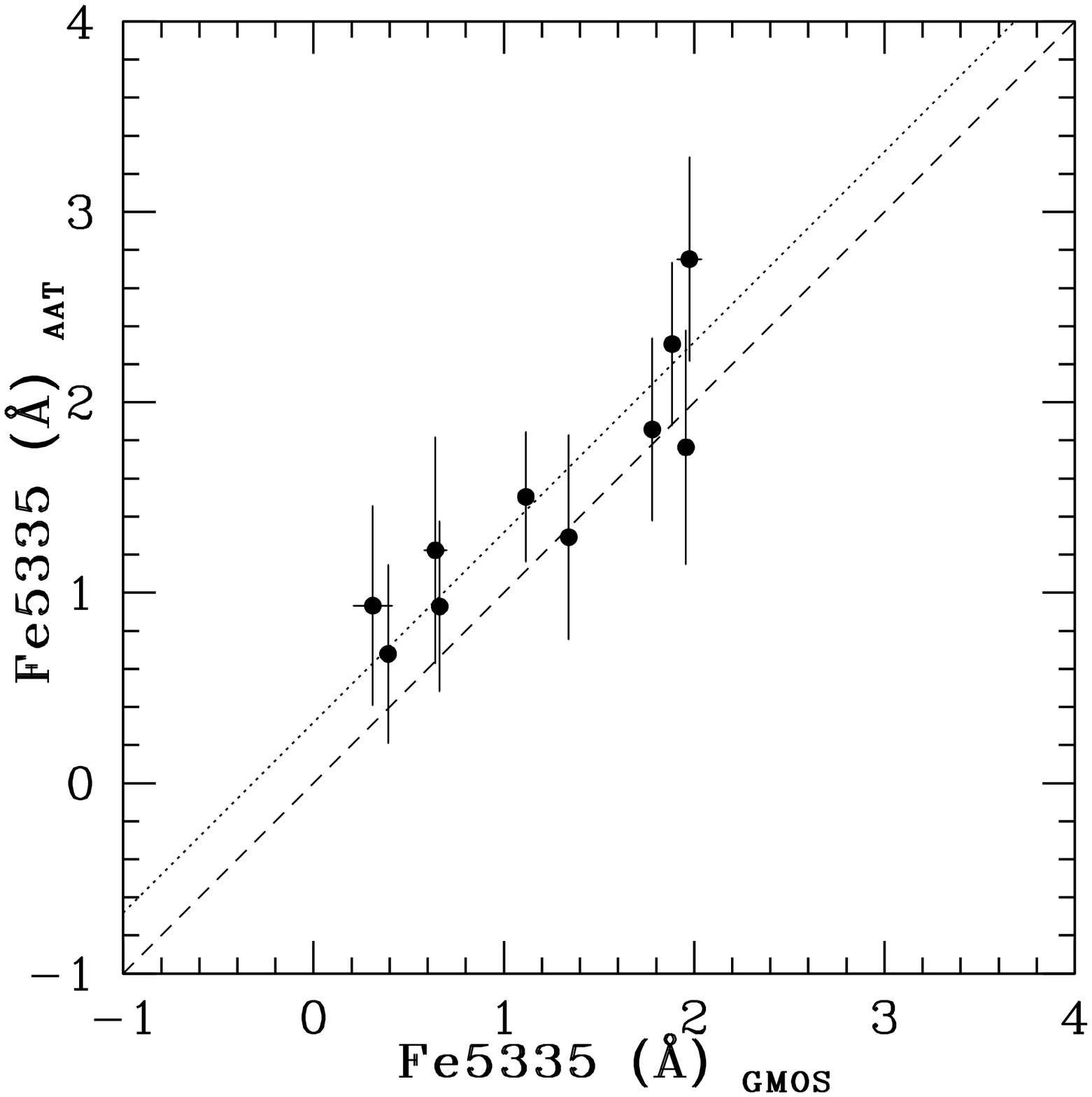,width=0.2\linewidth,clip=} &
\epsfig{file=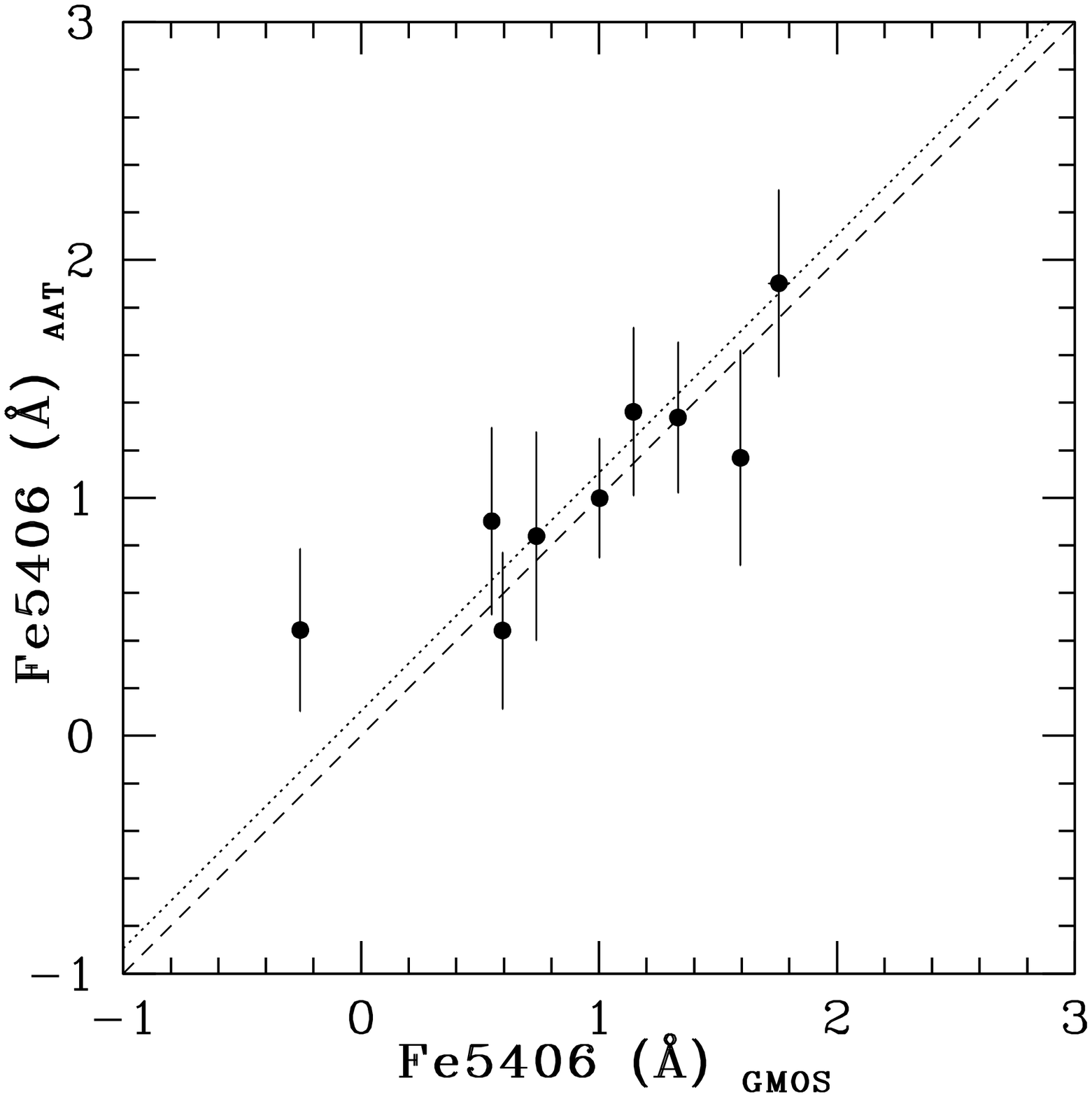,width=0.2\linewidth,clip=} \\

\end{tabular}
\caption{The same as Figure~\ref{fig:cal05} but for the calibration of the
  GMOS 2007 dataset to the AAT study by \cite{beasley08}.} 
\label{fig:cal07}
\end{figure}

\begin{figure}
\centering
\begin{tabular}{cccc}
\epsfig{file=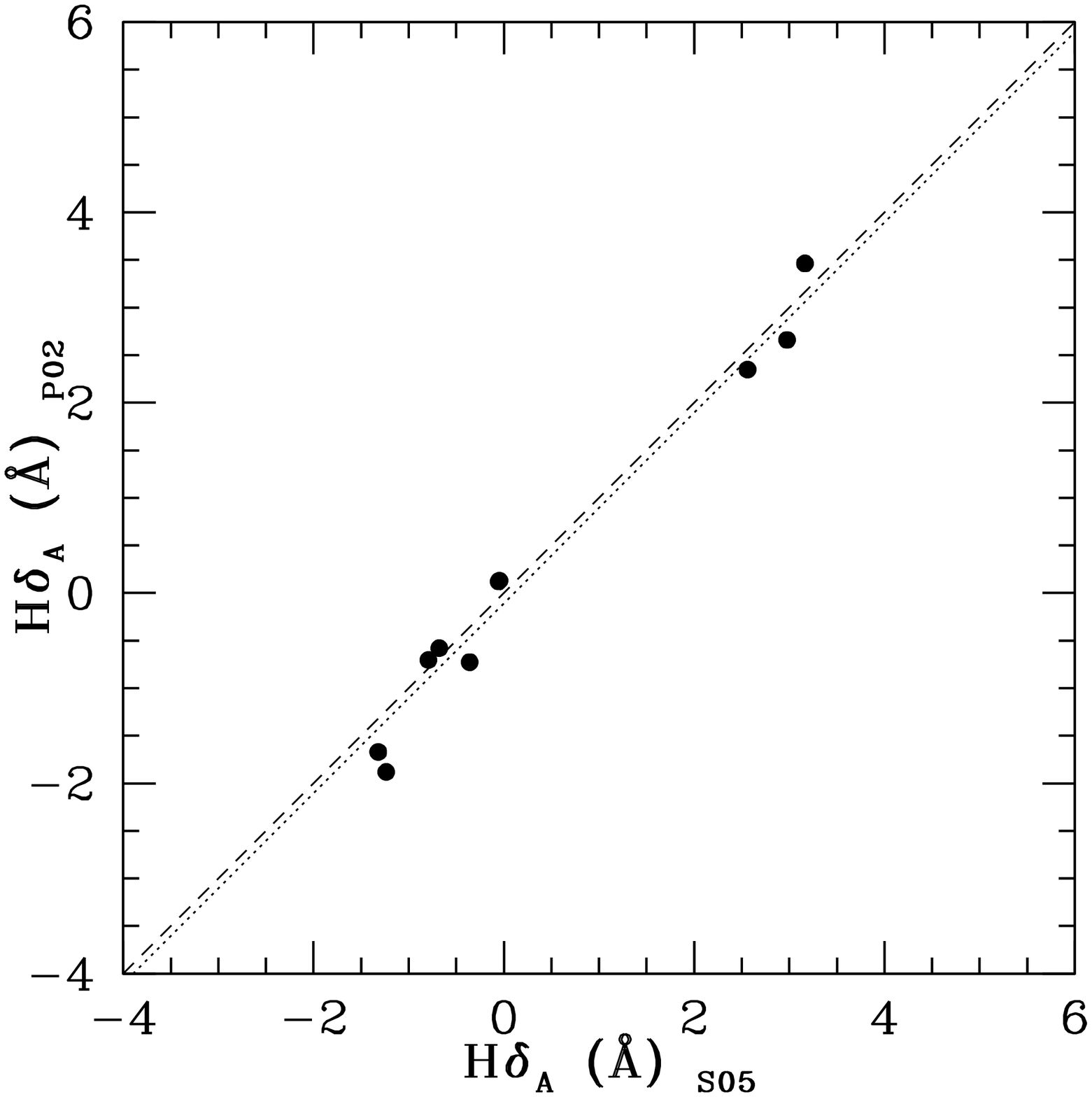,width=0.2\linewidth,clip=} &
\epsfig{file=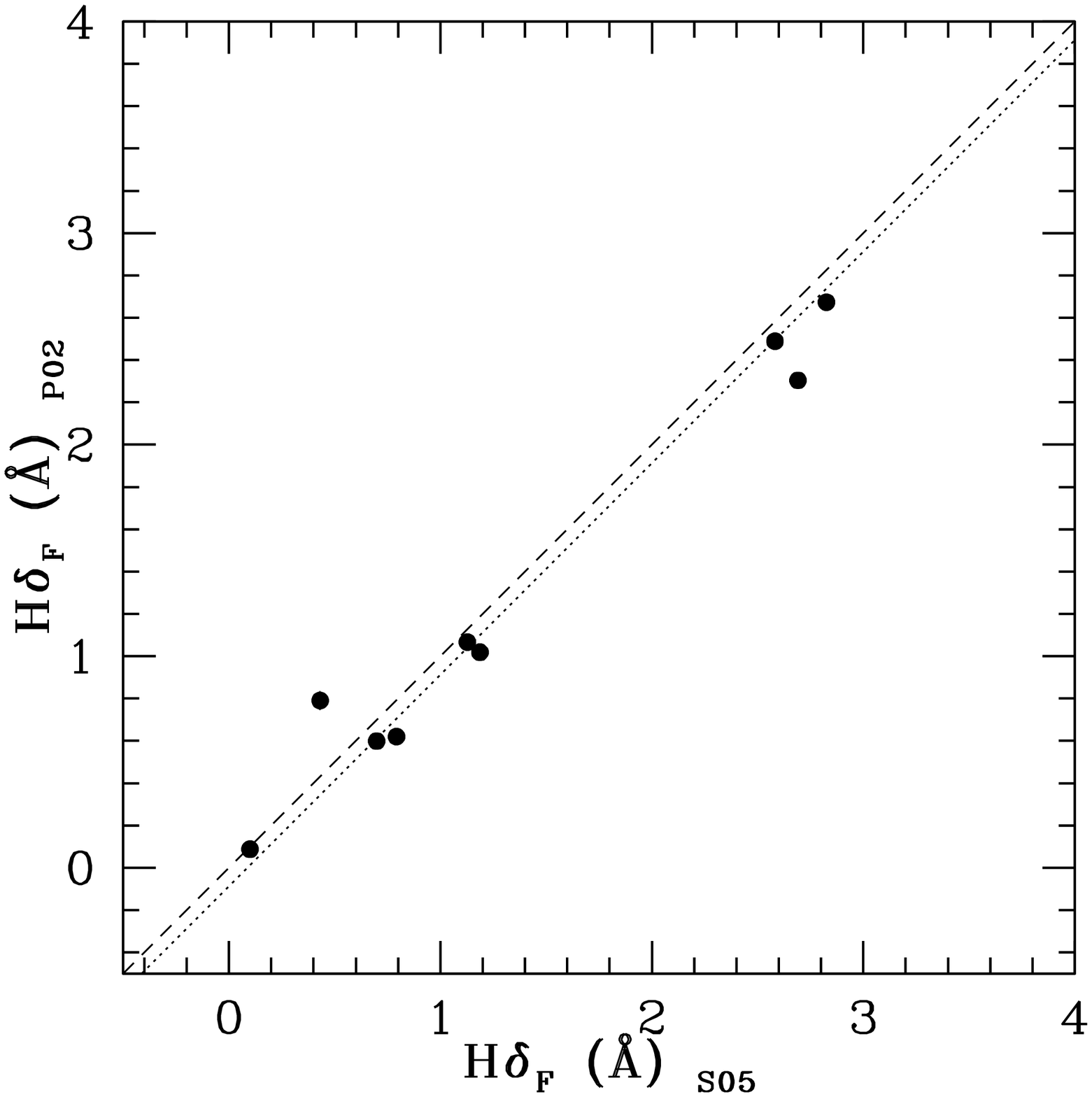,width=0.2\linewidth,clip=} &
\epsfig{file=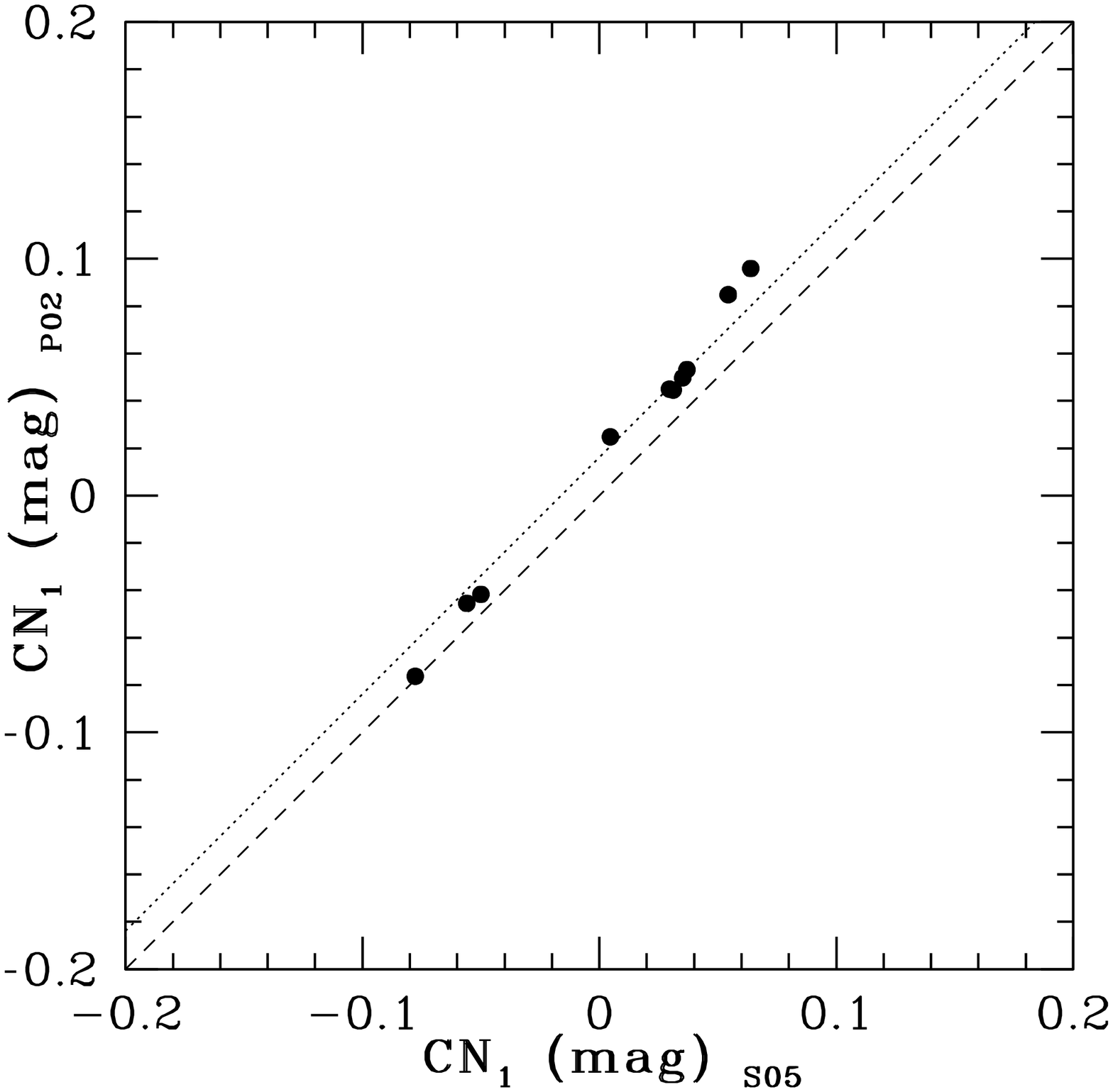,width=0.2\linewidth,clip=} &
\epsfig{file=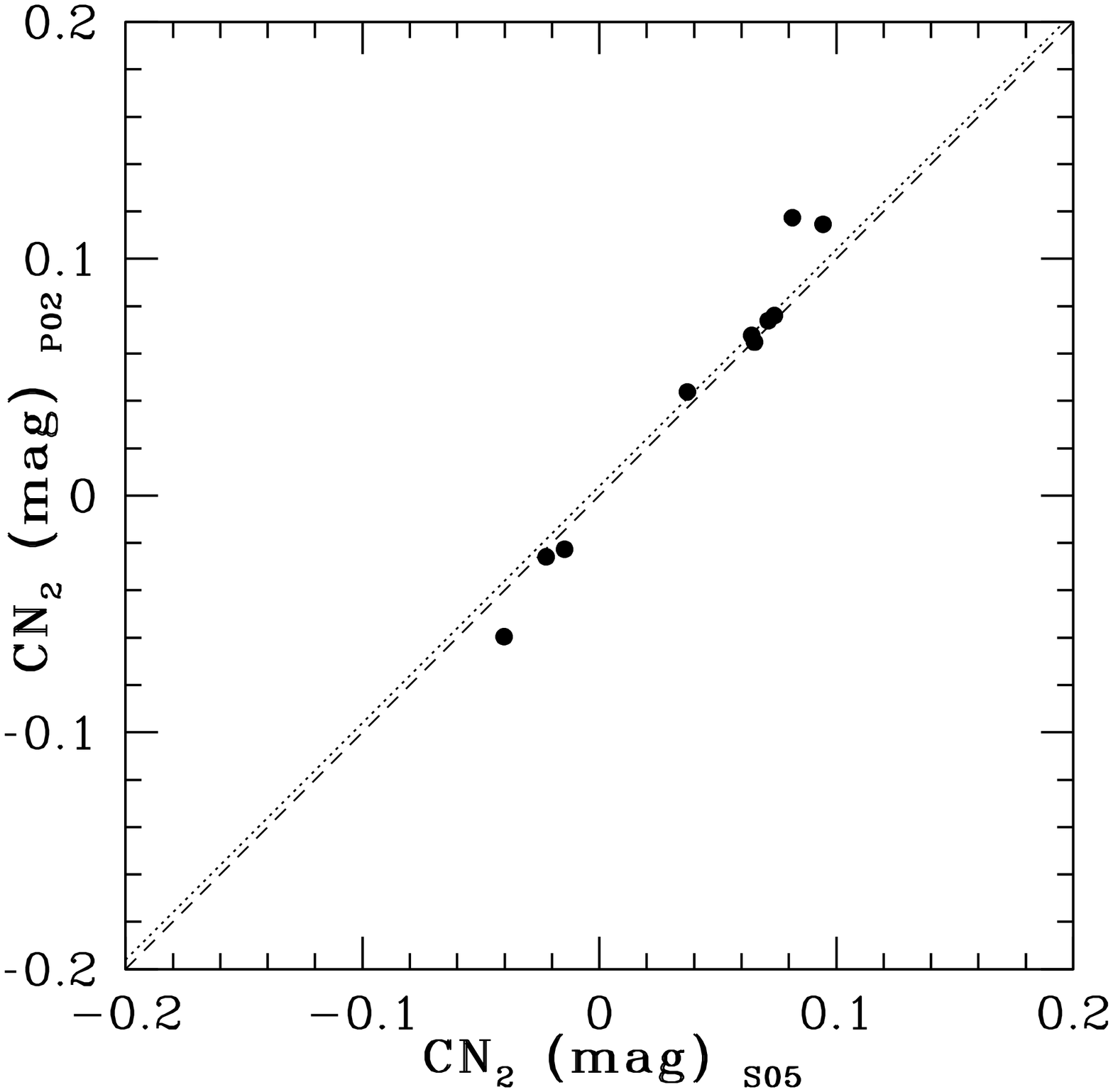,width=0.2\linewidth,clip=} \\
\epsfig{file=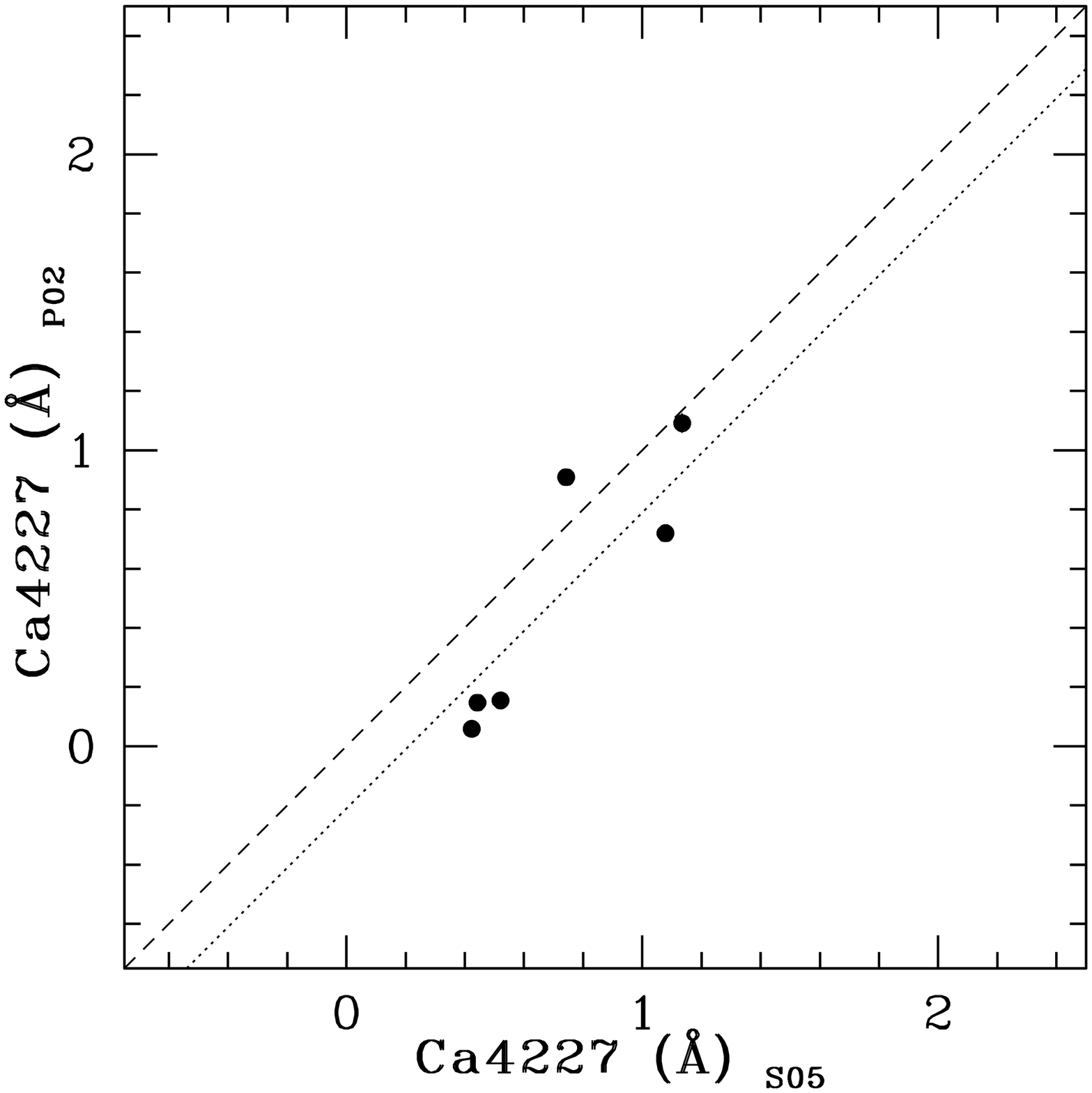,width=0.2\linewidth,clip=} &
\epsfig{file=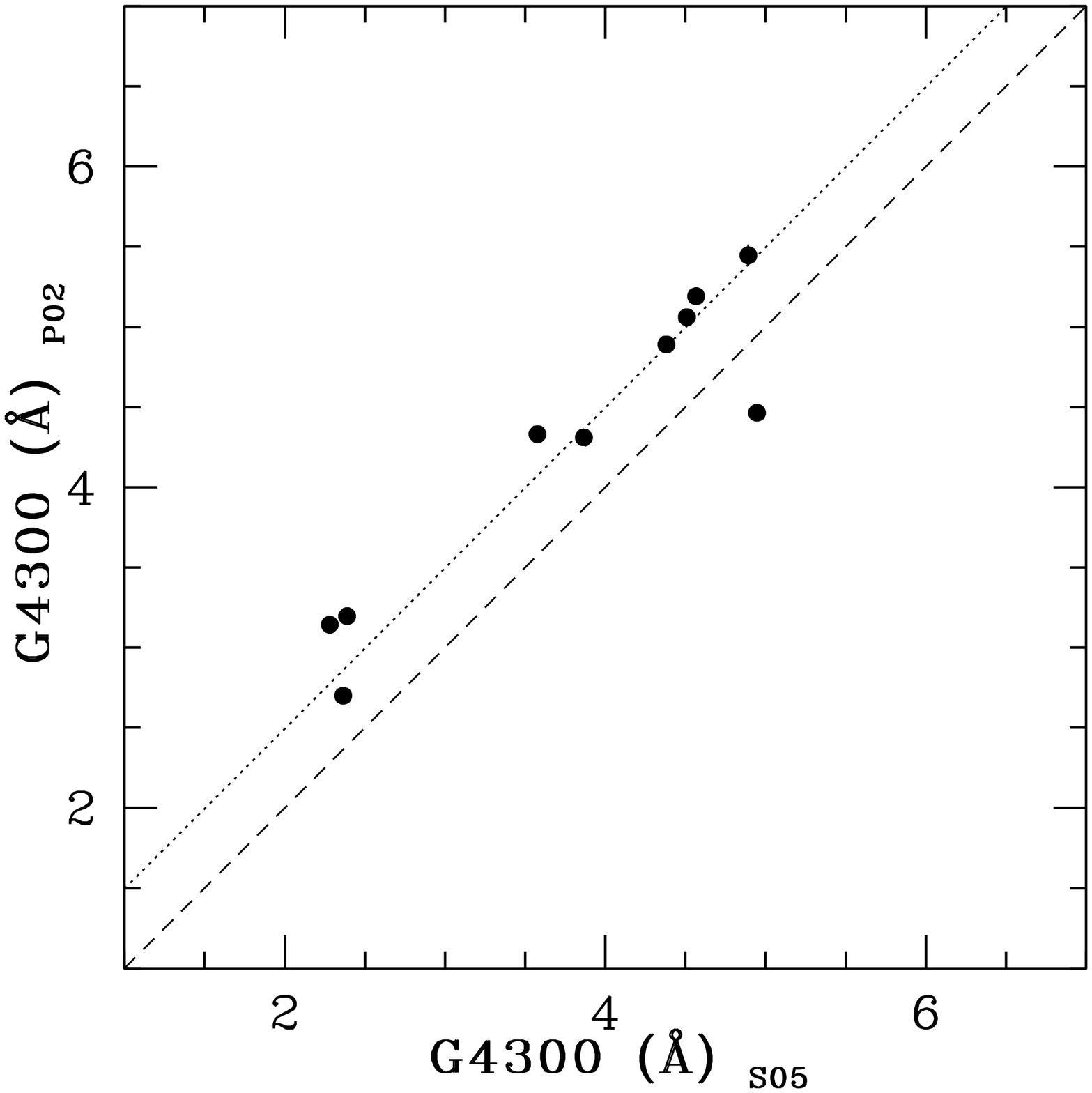,width=0.2\linewidth,clip=} &
\epsfig{file=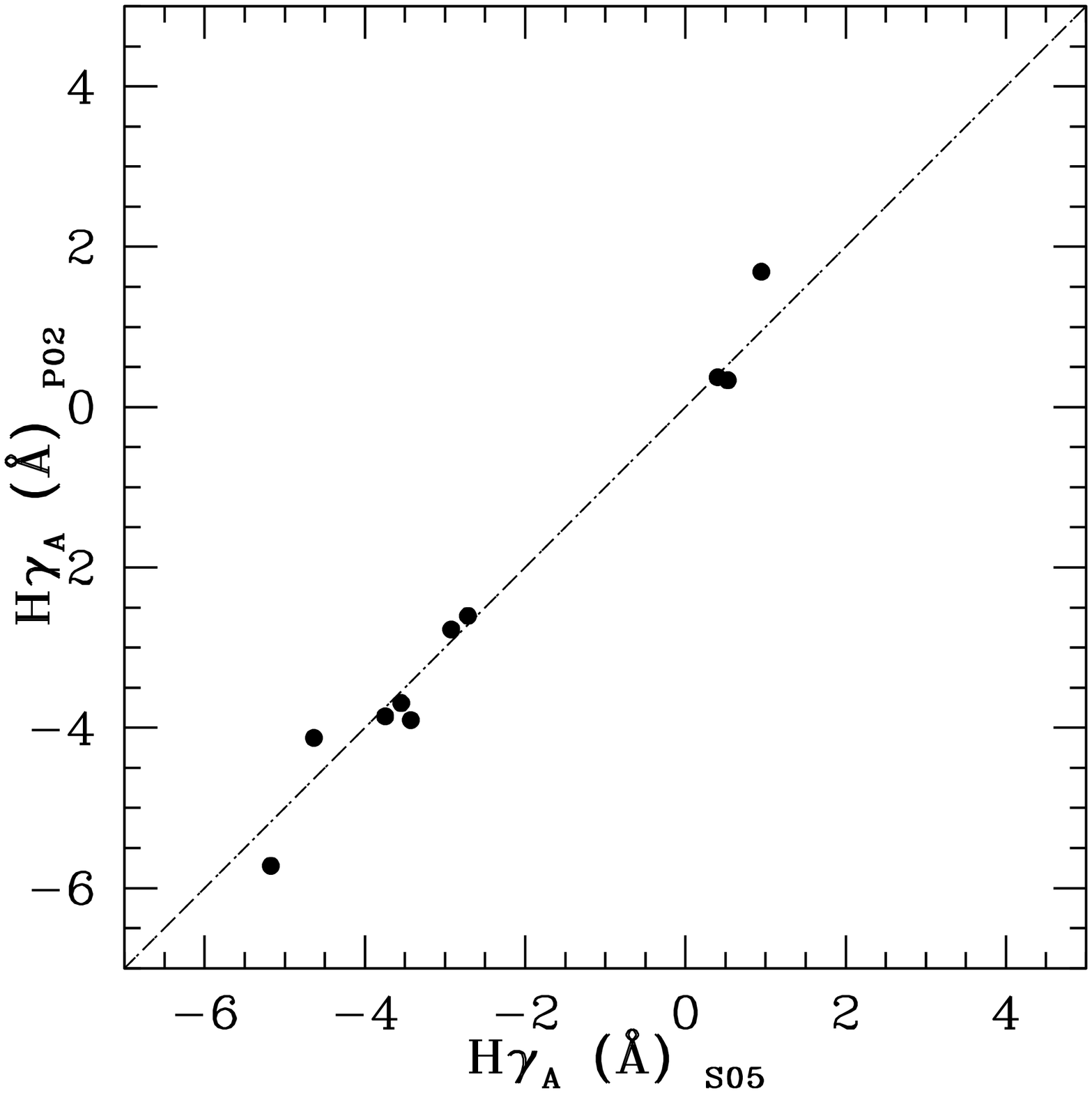,width=0.2\linewidth,clip=} &
\epsfig{file=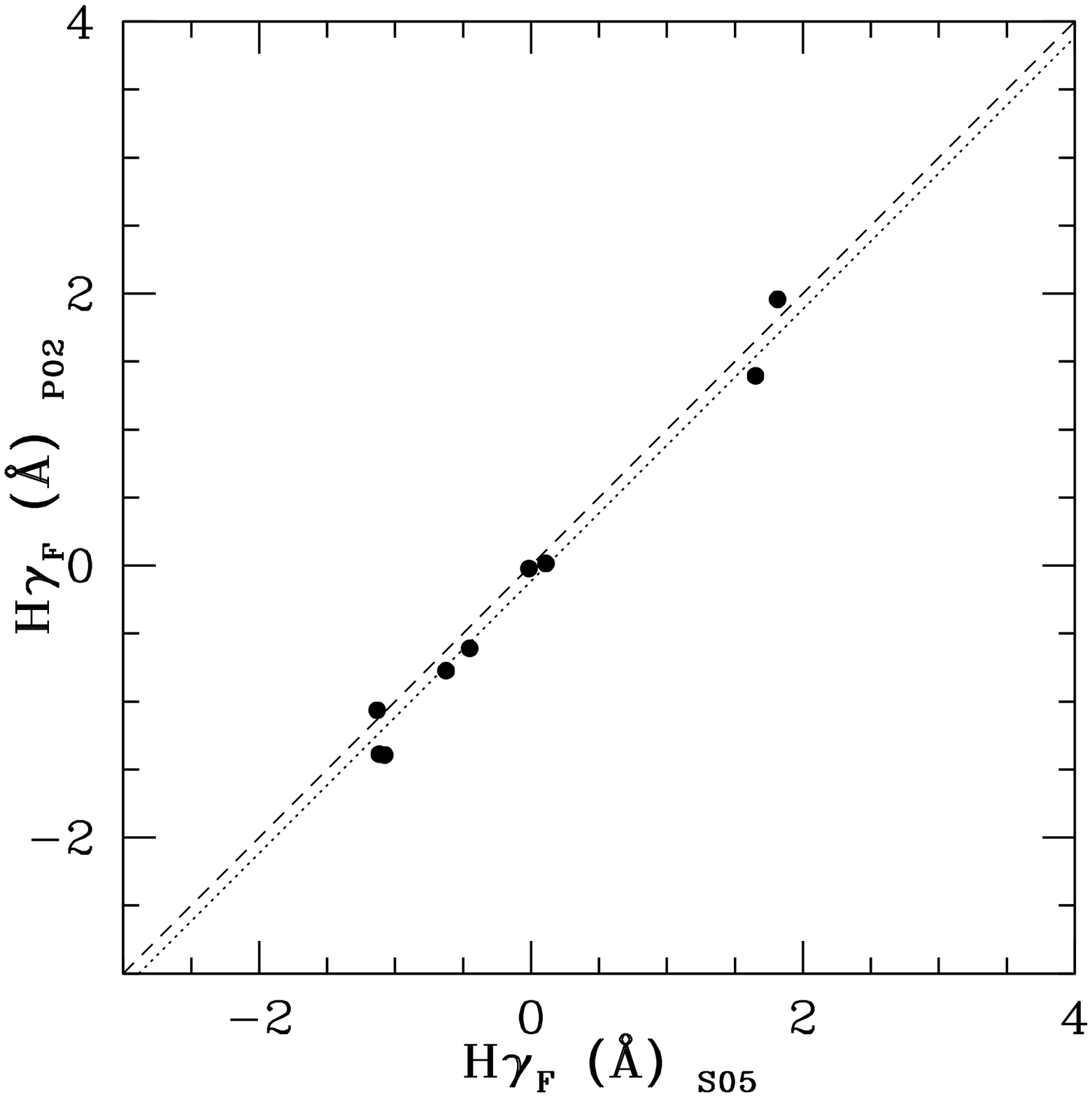,width=0.2\linewidth,clip=} \\
\epsfig{file=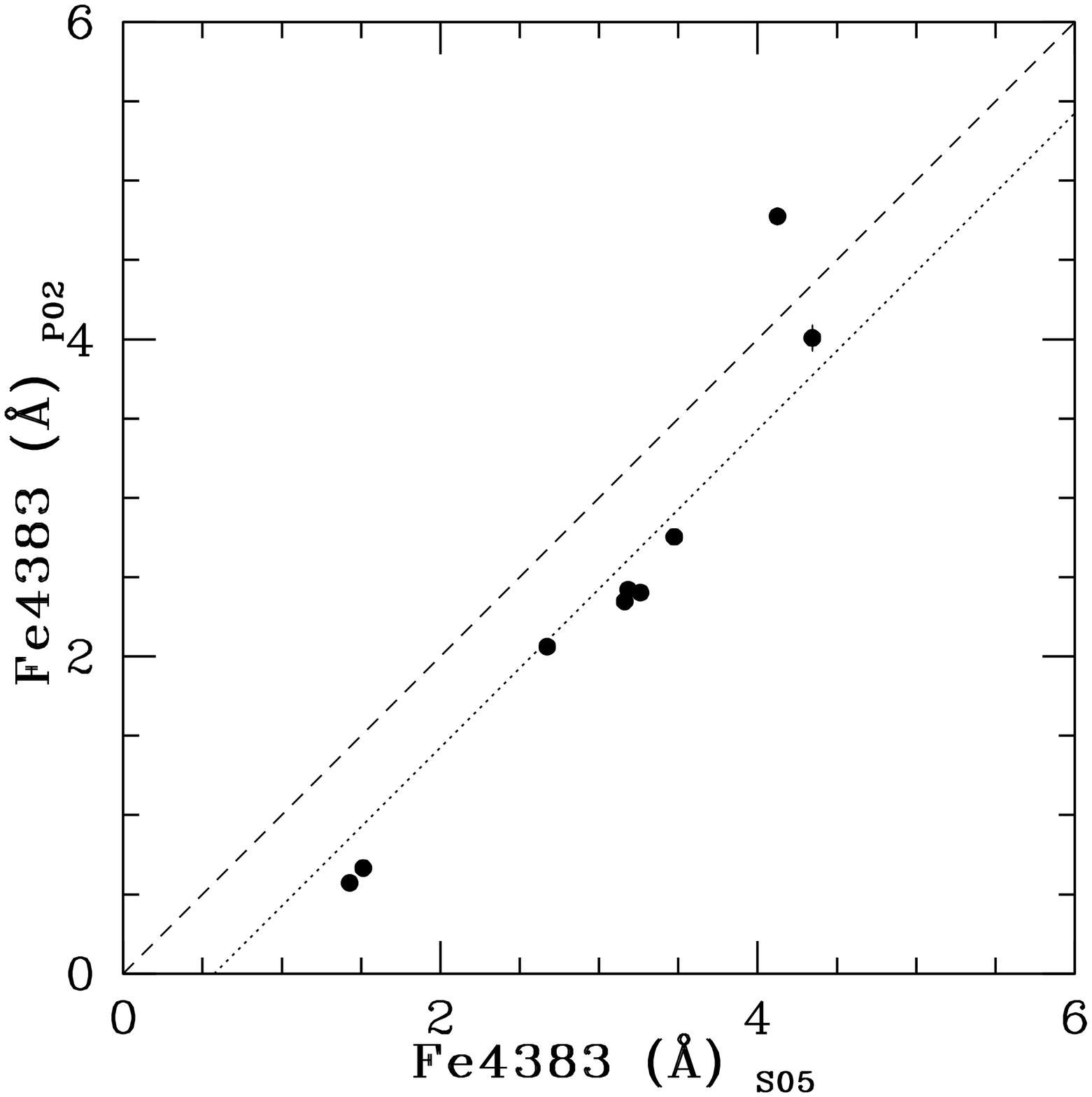,width=0.2\linewidth,clip=} &
\epsfig{file=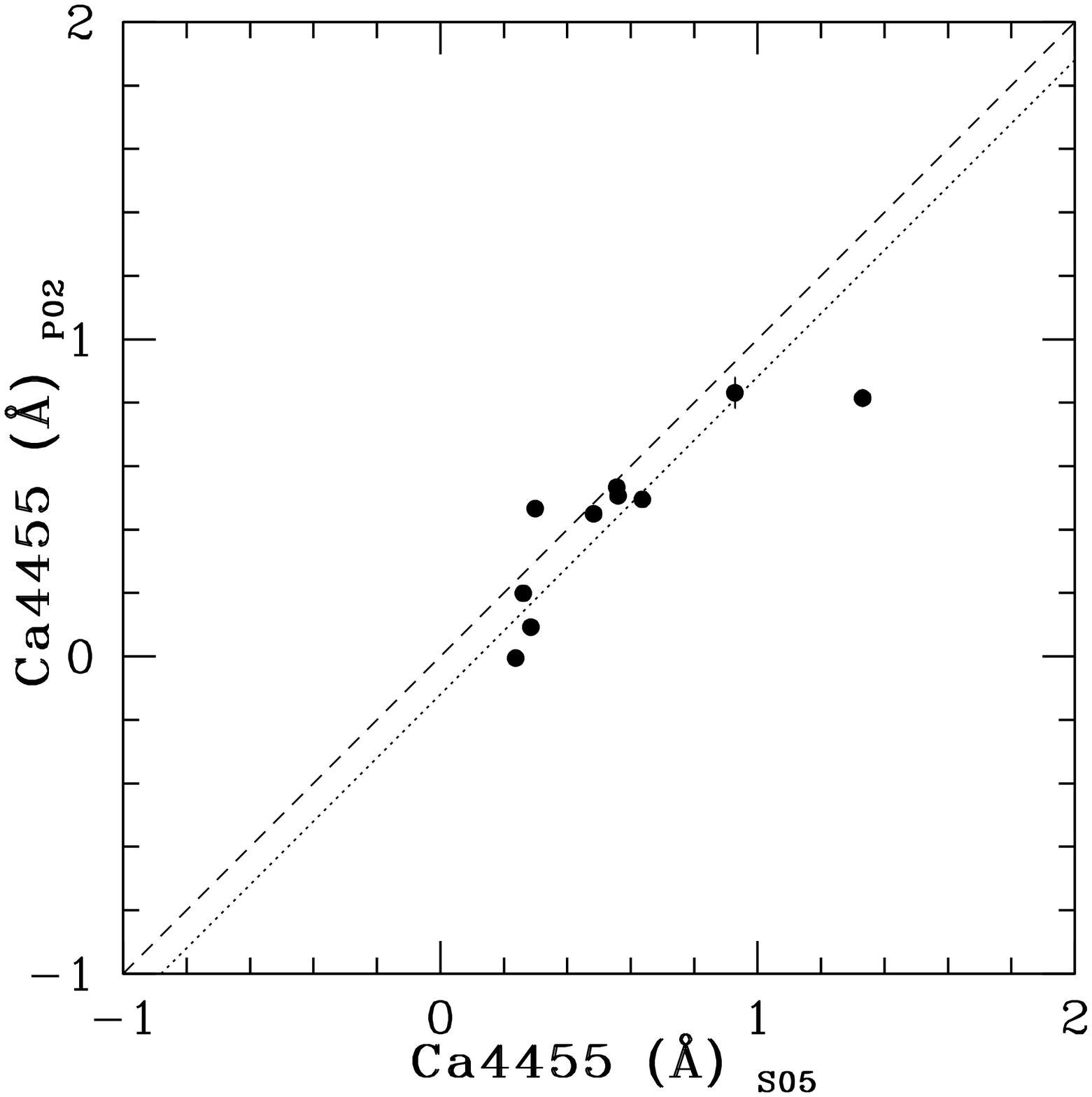,width=0.2\linewidth,clip=} &
\epsfig{file=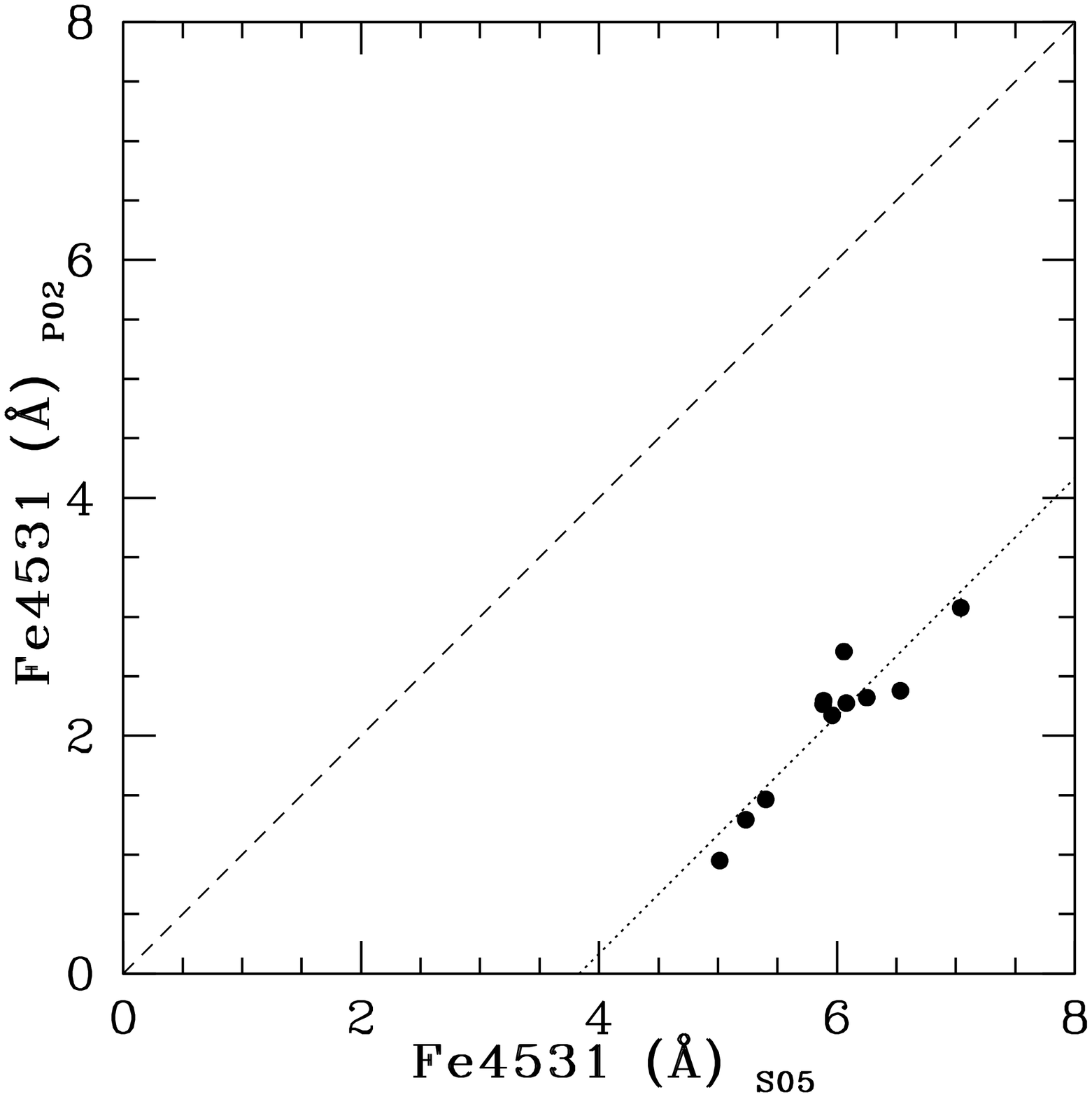,width=0.2\linewidth,clip=} &
\epsfig{file=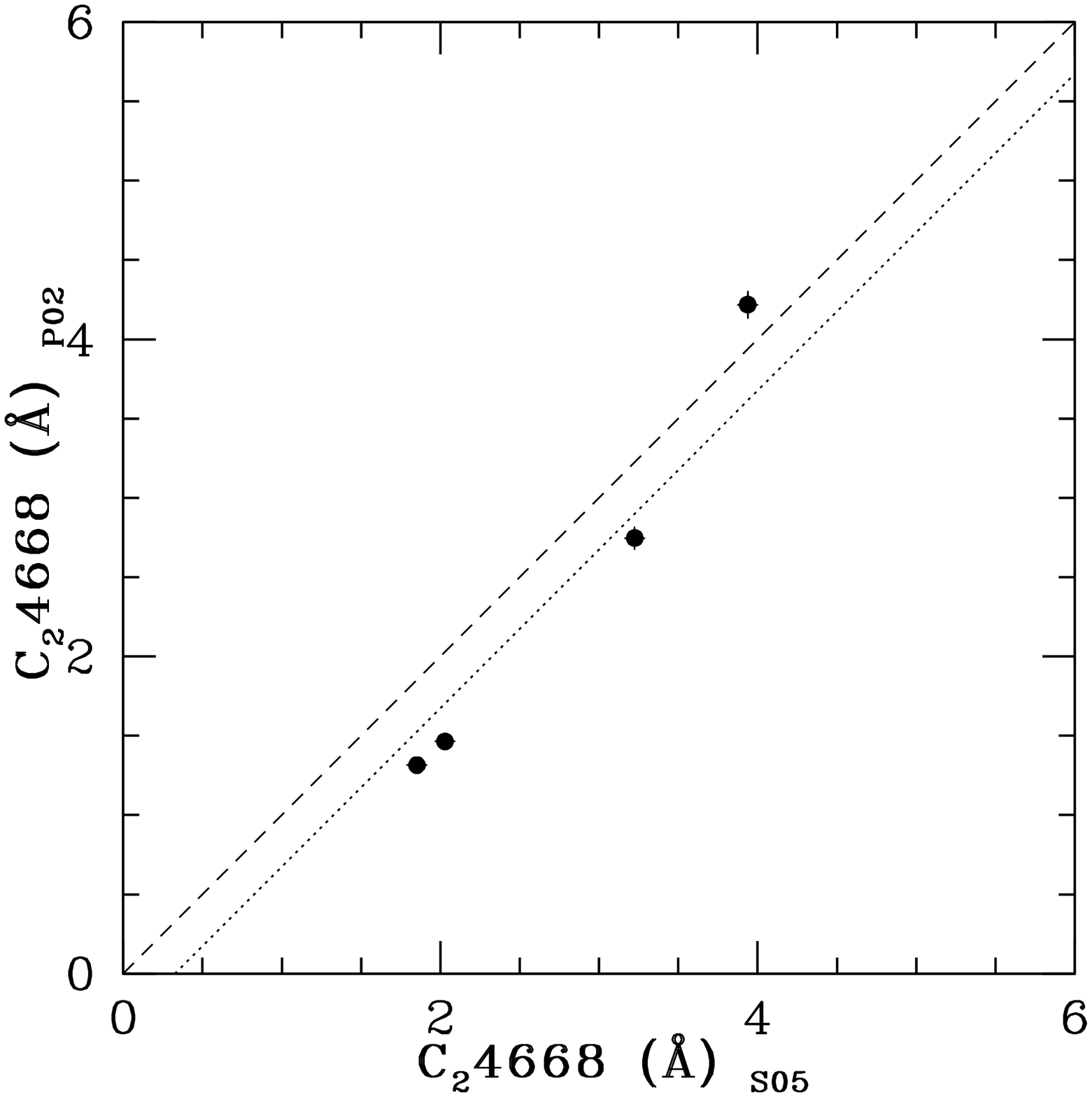,width=0.2\linewidth,clip=} \\
\epsfig{file=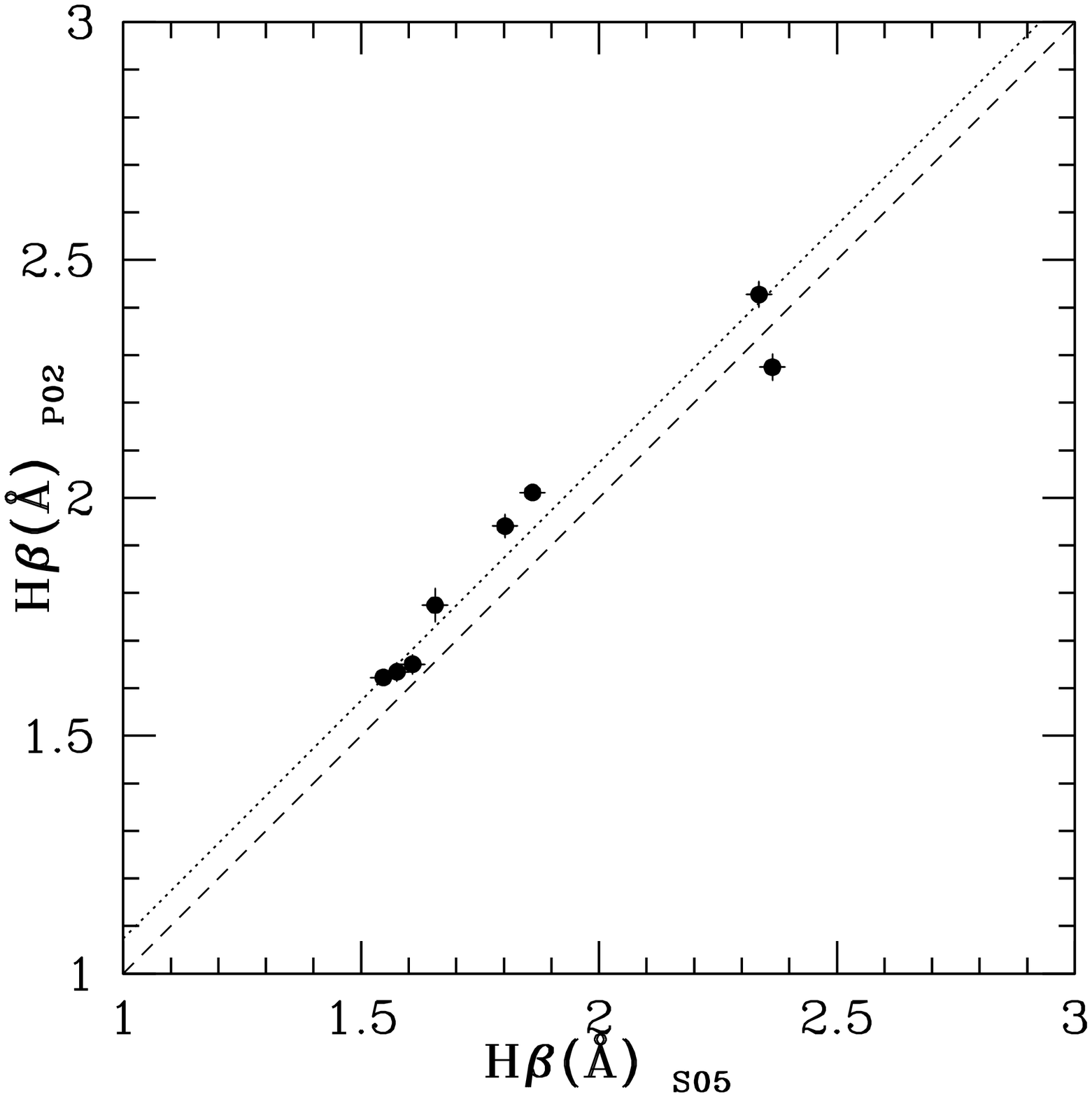,width=0.2\linewidth,clip=} &
\epsfig{file=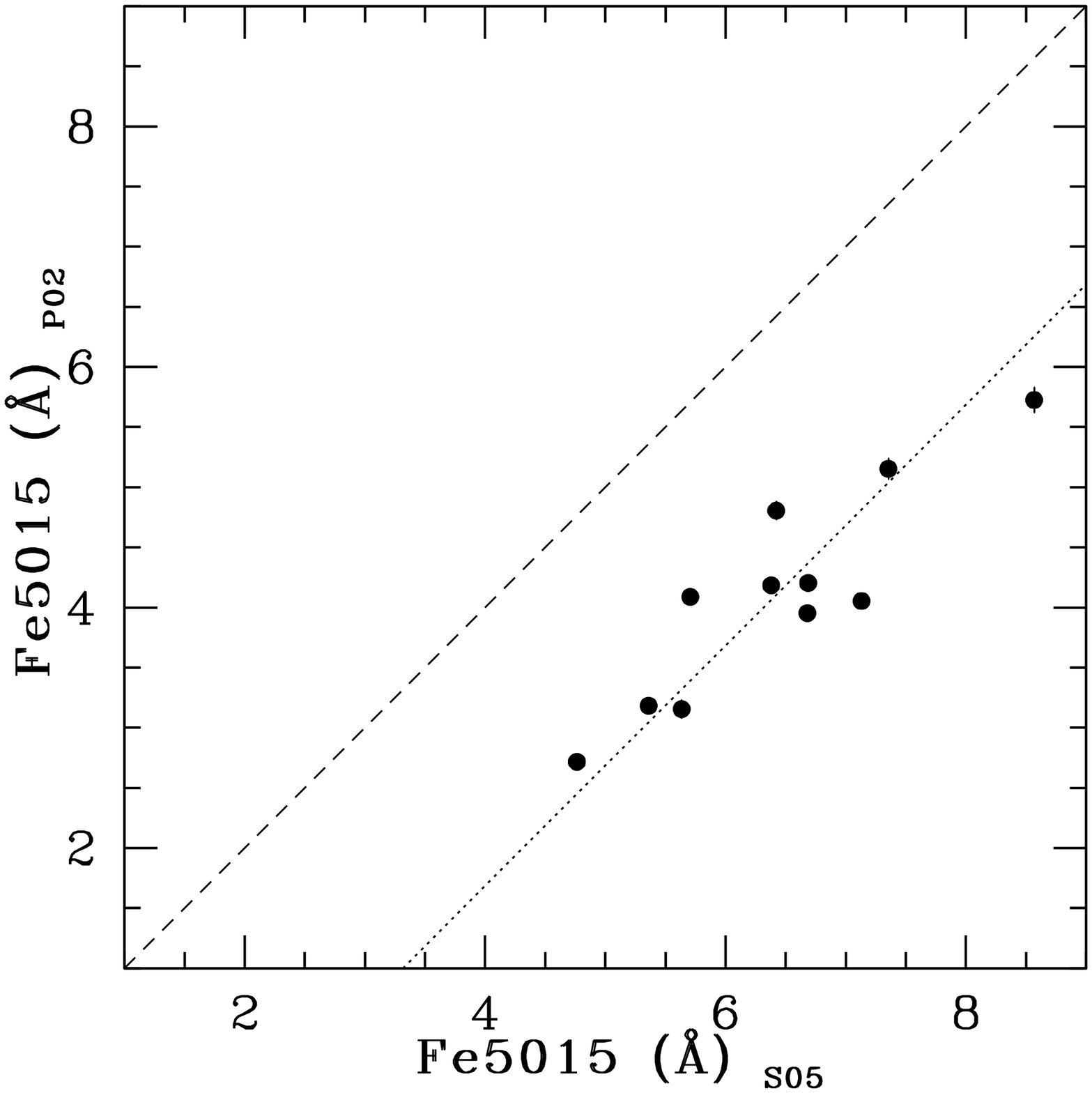,width=0.2\linewidth,clip=}&
\epsfig{file=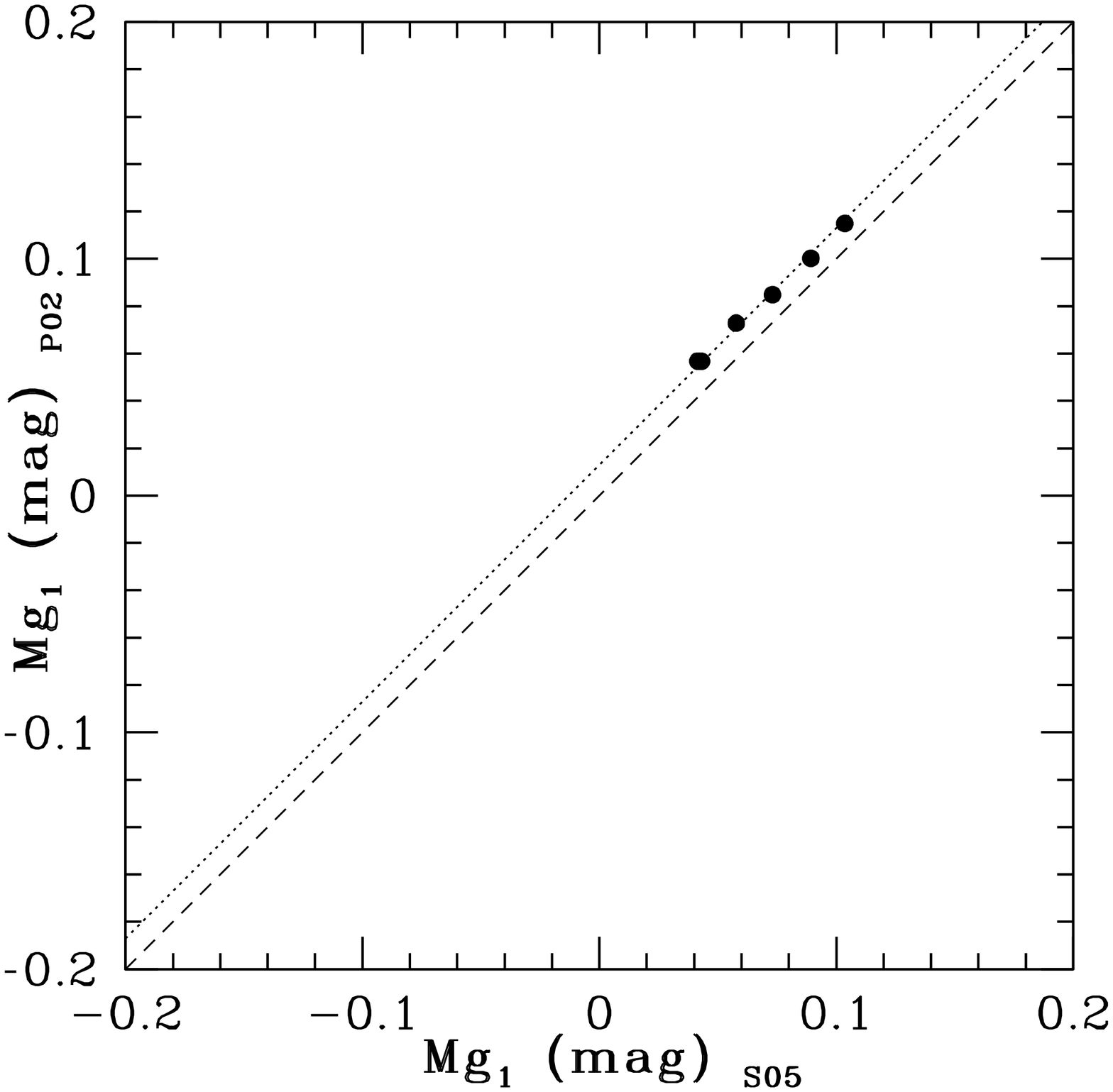,width=0.2\linewidth,clip=}&
\epsfig{file=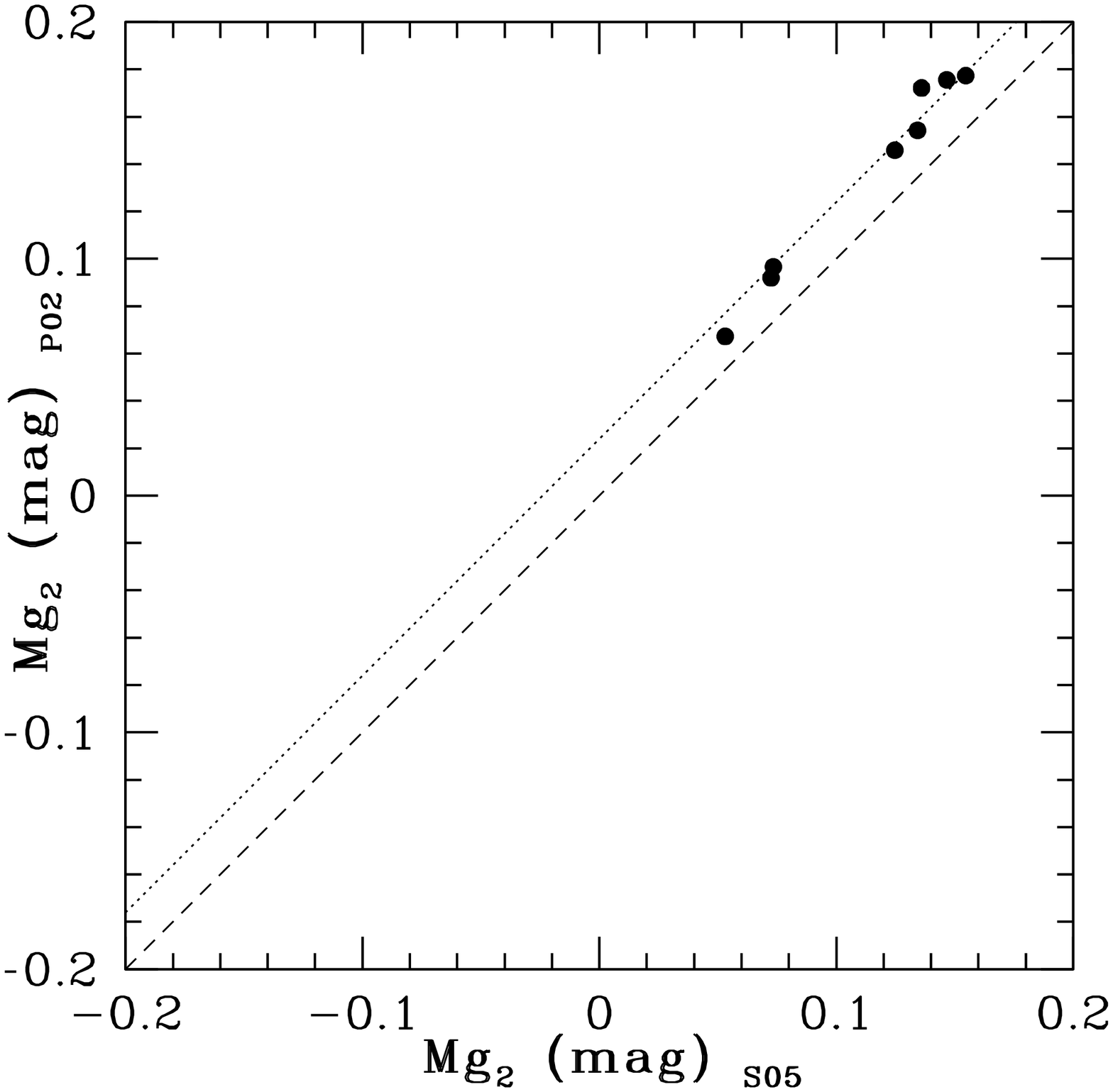,width=0.2\linewidth,clip=} \\
\epsfig{file=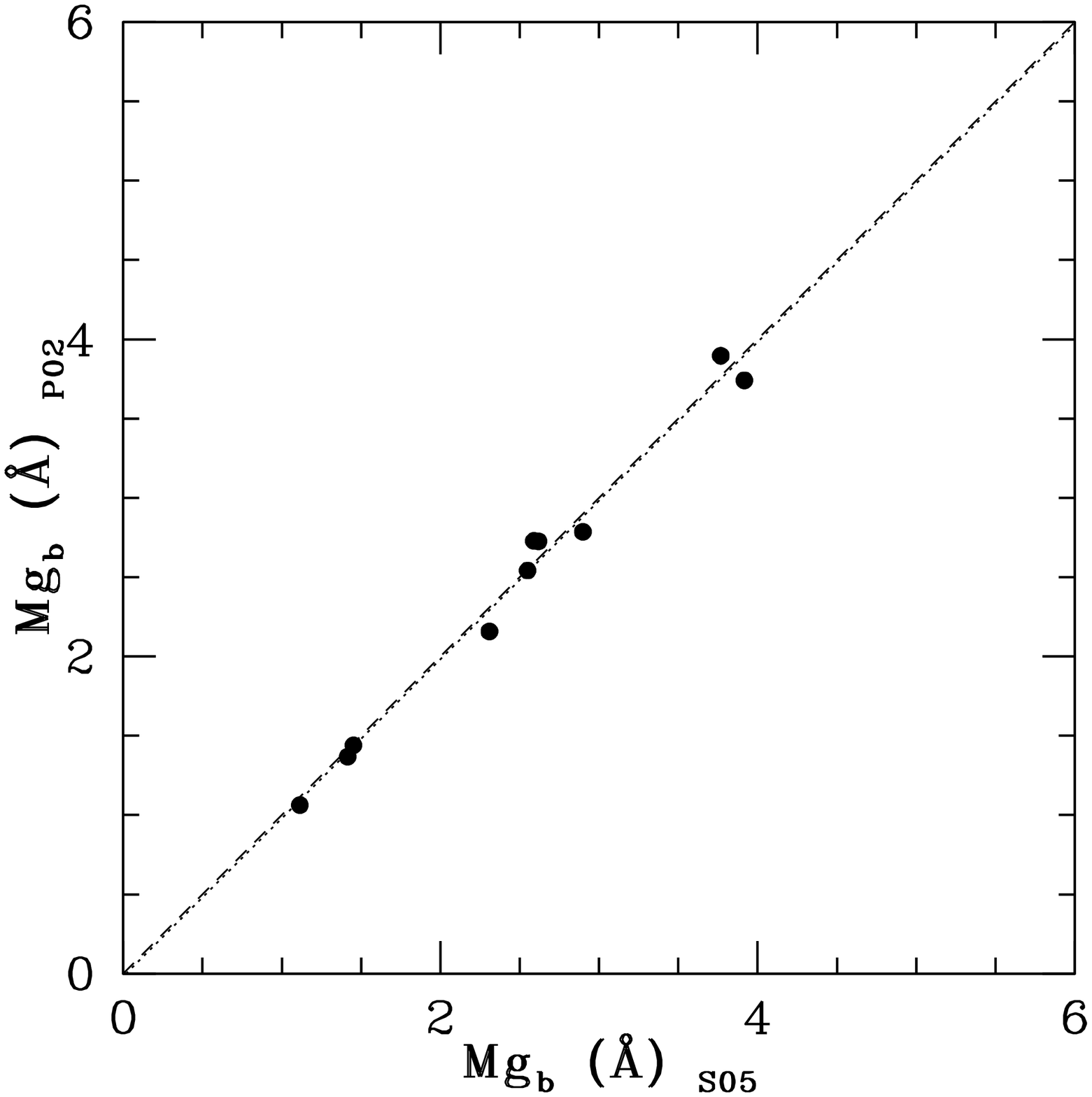,width=0.2\linewidth,clip=} &
\epsfig{file=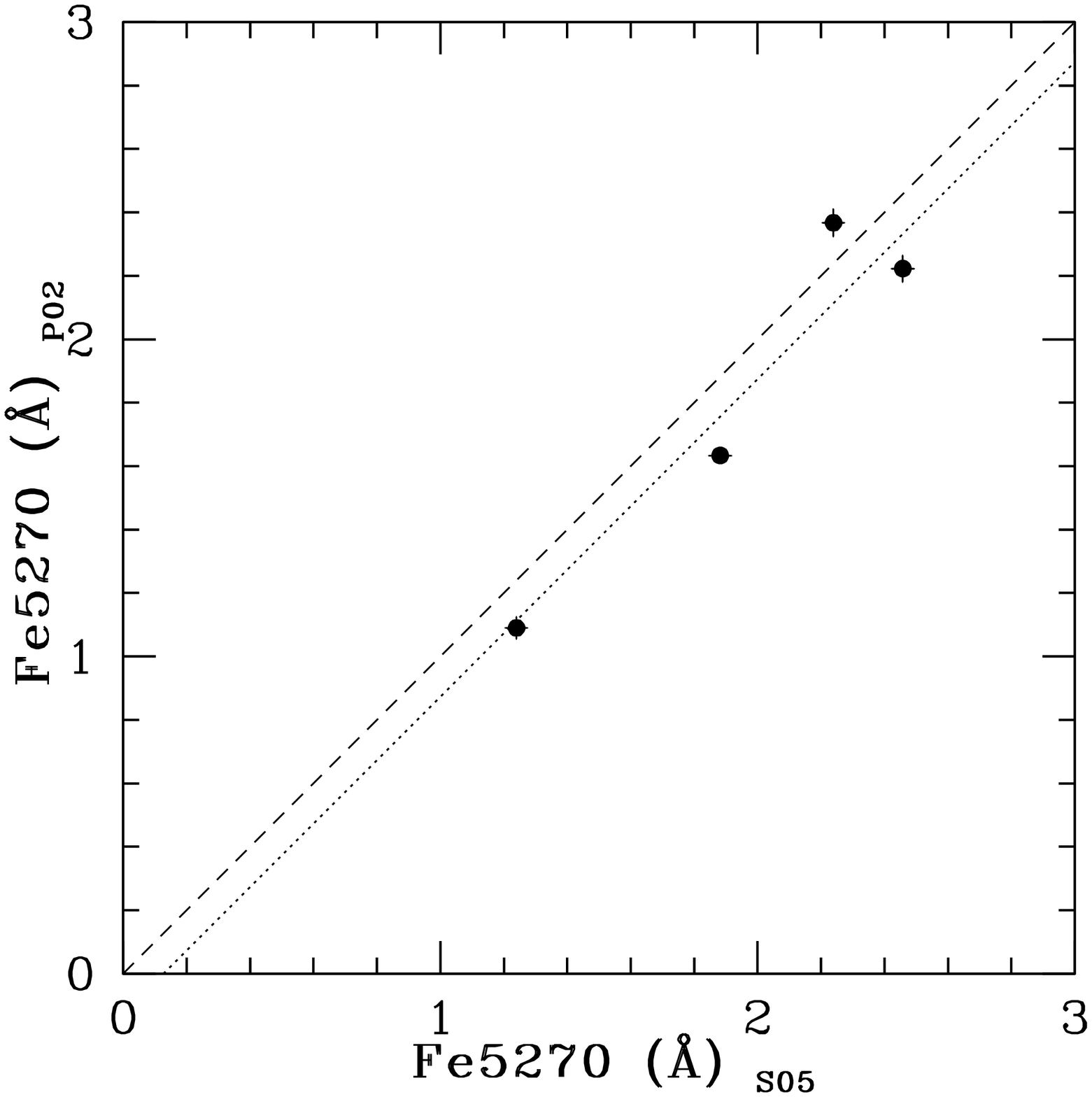,width=0.2\linewidth,clip=} &
\epsfig{file=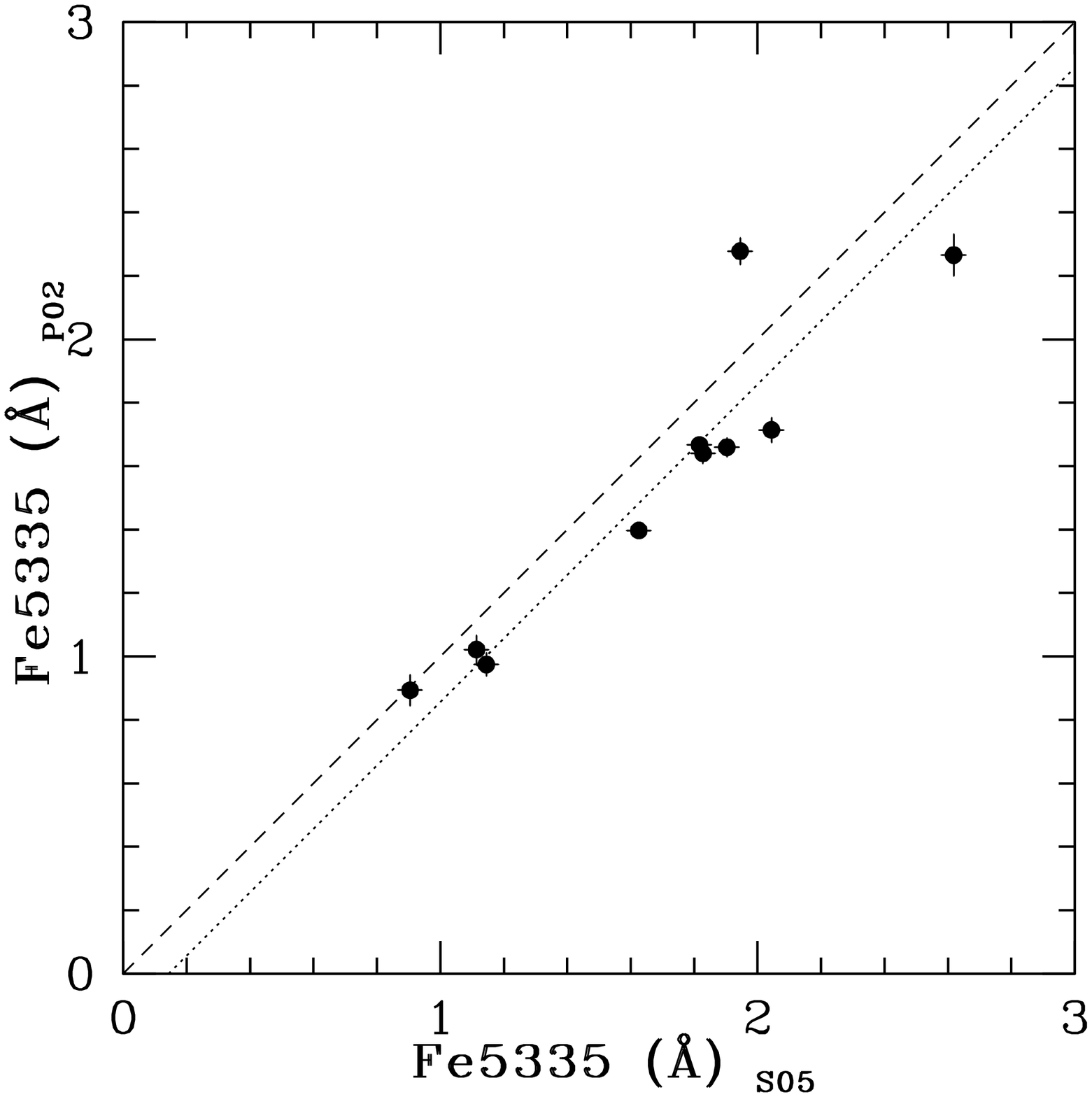,width=0.2\linewidth,clip=} &
\epsfig{file=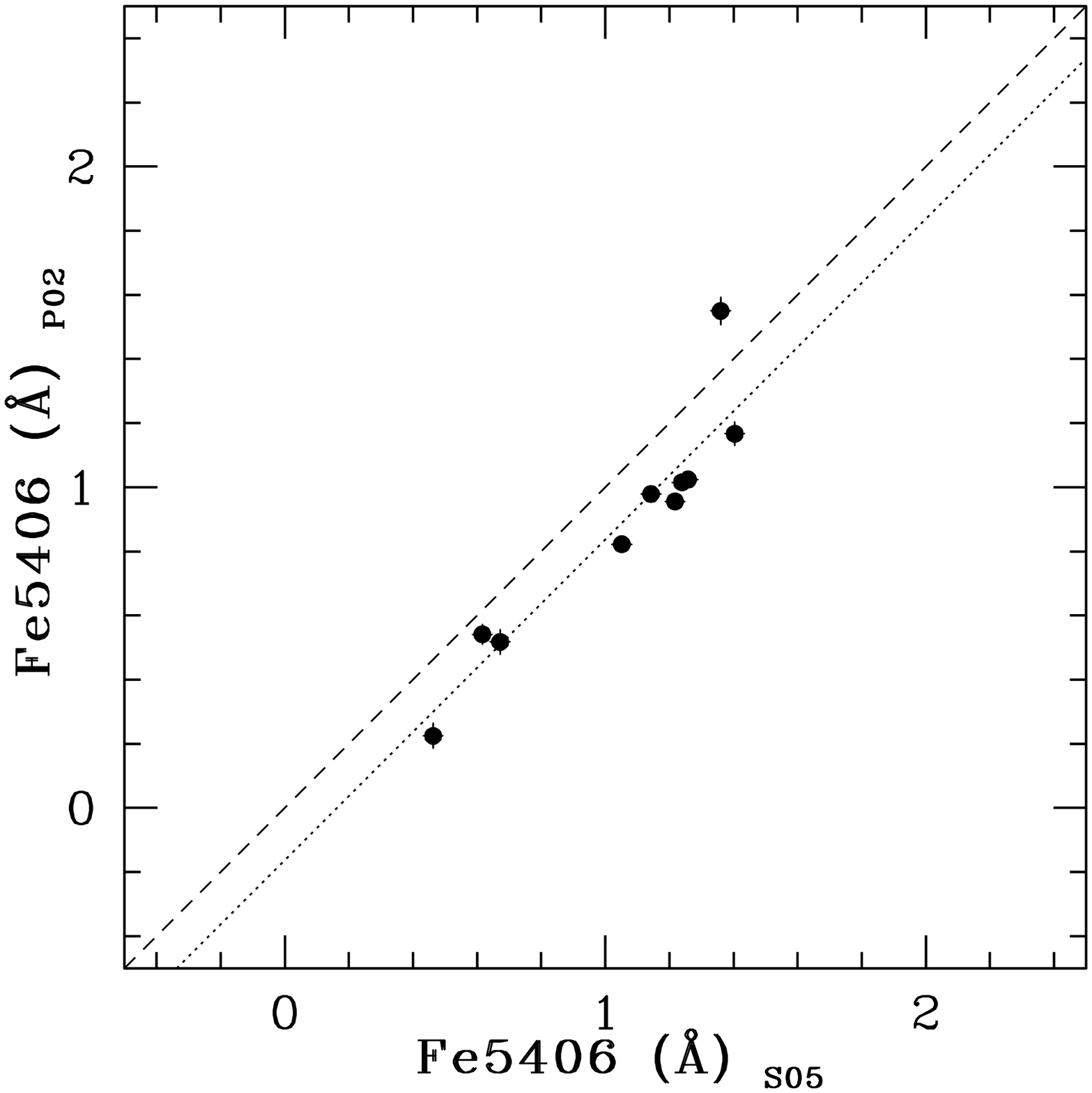,width=0.2\linewidth,clip=} \\

\end{tabular}
\caption{The same as Figure~\ref{fig:cal05} but for the calibration of the
  \cite{schiavon05} dataset to the \cite{puzia02} dataset for the Milky Way GCs.  The indices Fe4383 and Fe5015 were
  not used for any comparisons or calculations because of their
  unreliable index measurements caused by defects in the \cite{schiavon05}
  spectra.} 
\label{fig:calMW}
\end{figure}

\begin{figure}
\centering
\begin{tabular}{ccc}
\epsfig{file=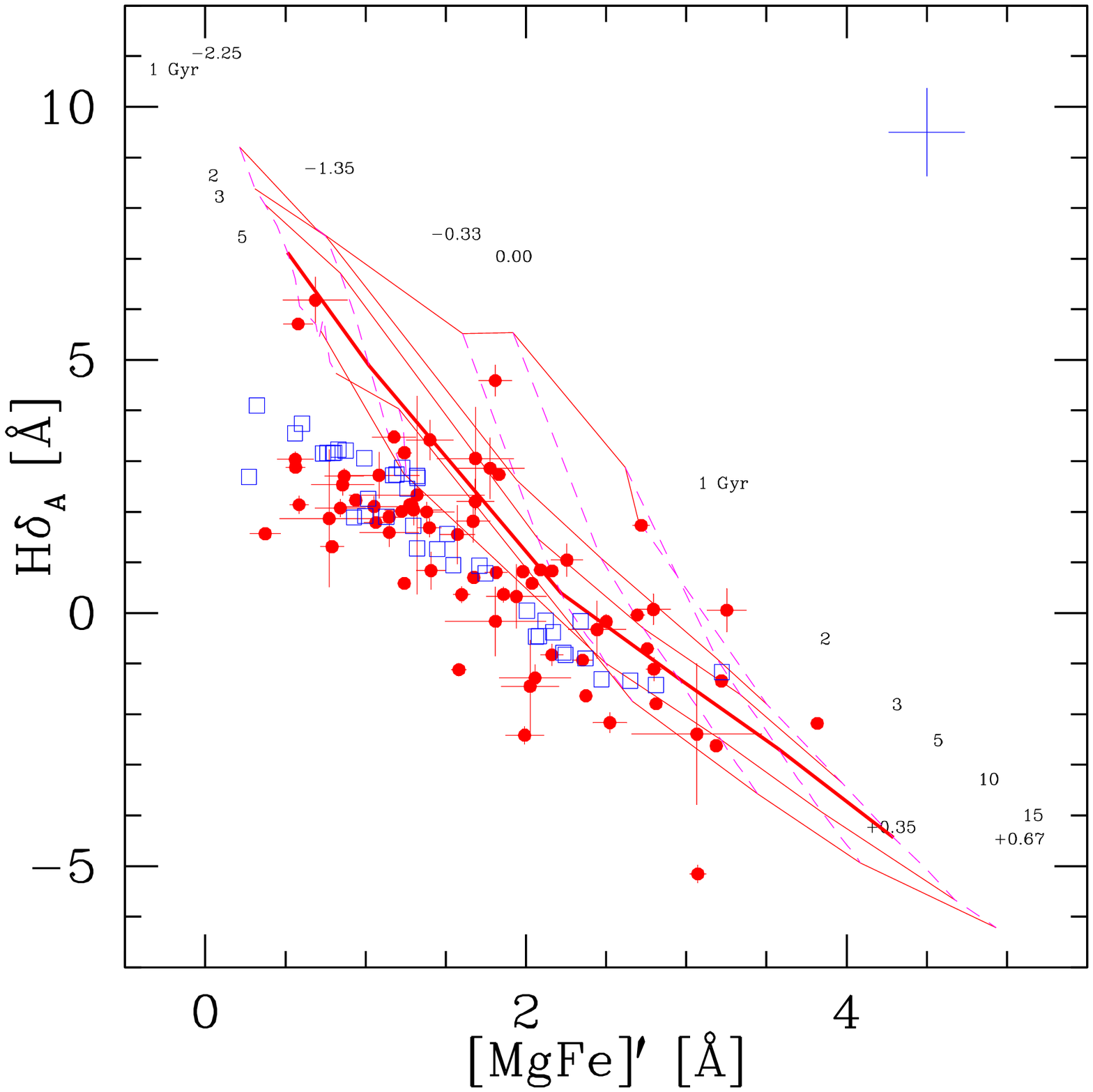,width=0.3\linewidth,clip=} &
\epsfig{file=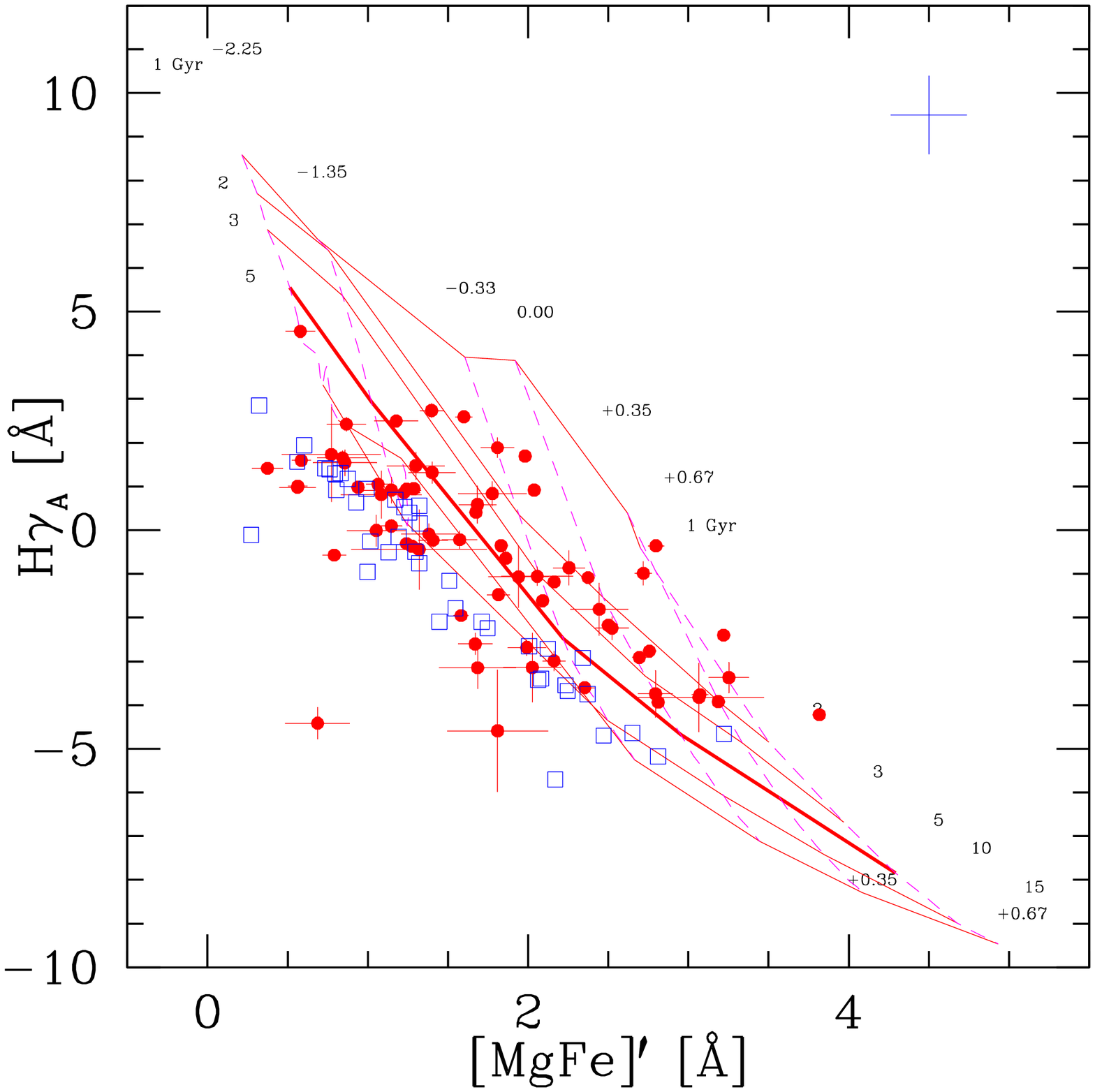,width=0.3\linewidth,clip=} &
\epsfig{file=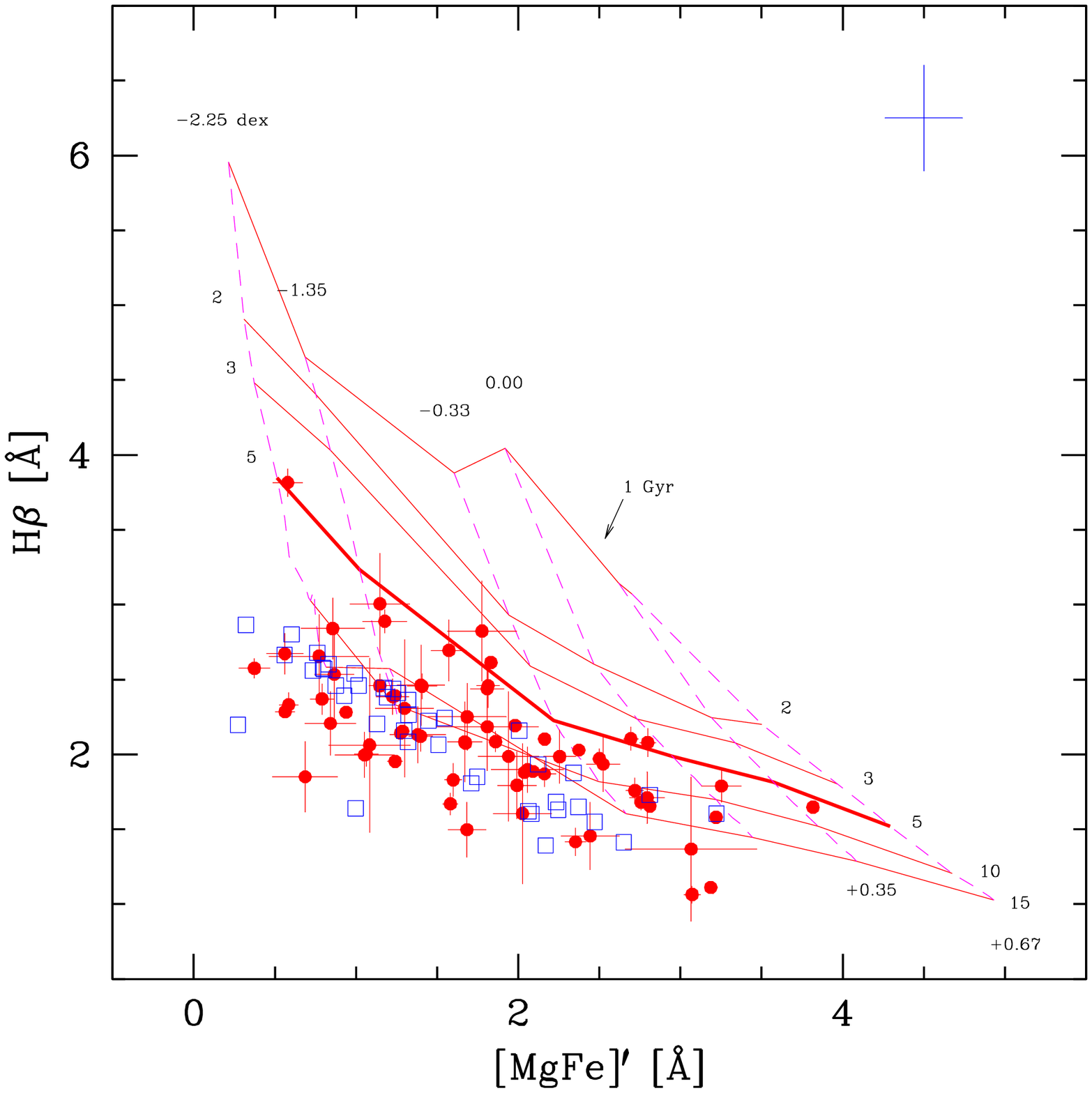,width=0.3\linewidth,clip=} \\
\epsfig{file=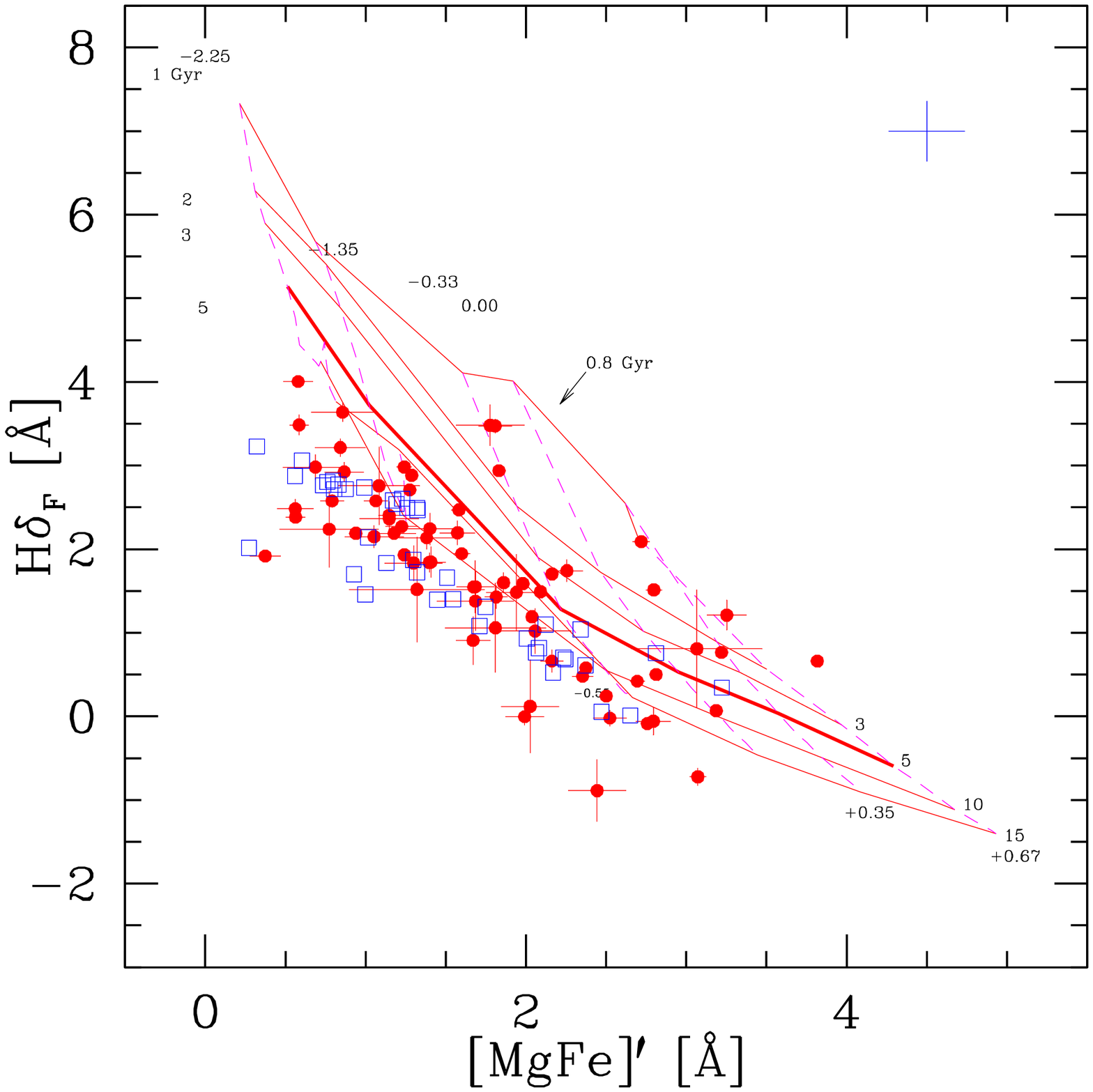,width=0.3\linewidth,clip=} &
\epsfig{file=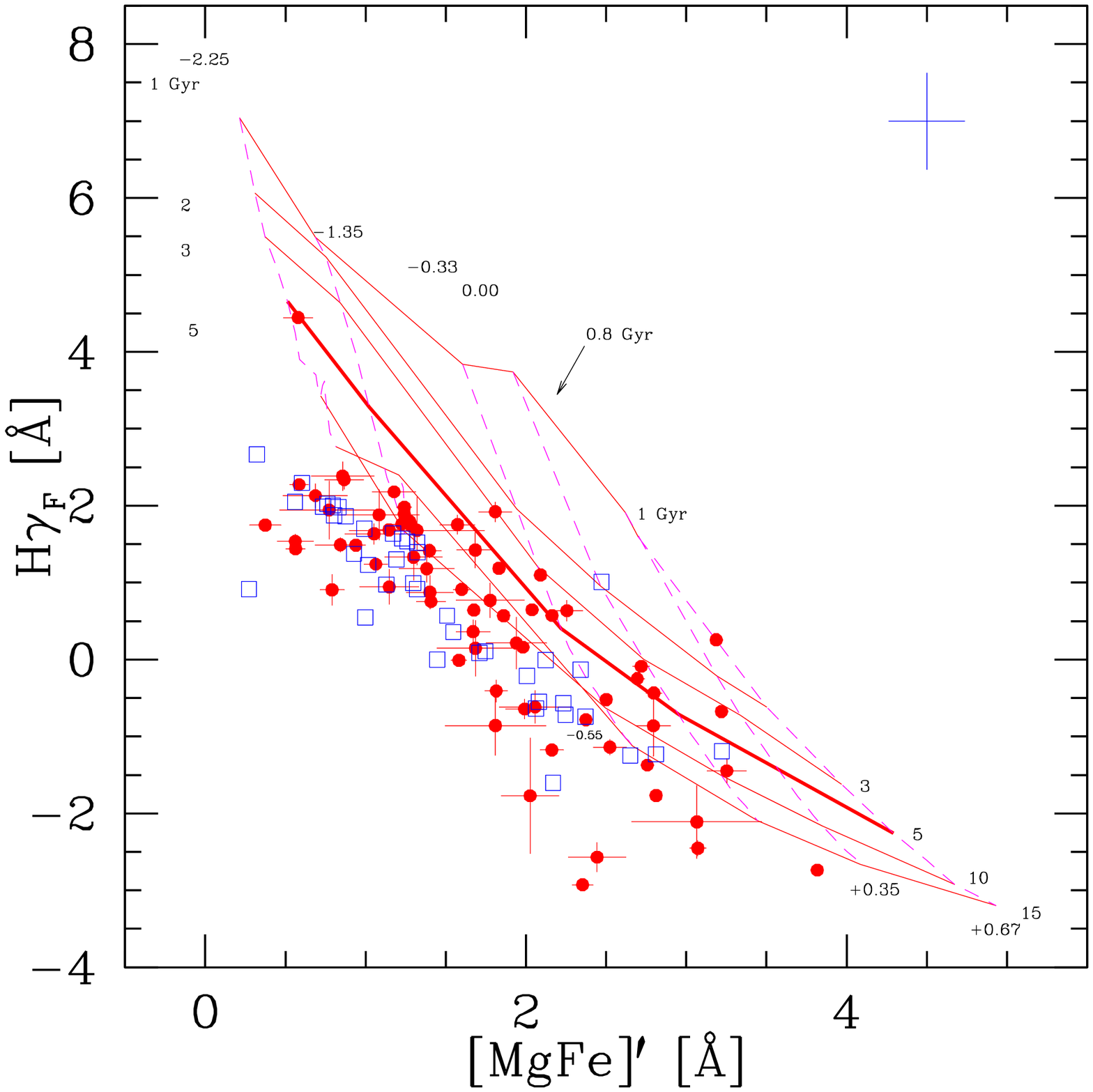,width=0.3\linewidth,clip=} &
\epsfig{file=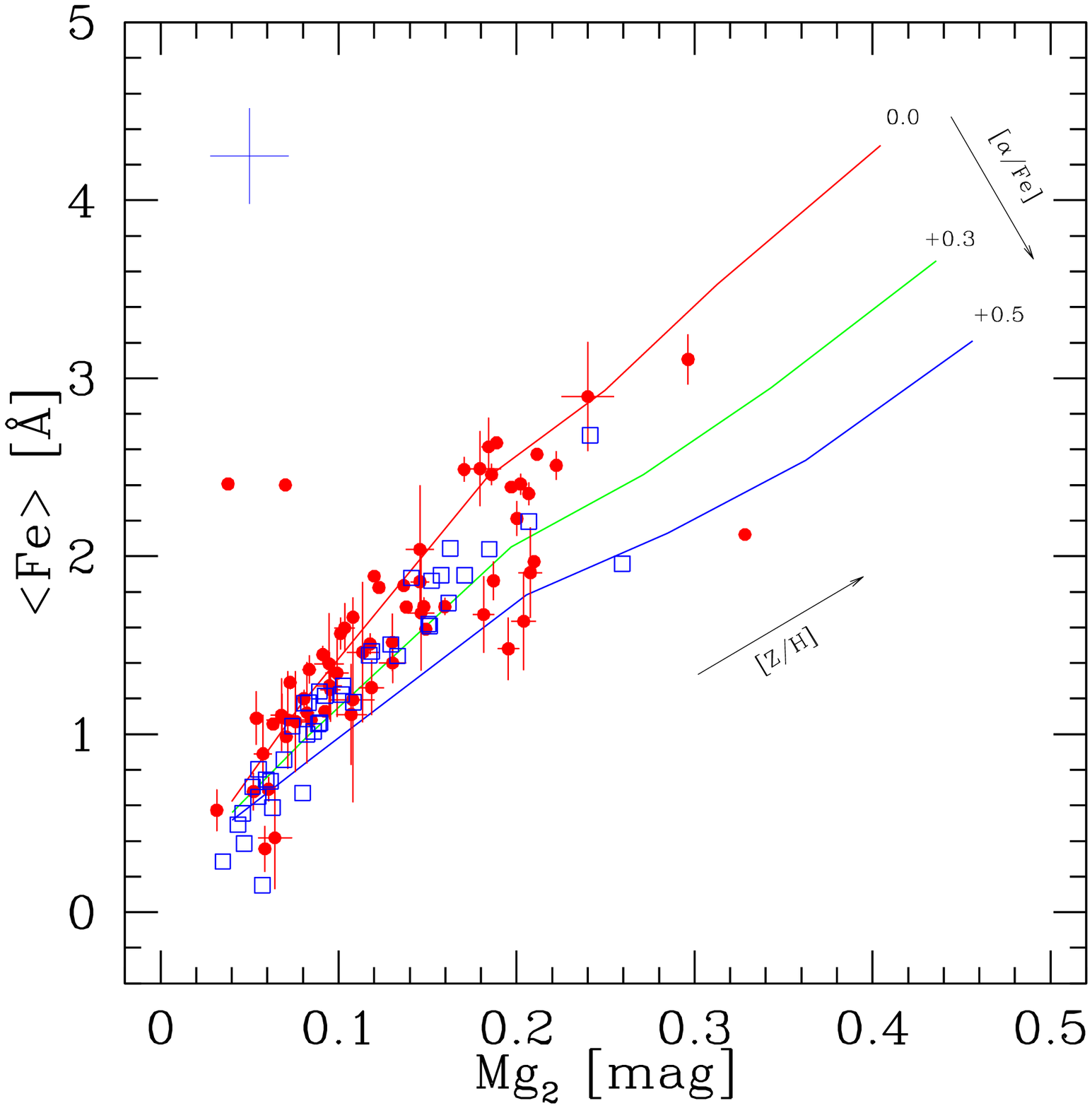,width=0.3\linewidth,clip=} \\

\end{tabular}
\caption{Diagnostic plots for all measurements (no duplicate
  measurements) of GCs in NGC 5128 with
  S/N$ > 30 \rm{\AA}^{-1}$ ({\it red circles}) and the
  Milky Way data \citep{schiavon05,puzia02} ({\it blue squares}).
  The SSP models are from \cite{tmb03,tmk04} and the Balmer line
  diagnostic plots are for the grids of [$\alpha$/Fe]$=0.0$ dex. The bootstrapping
  uncertainty is attached to each point and the systematic
  uncertainty is indicated by the {\it blue} cross in the corner.} 
\label{fig:diagnostic_hydrogen}
\end{figure}

\begin{figure}
\centering
\begin{tabular}{cc}
\epsfig{file=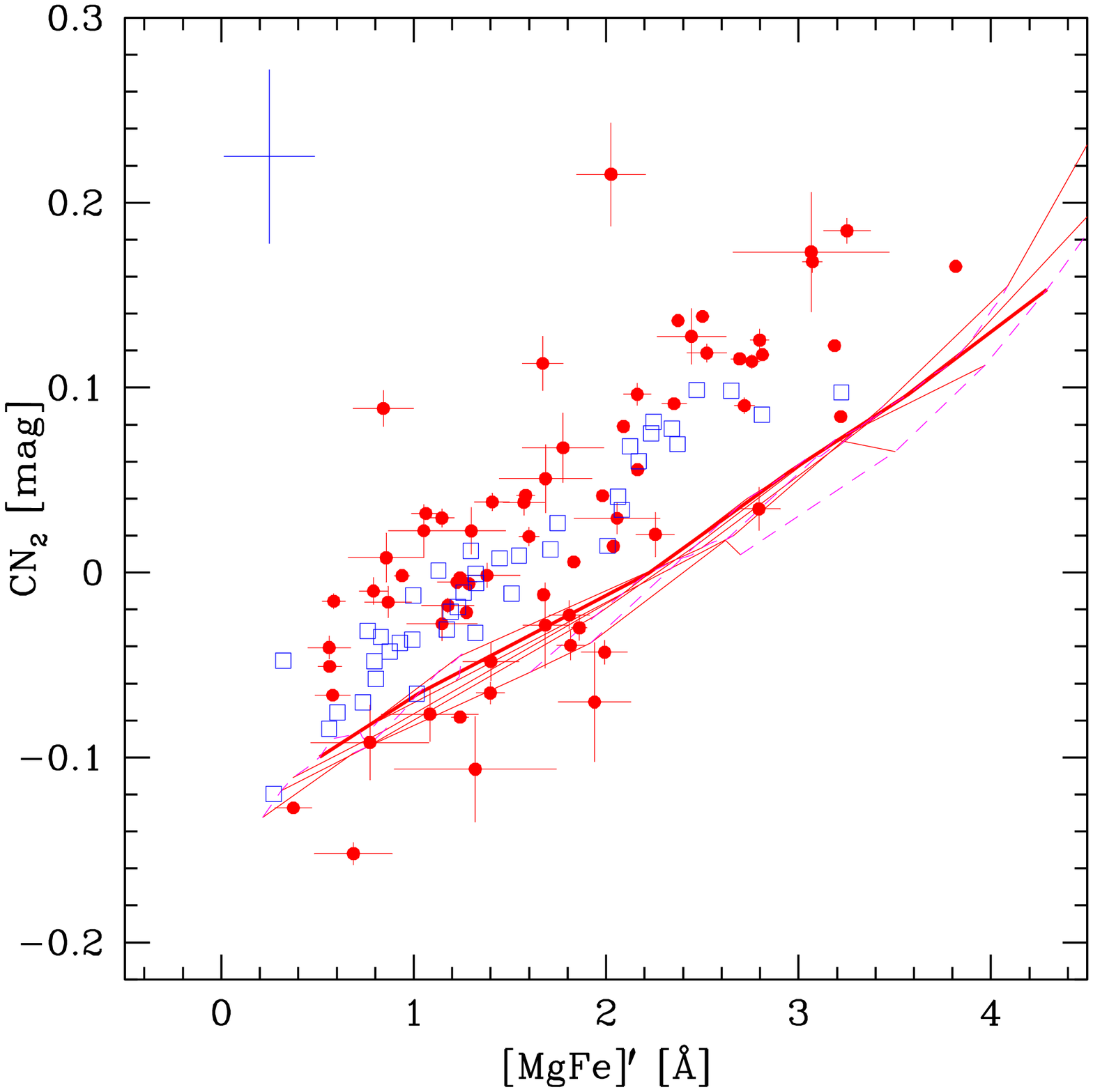,width=0.5\linewidth,clip=} &
\epsfig{file=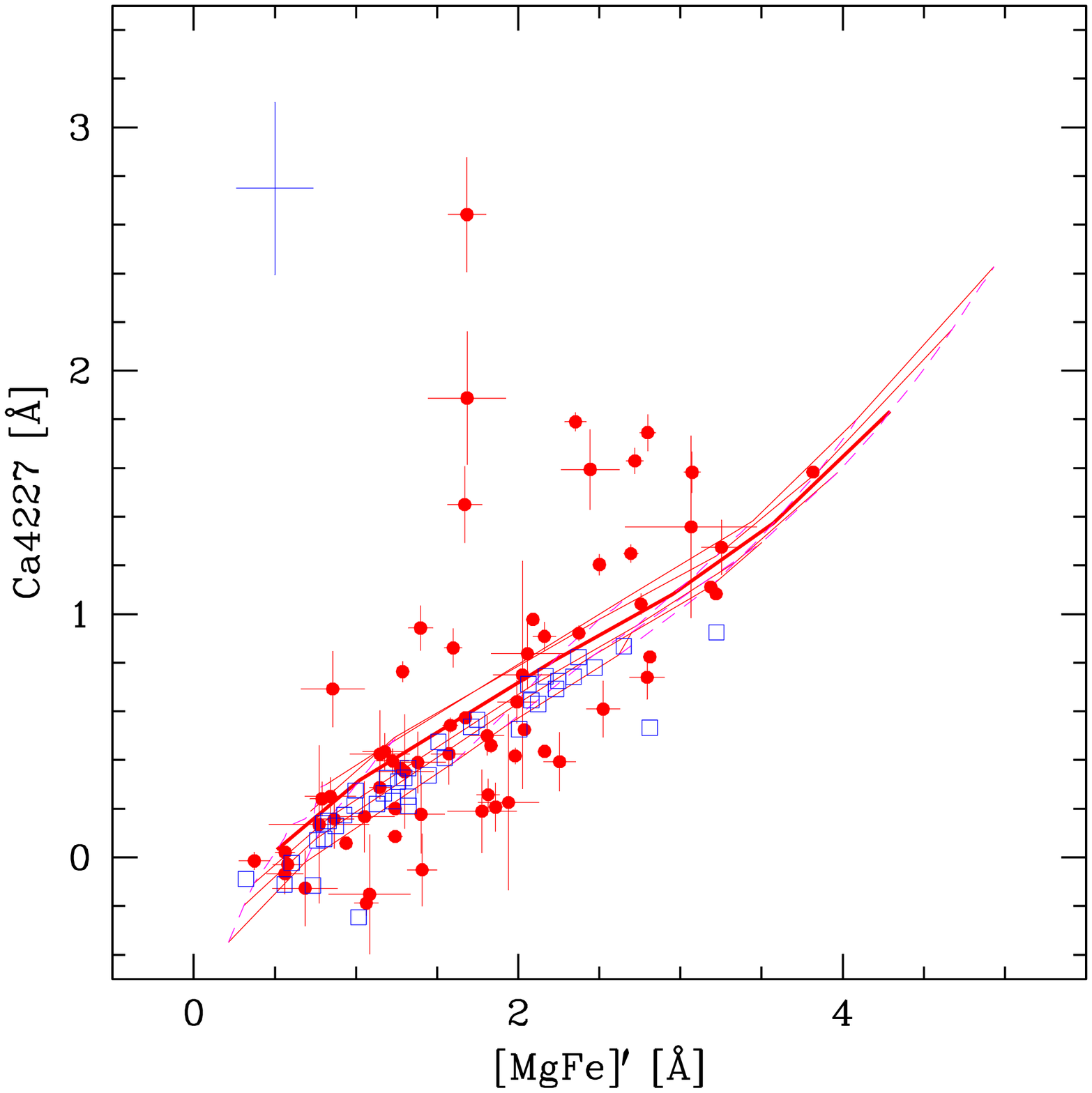,width=0.5\linewidth,clip=} \\
\epsfig{file=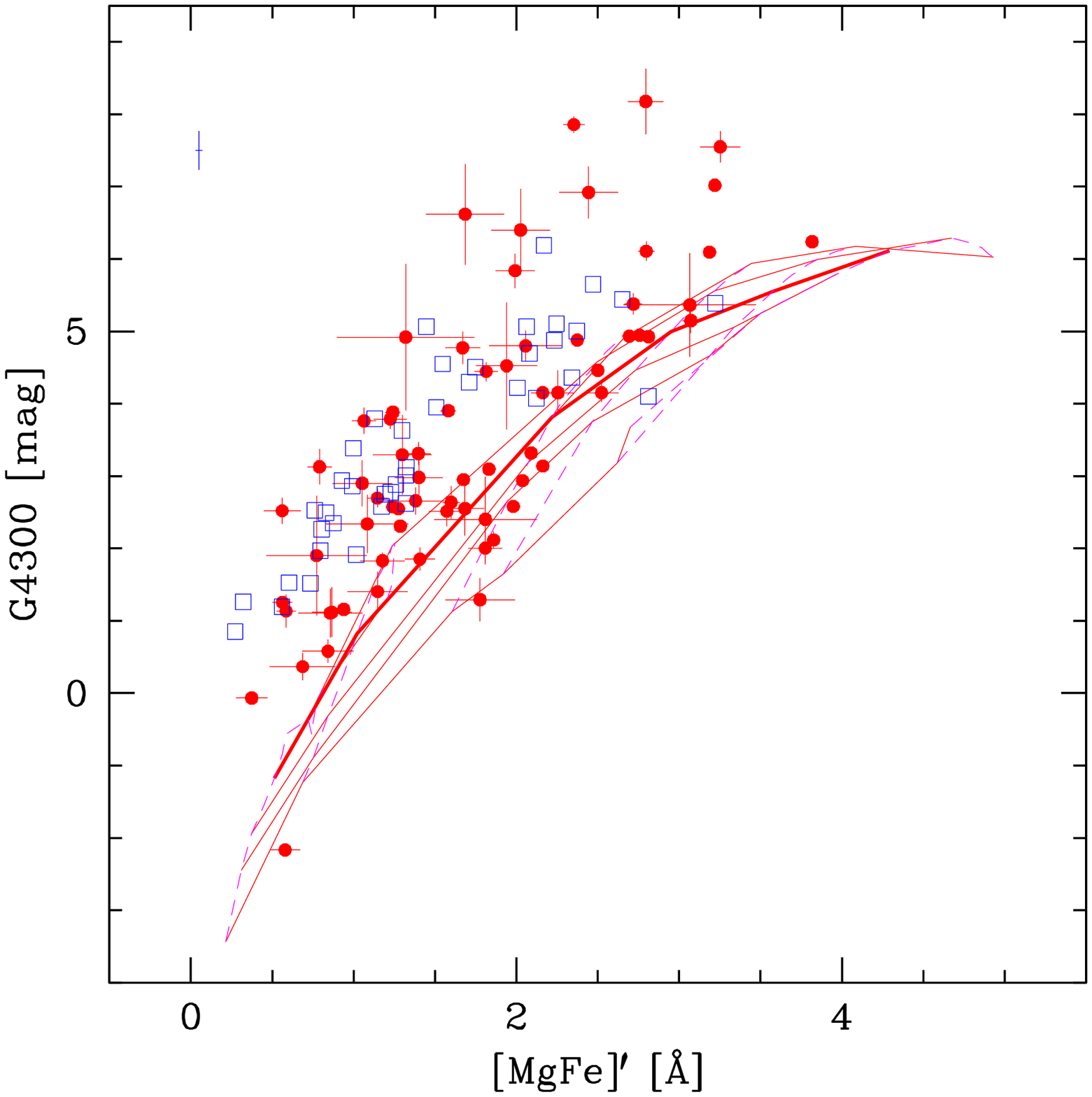,width=0.5\linewidth,clip=} &
\epsfig{file=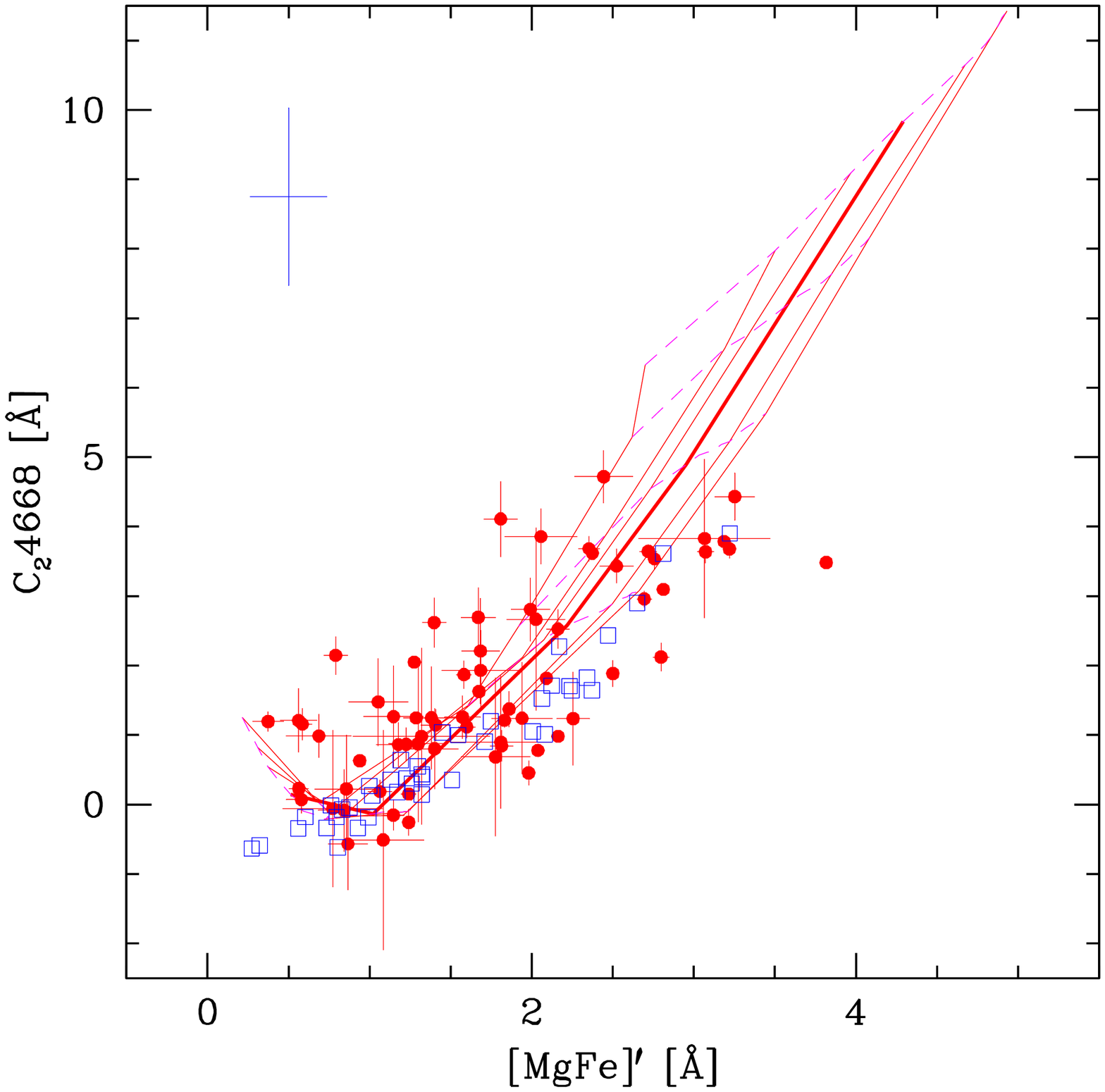,width=0.5\linewidth,clip=} \\

\end{tabular}
\caption{Diagnostic plots for calcium, carbon, and nitrogen sensitive
  elements of GCs in NGC 5128 with
  S/N$ > 30 \rm{\AA}^{-1}$ ({\it red circles}) and the
  Milky Way data \citep{schiavon05,puzia02} ({\it blue squares}).
  The SSP models are from \cite{tmb03,tmk04} and the Balmer line
  diagnostic plots are for the grids of [$\alpha$/Fe]$=0.0$ dex.  The bootstrapping
  uncertainty is attached to each point and the systematic
  uncertainty is indicated by the {\it blue} cross in the corner.} 
\label{fig:diagnostic_cn}
\end{figure}

\begin{figure}
\plotone{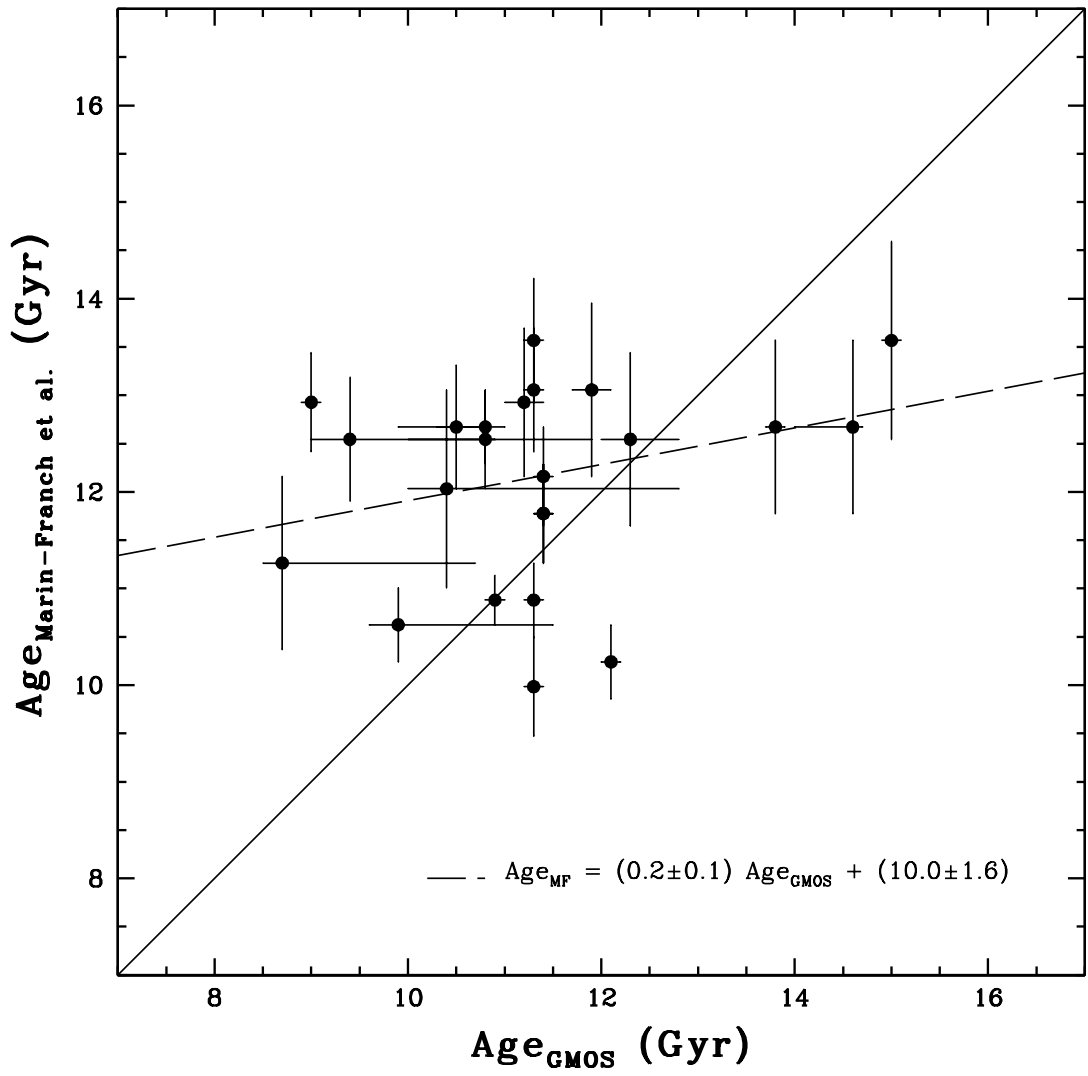}
\caption{A comparison of the ages of 24 Milky Way GCs in common
  between this study and that of relative age main sequence fitting of
  \cite{marin-franch09}.  A 1:1 {\it solid line} is shown as well as
  the least squared fit ({\it dashed line}) with a slope and intercept of $0.2\pm0.1$ and $10.0\pm1.6$ and an rms$=0.97$. } 
\label{fig:MW_age_comp}
\end{figure}
       
\begin{figure}
\plotone{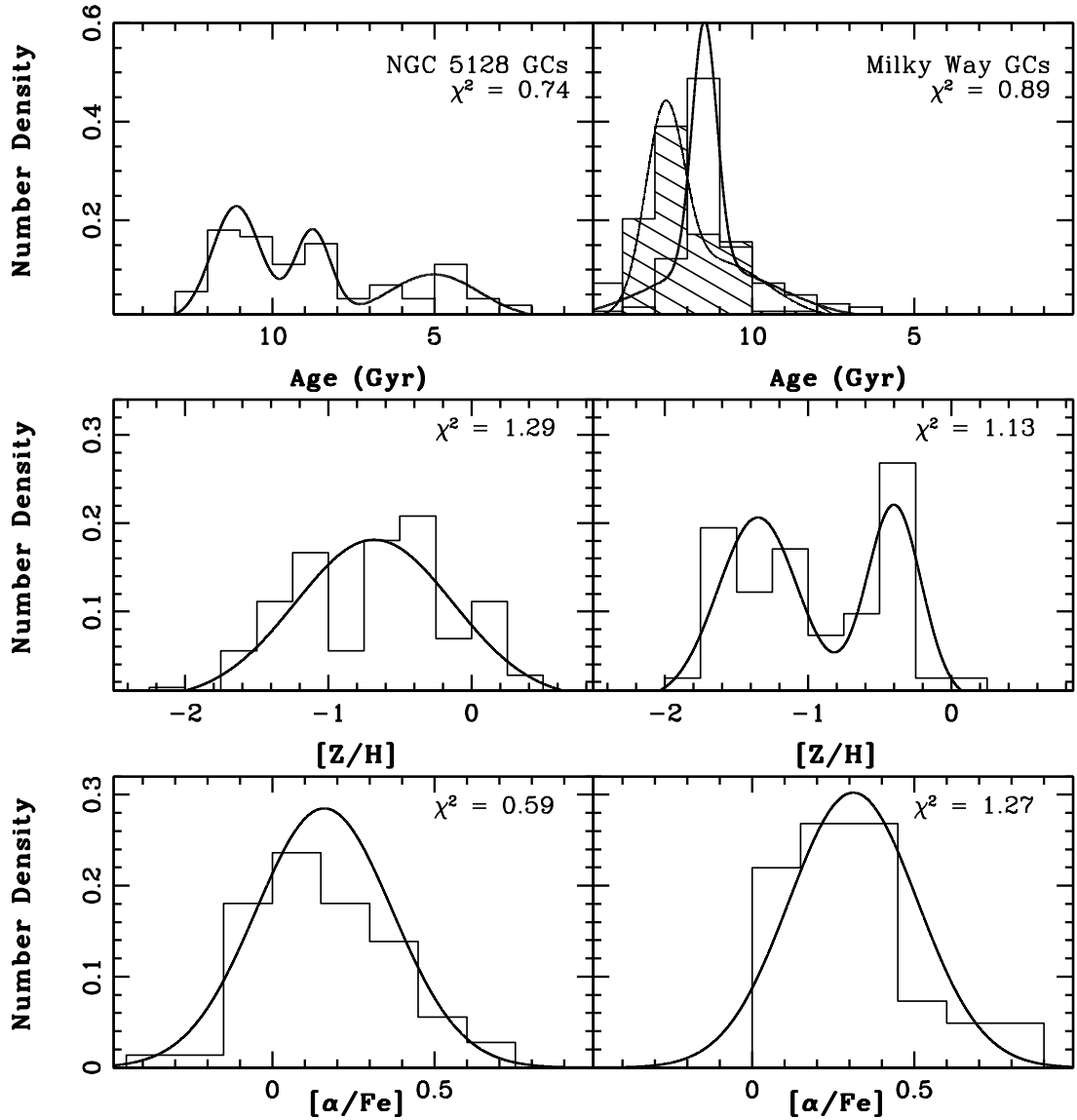}
\caption{The  age   ({\it  top  panels}),   metallicity  ({\it  middle
panels}),  and   [$\alpha$/Fe]  ({\it  bottom   panels})  distribution
functions for 72 GCs in NGC 5128 ({\it left}) and for 41 Milky Way GCs
({\it right}) derived from the SSP models of \cite{tmb03} with varying
$\alpha$-elemental  abundance \citep{tmk04}.   The  summed best fit Gaussian
distributions are
overplotted   ({\it  solid  line}),   with  reduced   $\chi^2$  values
indicated.  The fits are listed
in Table~\ref{tab:rmixfits}.   The cross-hatched  histogram in the  Milky Way
age distribution  is the result from  \cite{marin-franch09}, with best
fit    bimodal    distribution    ({\it    dot   dashed    line}    of
(0.44$\pm0.09$,11.07$\pm0.63$,1.52$\pm0.25$)                        and
(0.56$\pm0.09$,12.71$\pm0.20$,0.59$\pm0.22$)  with  reduced $\chi^2  =
0.21$.  }
\label{fig:agemetafe}
\end{figure}

\begin{figure}
\centering
\begin{tabular}{cc}
\epsfig{file=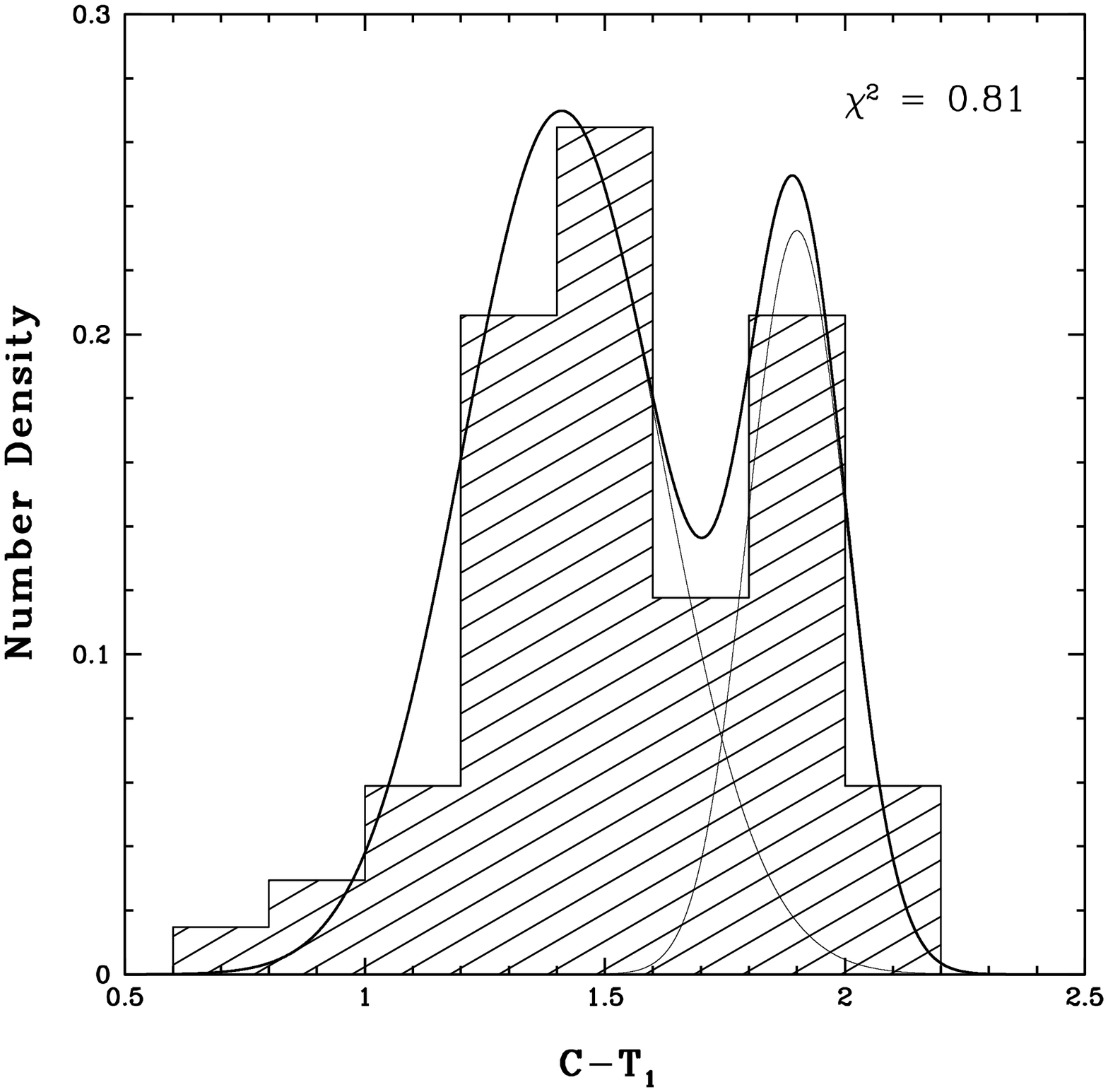,width=0.5\linewidth,clip=} &
\epsfig{file=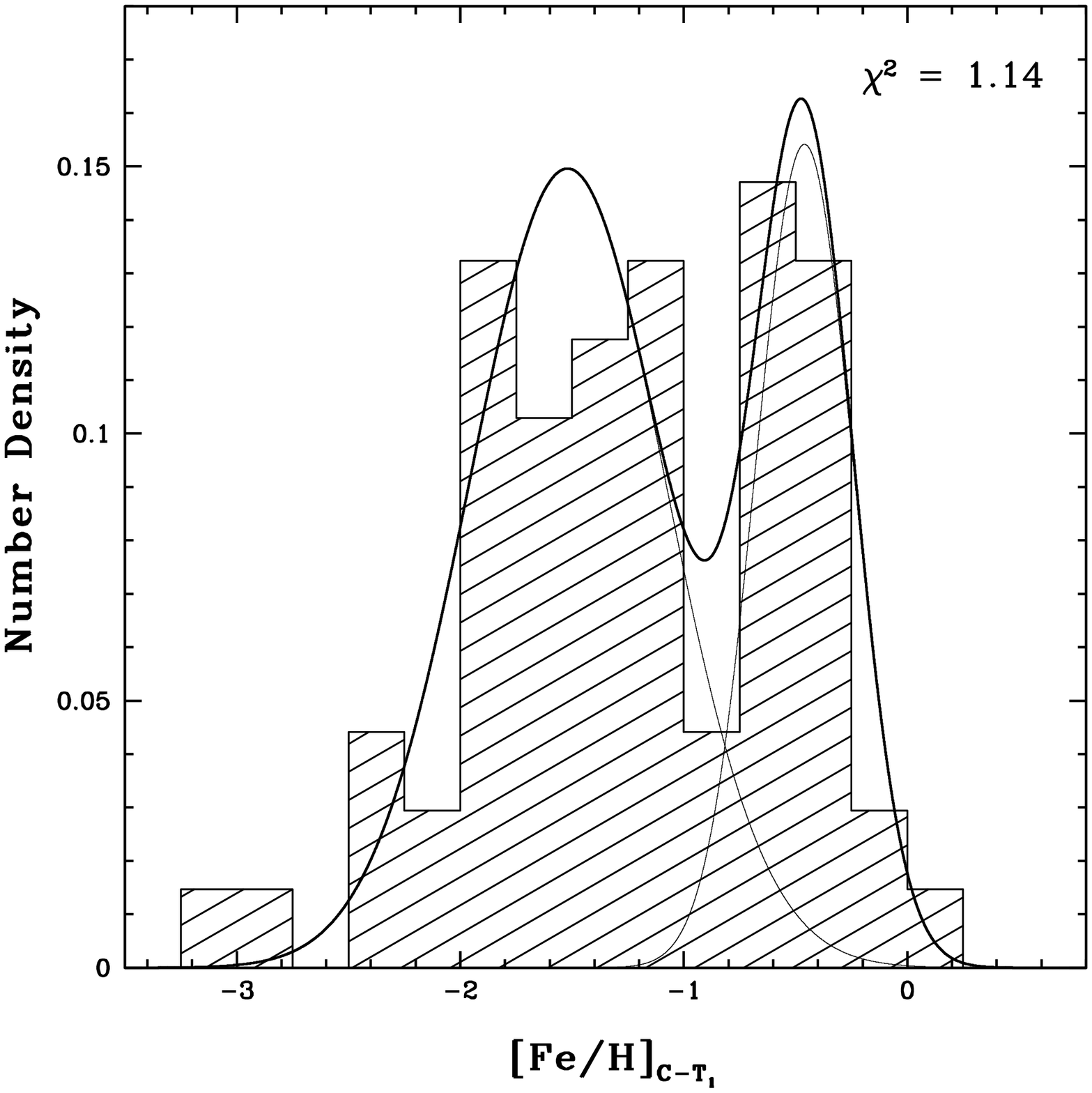,width=0.5\linewidth,clip=} \\
\epsfig{file=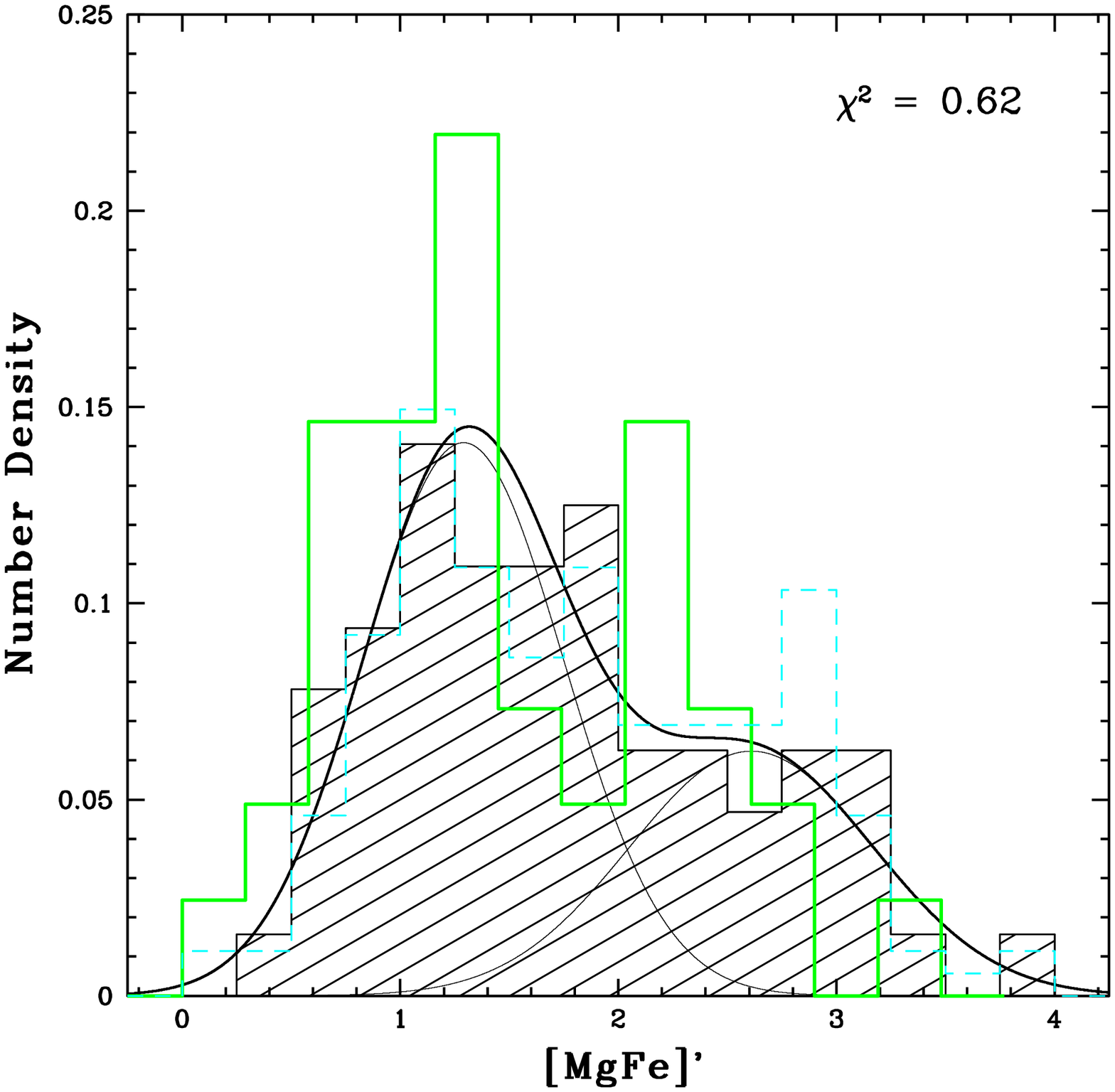,width=0.5\linewidth,clip=} &
\epsfig{file=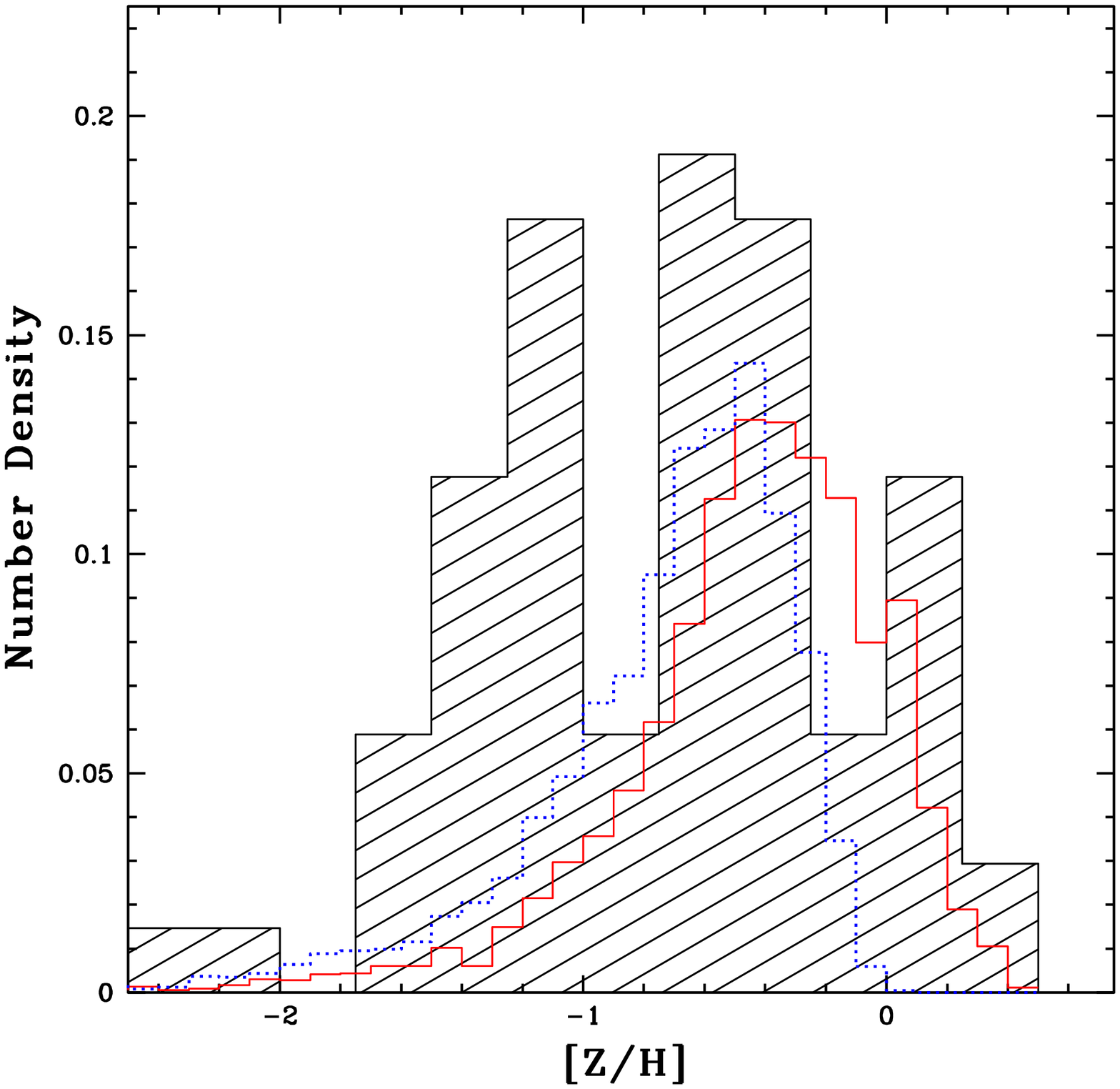,width=0.5\linewidth,clip=} \\

\end{tabular}
\caption{Various distributions  of 68 GCs in NGC  5128 with determined
  metallicity and  available (C-T$_1$) from  \cite{harris04}, with the
  best fit Gaussian distributions and their summation overplotted.{\it
  Top left:} The color distribution, (C-T$_1$), with best fit Gaussian
  distributions.   The  fitted  parameters  are (proportion  of  total
  sample,  peak, sigma)  = (0.70$\pm0.08$,1.41$\pm0.04$,0.21$\pm0.04$)
  and  (0.30$\pm0.08$,1.90$\pm0.04$,0.10$\pm0.26$),   with  a
  $\chi^2_{red}  =   0.81$.  {\it   Top  right:}  The   metallicity,  [Fe/H]
  transformed from  the color index, (C-T$_1$) (see  text). The fitted
  parameters   are   (0.66$\pm0.08$,-1.52$\pm0.10$,0.44$\pm0.08$)  and
  (0.34$\pm0.08$,-0.46$\pm0.07$,0.22$\pm0.05$), with a $\chi^2_{red}
  =  1.14$. {\it  Bottom left:}  The distribution  of  the metallicity
  [MgFe]$^\prime$  index derived from  our measured  spectroscopy.  The
  fitted  parameters for a bimodal distribution are  (0.65$\pm0.33$,1.29$\pm0.29$,0.46$\pm0.15$)
  and  (0.35$\pm0.33$,2.61$\pm0.73$,0.56$\pm0.35$),   with  a 
  $\chi^2_{red} = 0.62$. A unimodal distribution has a  $\chi^2_{red} = 1.02$. The {\it open thick green} histogram is for the 41 Milky
  Way  GCs used  in  this study  \citep{schiavon05,puzia02}. The  {\it
  hatched blue} histogram  is for the combination of  all available NGC
  5128  data   from  this  study  with   S/N$  >  30   \AA$  and  from
  \cite{beasley08}.  No duplicate GCs are used in this plot, totalling
  174 GCs.  {\it Bottom right:}  The metallicity obtained from the SSP
  models \citep{tmb03,tmk04}.  The best fit distributions are shown in
  Fig.~\ref{fig:agemetafe}.   Overplotted  are  the  distributions  of  the
  stellar halo in  NGC 5128 for the inner halo, 8  kpc ({\it red open histogram}) and
  the combination  of two outer fields  at 21 and 31  kpc ({\it dashed blue histogram})
  \citep{harris00,harris02,rejkuba05}.}
\label{fig:cmd}
\end{figure}

\begin{figure}
\plotone{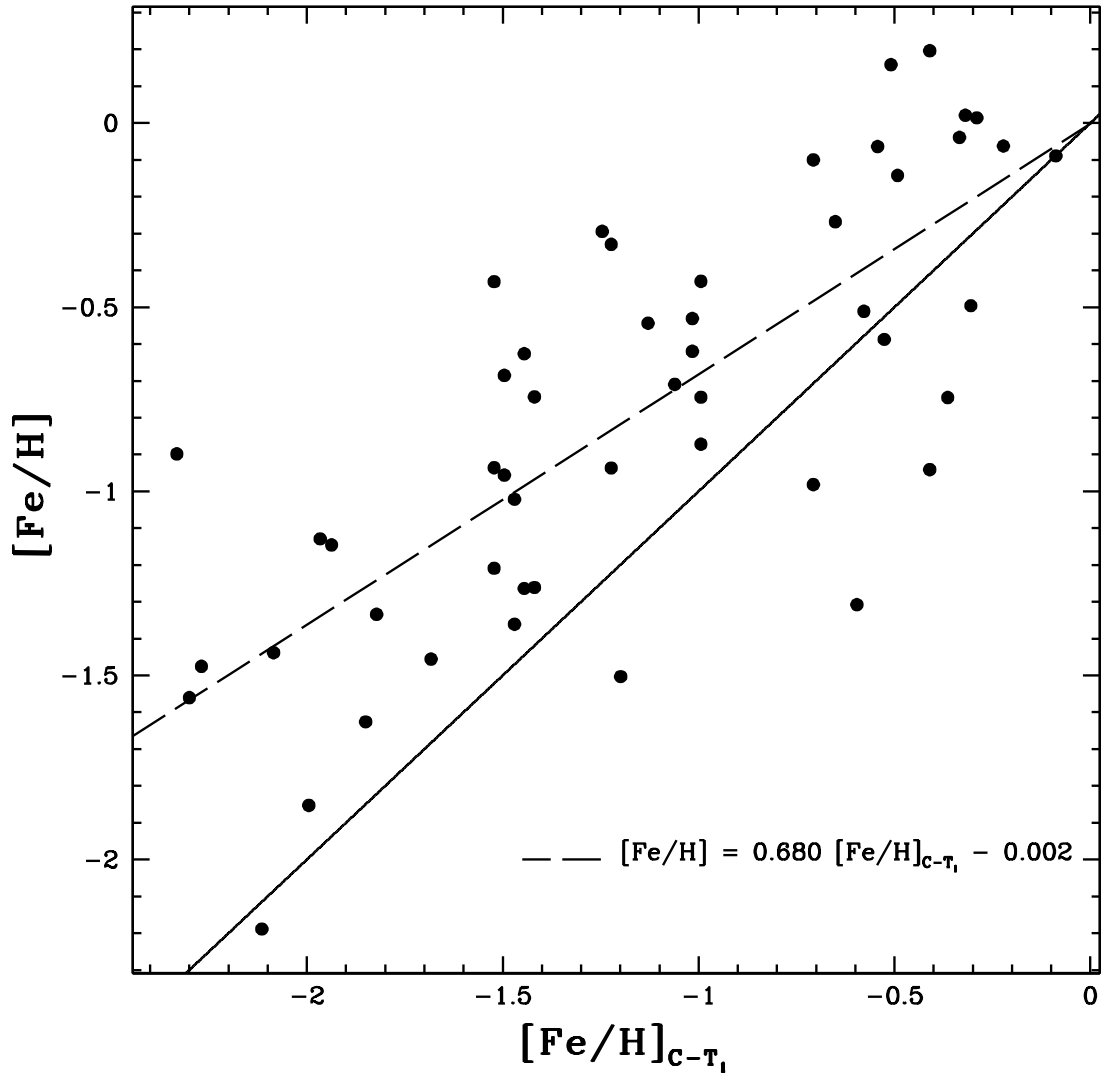}
\caption{Metallicity, [Fe/H]$_{C-T_1}$, derived from the color
  (C-T$_1$) compared to the SSP model determined [Fe/H] value for the
  GCs in NGC 5128.  The 1:1 line is shown as a {\it solid line}.  The
  best fit line is shown as a {\it dashed line} after 5 points were
  removed based on the Chauvenet Criteria \citep{parratt61}. It
  appears there is not a direct one-to-one correlation either due to
  potential non-linearity in the color-to-metallicity conversion or
  due to the model interpolation.} 
\label{fig:ZH_FeH}
\end{figure}
       
\begin{figure}
\centering
\begin{tabular}{c}
\epsfig{file=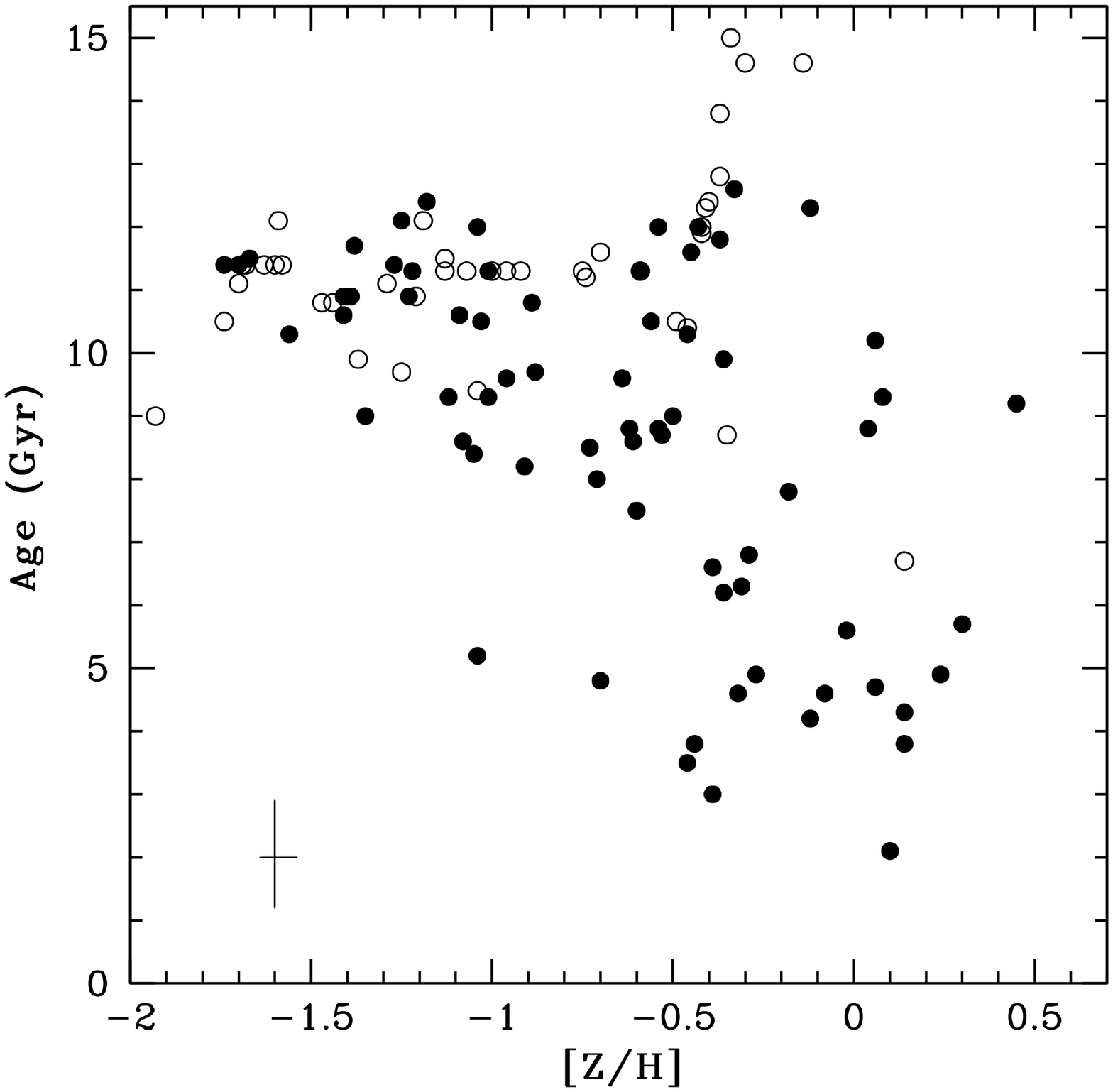,width=0.4\linewidth,clip=} \\
\epsfig{file=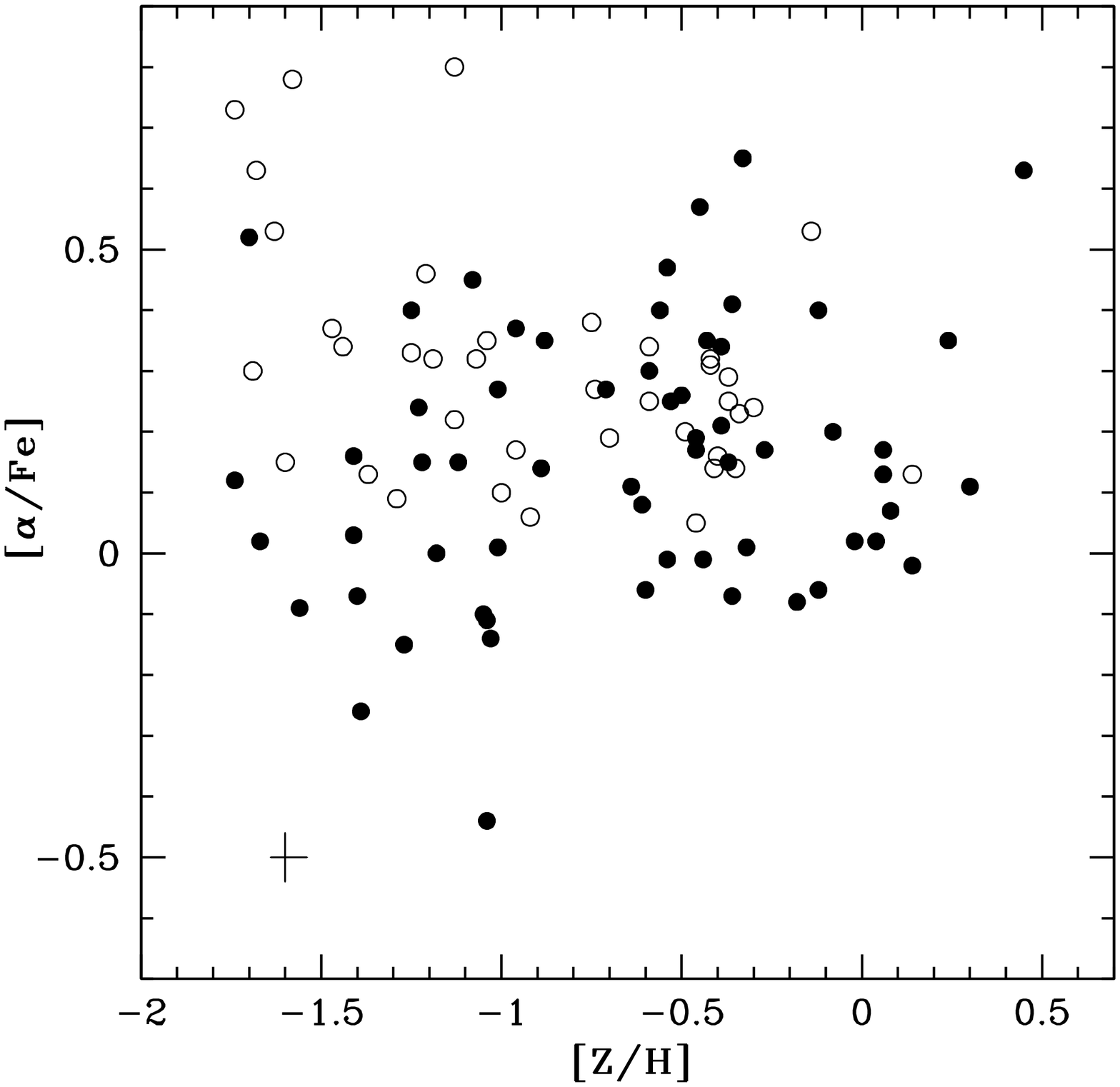,width=0.4\linewidth,clip=} \\
\epsfig{file=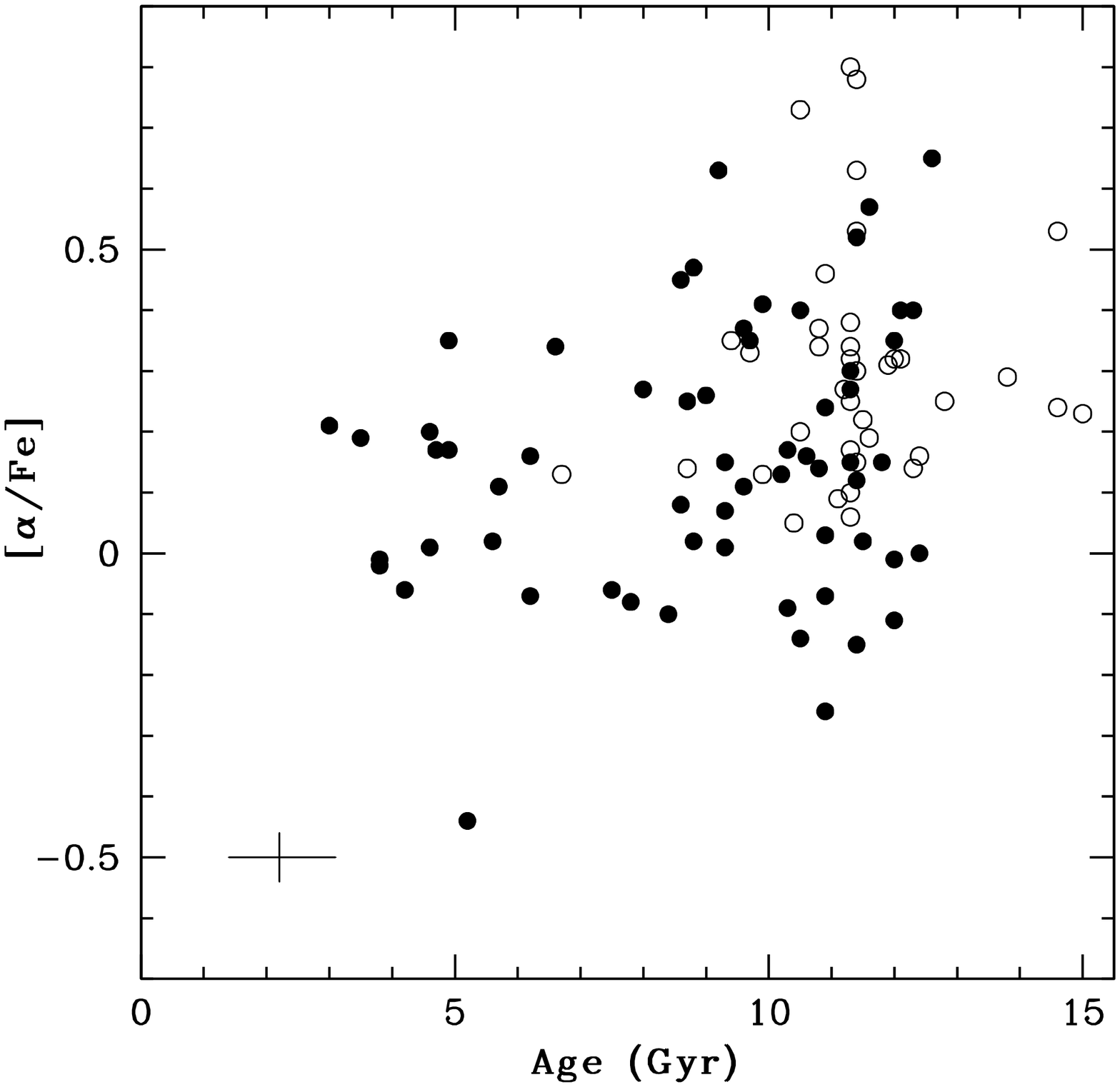,width=0.4\linewidth,clip=} \\

\end{tabular}
\caption{Metallicity as a function of age ({\it top}) and
  $\alpha$-to-Fe abundance ratio ({\it middle}), and age as a function
of $\alpha$-to-Fe abundance ratio ({\it bottom}) obtained from the SSP
models of \cite{tmb03} with varying $\alpha$-elemental abundance
\citep{tmk04}.  The 72 GCs in NGC 5128 are {\it solid circles} and the
41 Milky Way GCs are {\it open circles}.  The mean uncertainties are
displayed in the lower left of each plot.} 
\label{fig:agemetafe_comp}
\end{figure}

\begin{figure}
\plotone{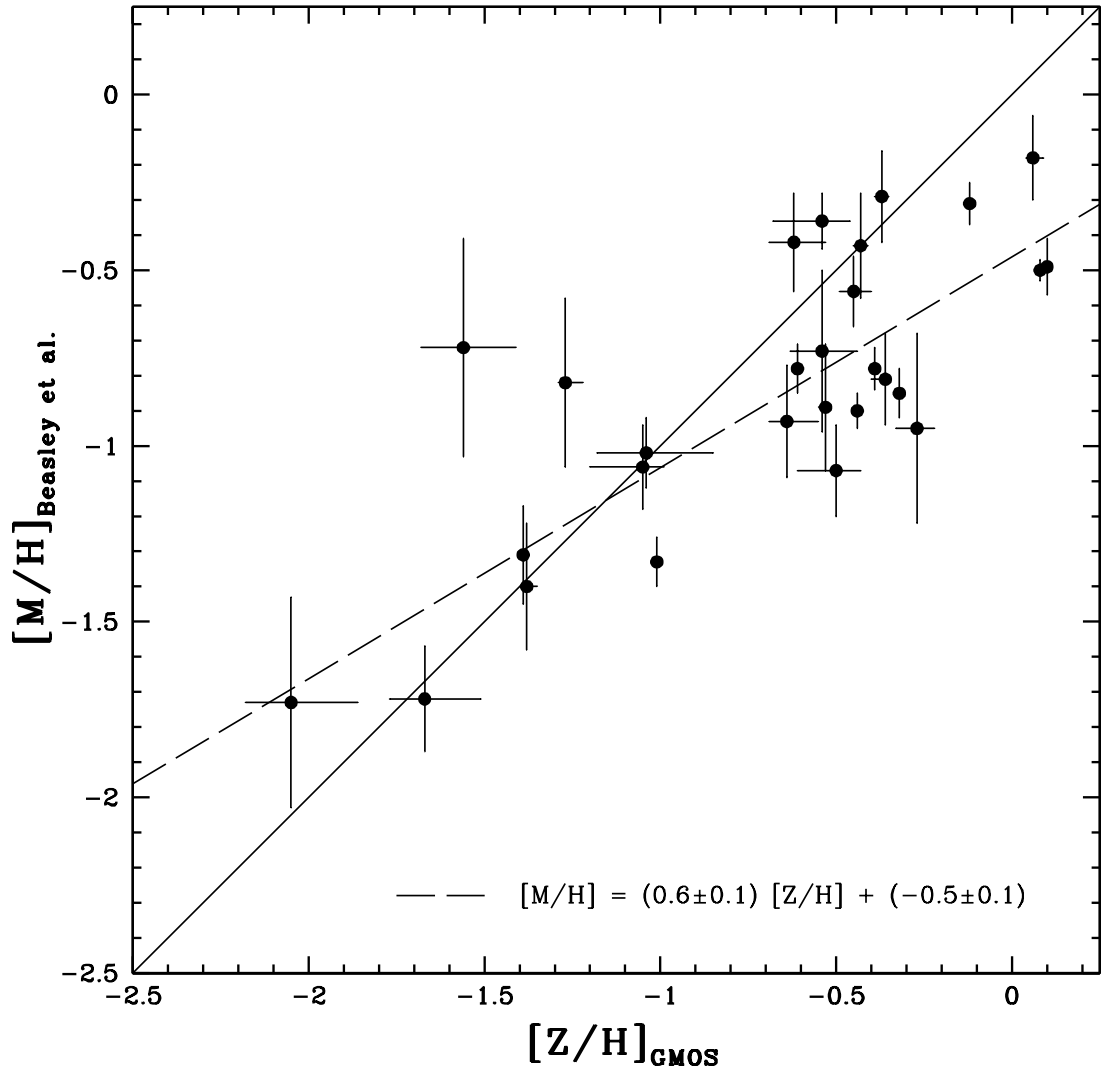}
\caption{Metallicity, [Z/H] derived from the SSP models, 
compared to empirically derived metallicities of GCs 
from \cite{beasley08}.  The 1:1 line is shown as a {\it solid line}.  The
  best fit line is shown as a {\it dashed line}  with a slope and
  intercept of $0.6\pm0.1$ and $-0.5\pm0.1$ and an rms$=0.25$.} 
\label{fig:mh_zh}
\end{figure}
       
\end{document}